\documentclass{aastex701}
\usepackage{graphicx}
\usepackage{txfonts}
\usepackage{caption}
\usepackage{stackengine}
\usepackage{hyperref}
\usepackage{color}
\hypersetup{colorlinks=true, linkcolor=blue, filecolor=blue, urlcolor=blue, citecolor=blue}
\usepackage{gensymb}
\usepackage{textgreek}
\usepackage{upgreek}
\begin{document}
	\title{Hard X-ray Sources in Fermi UFOs}

	\correspondingauthor{Alessandro Paggi}
	\author[0000-0002-5646-2410,gname='Alessandro',sname='Paggi']{Alessandro Paggi}
	\affiliation{Institute of Astrophysics, Foundation for Research and Technology - Hellas, Voutes, 7110, Heraklion, Greece}
	\affiliation{INAF-Istituto di Radioastronomia, via Gobetti 101, 40129, Bologna, Italy}	
	\email[show]{apaggi@ia.forth.gr}
	\author[0000-0001-9200-4006,gname='Ioannis',sname='Liodakis']{Ioannis Liodakis}
	\affiliation{Institute of Astrophysics, Foundation for Research and Technology - Hellas, Voutes, 7110, Heraklion, Greece}
	\email{liodakis@ia.forth.gr}
	\author[0000-0003-4453-3776,gname='John',sname='Antoniadis']{John Antoniadis}
	\affiliation{Institute of Astrophysics, Foundation for Research and Technology - Hellas, Voutes, 7110, Heraklion, Greece}
	\email{jantoniadis@ia.forth.gr}
	\author[0000-0002-4945-5079,gname='Chien-Ting',sname='Chen']{Chien-Ting Chen}
	\affiliation{Science and Technology Institute, Universities Space Research Association, Huntsville, AL 35805, USA}
	\email{chien-ting.chen@nasa.gov}
	\author[0000-0003-4420-2838,gname='Steven',sname='Ehlert']{Steven R. Ehlert}
	\affiliation{NASA Marshall Space Flight Center, Huntsville, AL 35812, USA}
	\email{steven.r.ehlert@nasa.gov}
	\author[0000-0003-3270-7644,gname='Daniel',sname='Gruen']{Daniel Gruen}
	\affiliation{University Observatory, Faculty of Physics, Ludwig-Maximilians-Universit\"{a}t, Scheinerstr. 1, 81679 Munich, Germany}
	\affiliation{Excellence Cluster ORIGINS, Boltzmannstr. 2, 85748 Garching, Germany}
	\email{daniel.gruen@lmu.de}
	\author[0000-0002-3638-0637,gname='Philip',sname='Kaaret']{Philip Kaaret}
	\affiliation{NASA Marshall Space Flight Center, Huntsville, AL 35812, USA}
	\email{philip.kaaret@nasa.gov}
	\author[0000-0001-6607-8933,gname='Ignacio',sname='De La Calle Perez']{Ignacio De La Calle Perez}
	\affiliation{Quasar Science Resources S.L. for the European Space Agency (ESA), European Space Astronomy Centre (ESAC), Camino Bajo del Castillo s/n, Villanueva de la
		Canada, 28692, Madrid, Spain}
	\email{Ignacio.de.la.Calle@ext.esa.int}
	\author[0000-0001-7920-4134,gname='Elena',sname='Jimenez Bailon']{Elena Jimenez Bailon}
	\affiliation{Instituto de Astronom\'{i}a, Universidad Nacional Aut\'{o}noma de M\'{e}xico}
	\affiliation{Quasar Science Resources S.L. for the European Space Agency (ESA), European Space Astronomy Centre (ESAC), Camino Bajo del Castillo s/n, Villanueva de la Canada, 28692, Madrid, Spain}
	\email{Elena.Jimenez.Bailon@ext.esa.int}
	\author[0009-0008-7675-4155,gname='Mykhailo',sname='Ilin']{Mykhailo Ilin}
	\affiliation{University Observatory, Faculty of Physics, Ludwig-Maximilians-Universit\"{a}t, Scheinerstr. 1, 81679 Munich, Germany}
	\email{m.ilin@campus.lmu.de}

	\begin{abstract}
	Identification and/or association of unidentified \(\gamma\)-ray sources with lower-energy counterparts represents a key challenge in modern astronomy, due to the relatively large positional uncertainty provided by \(\gamma\)-ray detectors.
	
	We selected Unidentified Fermi Objects (UFOs) positionally compatible with hard X-ray sources in the latest Palermo \textit{Swift}-BAT hard X-ray Catalog and in the SRG/ART-XC all-sky X-ray survey Catalog, to identify lower-energy sources and possibly associate them to the UFOs.
	
	We found 17 UFOs with overlapping hard X-ray sources. We then collected soft X-ray data from \textit{Swift}-XRT, \textit{Chandra}-ACIS, \textit{XMM-Newton}-EPIC, and eROSITA, identified 16 soft X-ray counterparts to the hard X-ray sources, and associate 15 with known astronomical objects, classified as: blazars/blazars candidates (2 sources), Seyfert galaxies (5 sources), X-ray binaries (2 sources), generic X-ray sources (1 source), cataclysmic variables (2 sources), and variable stars (3 sources).
	
	Blazars and Seyfert galaxies are likely lower-energy counterparts to the UFOs, since their mid-IR colors and broad-band spectral energy distributions suggest significant jetted, non-thermal emission. X-ray binaries can be potential lower-energy counterparts to the UFOs, since this class of sources has been already observed to emit \(\gamma\)-rays. The generic X-ray source has been proposed as a pulsar candidate, and we therefore suggest that it can be the lower-energy counterpart to the UFO. Cataclysmic variables have been suggested as potential \(\gamma\)-ray emitters so, if confirmed, the 2 sources classified as cataclysmic variables would represent the first \(\gamma\)-ray emitting sources of this kind. Finally, we consider the association of the 3 variable stars with the UFOs unlikely.
	\end{abstract}
	
	
	\section{Introduction}\label{sec:introduction}
	
	Since the launch in April 1991 of the  \textit{Compton Gamma Ray Observatory} (\textit{CGRO}) with its Energetic Gamma Ray Experiment Telescope \citep[EGRET;][]{1980ITNS...27..364H, 1988SSRv...49...69K, 1993ApJS...86..629T, 1999ApJS..123..203E}, one of the most important challenges has been the study of unidentified/unassociated \(\gamma\)-ray sources. In fact, about \(46 \%\) of the 188 sources listed in the revised EGRET catalog \citep{2008A&A...489..849C} are unassociated to lower-energy counterparts.
	
	The \textit{Fermi} Gamma-ray Space Telescope, launched in June 2008 dramatically increased the number of detected \(\gamma\)-ray sources. As a matter of fact, the latest release of the Large Area Telescope (LAT) 14-year Source Catalog \citep[4FGL-DR4,][]{2023arXiv230712546B, 2022ApJS..260...53A} lists 7194 \(\gamma\)-ray sources detected at significance \(>4\sigma\). However, \(\sim 34\%\) of the sources listed in 4FGL-DR4 catalog (2423 sources) are not yet identified or likely associated with a lower-energy counterpart. These Unidentified Fermi Objects (UFOs) are almost equally found at low (\(|b| < 15 \degree\)) (1352 sources, \(\sim 56\%\)) and high (\(|b| \geq 15 \degree\)) Galactic latitudes (1071 sources, \(\sim 44\%\)).
	
	In general, the \(\gamma\)-ray sky is dominated by blazars \citep{2010ApJS..188..405A}, a rare class of radio-loud active galactic nuclei (AGNs), launching relativistic jets closely aligned with our line of sight \citep{1978PhyS...17..265B, 2019ARA&A..57..467B, 2019NewAR..8701541H}. As a matter of fact, blazars (or blazar candidates) represent \(\sim 71\%\) of the high Galactic latitude 4FGL sources, and \(\sim 55\%\) of the whole catalog. Therefore, we expect UFOs outside the Galactic plane to be dominated by blazars, although a contribution from Galactic sources \citep[e.g.,][]{2001ASSL..267...17R} including pulsars \citep[e.g.,][]{2012ApJS..199...31N} is also expected. On the other hand, we expect the majority of low Galactic latitude UFOs to be associated with Galactic sources, like molecular clouds, supernova remnants, massive stars, OB associations, pulsars, transitional pulsars, pulsar wind nebulae, or X-ray binaries \citep[see e.g.,][]{1996ApJ...462L..35K, 1999APh....11..277G, 2010PASJ...62..769C, 2010ApJ...724L.207T, 2013Sci...339..807A, 2013A&A...553A..34D, 2013A&A...550A..89D}.

	Several attempts towards characterization and possibly identification/association of UFOs have been carried out, through observations in radio bands \citep[e.g.,][]{2010ApJ...718..587M, 2013MNRAS.432.1294P}, near infrared \citep[e.g.,][]{2005ApJ...631..163S, 2010ApJ...708..584E, 2010ApJ...717.1181P, 2012ApJ...753...30S, 2013ApJS..206...12D}, and soft X-rays \citep[e.g.,][]{2019ApJ...887...18K, 2021ApJ...908..177K}, as well as with statistical approaches \citep[e.g.][]{2012ApJ...753...83A} and machine learning methods \citep{2019MNRAS.486.3415L, 2023MNRAS.525.1731C, 2023ApJ...943..167K}. Recently, \citet{2024A&A...684A.208M} built a probabilistic catalog of possible pulsar and/or blazar type X-ray counterparts to UFOs from the internal eRASS:4 single-band source catalog, using a Bayesian cross-matching scheme.
	
	The hard X-ray (that is, at energies \(\gtrsim 10 \text{ keV}\)) sky has been observed since the launch of the first X-ray telescope Uhuru in 1972 \citep{1979ApJ...230..540G}, and subsequently by the High Energy Astronomy Observatory (HEAO) A4 detector on board of the HEAO-1 mission \citep{1979SSI.....4..269R}, the INTErnational Gamma-Ray Astrophysics Laboratory \citep[INTEGRAL,][]{2003A&A...411L...1W} and its Imager on Board the INTEGRAL Satellite \citep[IBIS][]{2003A&A...411L.131U}, the Burst Alert Telescope \citep[BAT,][]{2004SPIE.5165..175B} on board of the Neil Gehrels \textit{Swift} Observatory \citep{2004ApJ...611.1005G}, and the and the Mikhail Pavlinsky Astronomical Roentgen Telescope–X-ray Concentrator \citep[ART-XC,][]{2021A&A...650A..42P} on board of the Spektrum-Roentgen-Gamma (SRG) orbital observatory \citep{2021A&A...656A.132S}.
	
	Association studies of hard X-ray and gamma-ray in AGNs have been performed in the past focusing on Seyfer galaxies \citep{2011ApJ...742...66T, 2012ApJ...747..104A, 2025NatAs...9.1086L}, while a cross-match between the 105 month \textit{Swift}-BAT catalog \citep{2018ApJS..235....4O} and the second release of the fourth Fermi-LAT Source Catalog \citep[4FGL-DR2,][]{2020arXiv200511208B} has been performed by \citet{2021ApJ...916...28T} finding 156 matches for point-like source and 31 extended sources.

	In this work we focus on UFOs with one (or more) hard X-ray counterparts listed in the 100-month Palermo BAT Catalog \citep[4PBC,][]{2014styd.confE.132C} and/or in SRG/ART-XC all-sky X-ray survey Catalog \citep{2024A&A...687A.183S}. We refine the position of these X-ray counterparts (with hard X-ray positional errors ranging from \(\sim 15\arcsec\) to a few arcminutes) making use of soft X-ray data either through archival pointed observations or through the SRG/eROSITA all-sky survey (with positional errors of a few arcseconds), and collect additional multi-frequency data for these sources in the attempt to characterize them and identify possible lower-energy counterparts to the UFOs.

	The paper is organized as follows: Sect. \ref{sec:sample_selection} is devoted to the UFO sample definition, Sect. \ref{sec:xray} describes the main data reduction procedure adopted for the \textit{Swift}-XRT, \textit{Chandra}-ACIS and \textit{XMM-Newton}-EPIC observations. In Sect. \ref{sec:sources} we illustrate the selection of soft X-ray counterparts to hard X-ray source, in Sect. \ref{sec:multilambda} we briefly discuss the collection of additional, multi-wavelength data for our selected sources, while in Sect. \ref{sec:discussion} we discuss our results and compare them with different, previous selections. Finally, Sect. \ref{sec:conclusion} is dedicated to our conclusions.

	\section{Sample selection}\label{sec:sample_selection}
	
	For this work we made use of the latest release of Fermi-LAT Source Catalog \citep[4FGL-DR4,][]{2023arXiv230712546B, 2022ApJS..260...53A}, the fourth and latest Palermo \textit{Swift}-BAT hard X-ray Catalog \citep[4PBC,][]{2014styd.confE.132C}, and the SRG/ART-XC all-sky X-ray survey Catalog \citep{2024A&A...687A.183S}.

	The 4FGL-DR4 is based on the first 14 years of all-sky Fermi-LAT data, and lists 7195 \(\gamma\)-ray sources in the \(50 \text{ MeV} - 1 \text{ TeV}\) energy range, with properties including coordinates, \(95\%\) uncertainty ellipses, fluxes in different energy bands, etc. The 4PBC covers \(\sim 50\%\) of the sky down to a \(15-150 \text{ keV}\) flux limit of \(\sim 5.4\times {10}^{-12} \text{ erg} \text{ cm}^{-2} \text{ s}^{-1}\). This catalog includes 1710 hard X-ray sources with properties including coordinates, \(95\%\) error radius, \(15-150 \text{ keV}\) flux and power-law slope. The SRG/ART-XC catalog lists sources detected in the \(4-12 \text{ keV}\) energy band down to a flux limit of \(\sim 4\times {10}^{-12} \text{ erg} \text{ s}^{-1} \text{ cm}^{-2}\) near the ecliptic plane and \(\sim 7\times {10}^{-13} \text{ erg} \text{ s}^{-1} \text{ cm}^{-2}\) near the ecliptic poles. This catalog includes 1545 hard X-ray sources with properties including coordinates, \(98\%\) error radius, and \(4-12 \text{ keV}\) flux.

	Since hard X-ray and \(\gamma\)-ray bands are very close, one would intuitively assume these emissions to be correlated. However, to quantitatively assess the reliability of a tentative association of \(\gamma\)-ray sources with lower energy counterparts through 4PBC and SRG/ART-XC hard X-ray catalogs, we first cross-matched the latter with the associated/identified sources in 4FGL-DR4 (excluding extended sources). To do this we used of the associated/identified 4FGL counterpart coordinates (\textsc{RA\_Counterpart} and \textsc{DEC\_Counterpart} columns in 4FGL-DR4) and their uncertainties (\textsc{Unc\_Counterpart} column in 4FGL-DR4) together with the hard X-ray sources coordinates and uncertainties as listed in the respective hard X-ray catalog. For the sake of uniformity, in the following we will consider uncertainty radii at \(99\%\) confidence level for all catalogs.
	
	We obtained 167 and 188 matches for 4PBC and SRG/ART-XC catalog, respectively. We then checked which hard X-ray sources were associated in the respective hard X-ray catalog with the same low-energy counterpart associated to the 4FGL source in 4FGL-DR4, finding that this happens for 151 (\(\sim 90\%\)) and 180 (\(\sim 96\%\)) sources in 4PBC and SRG/ART-XC catalog, respectively. We therefore conservatively estimate the hard X-ray selection to have a reliability of correctly pin-point the lower-energy counterpart of a \(\gamma\)-ray source of \(\sim 90\%\)  and \(\sim 96\%\) when using 4PBC and SRG/ART-XC catalog, respectively. In the following we refer to these samples of 151 and 180 \(\gamma\)-ray/hard X-ray sources as cross-match samples.
	
	Then, we investigate possible selection effects associated to a tentative association of \(\gamma\)-ray sources with lower energy counterparts through hard X-ray catalogs. We note that for associated/identified sources in 4FGL-DR4, high Galactic latitude (\(|b| > 15 \degree\)) sources are dominated by blazars (\(\sim 93.2\%\)), with a small contribution from pulsars (\(\sim 3.2\%\)). For low Galactic latitude (\(|b| < 15 \degree\)) associated sources in 4FGL-DR4 we still have a major contribution from blazars (\(\sim 53.8\%\)), followed by pulsars (\(\sim 17.4\%\)). Considering the 4FGL-DR4 4PBC and 4FGL-DR4 SRG/ART-XC cross-match samples we obtain similar percentages, with blazars and pulsars representing \(\sim 90\%\) and \(\sim 1.2 \%\) of the high Galactic latitude sources, respectively, while blazars and pulsars represent \(\sim 48.8\%\) and \(\sim 12.8 \%\) of the low Galactic latitude sources, respectively. We therefore do not expect tentative association of \(\gamma\)-ray sources with lower energy counterparts through hard X-ray catalogs to introduce significant selection effects with respect to specific source classes.
	
\begin{figure*}
		\centering
		\includegraphics[scale=0.5]{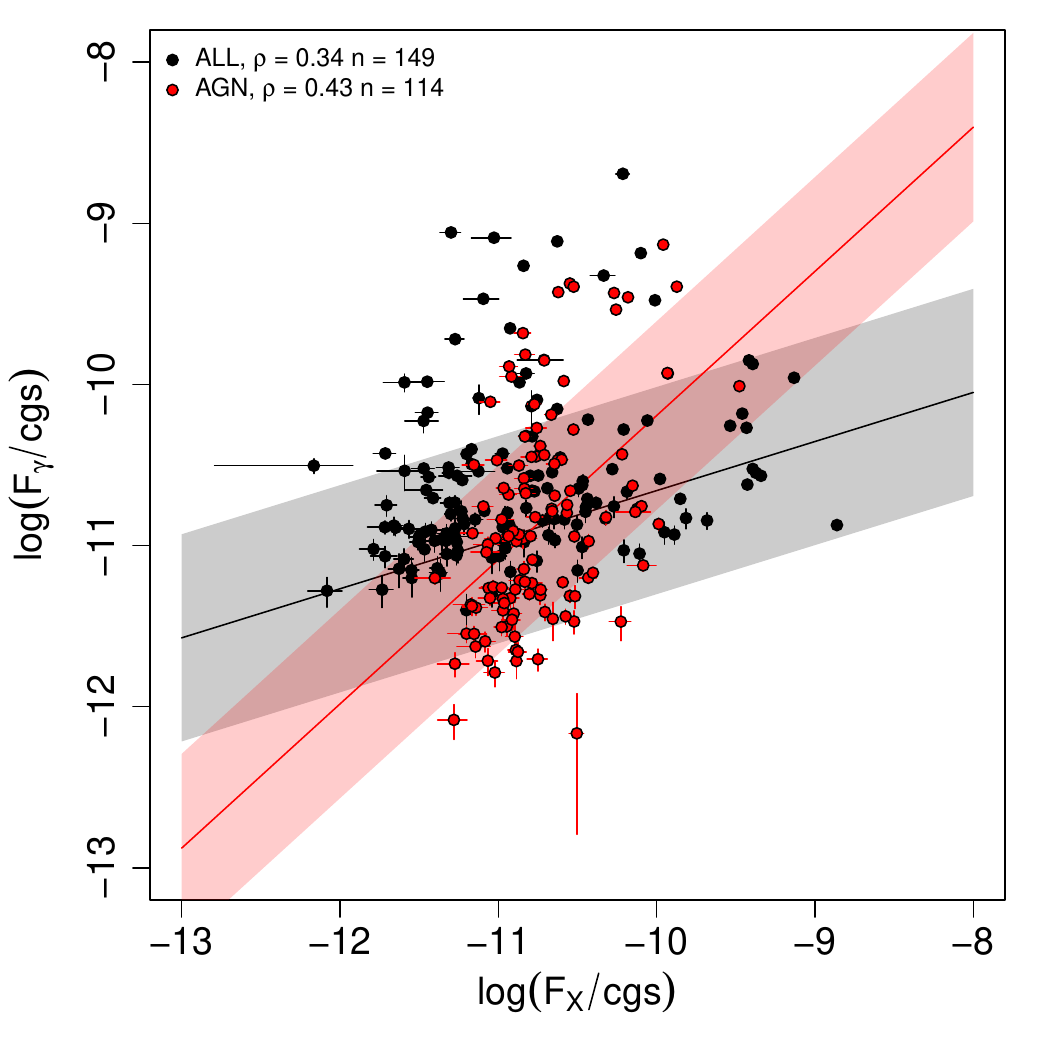}
		\includegraphics[scale=0.5]{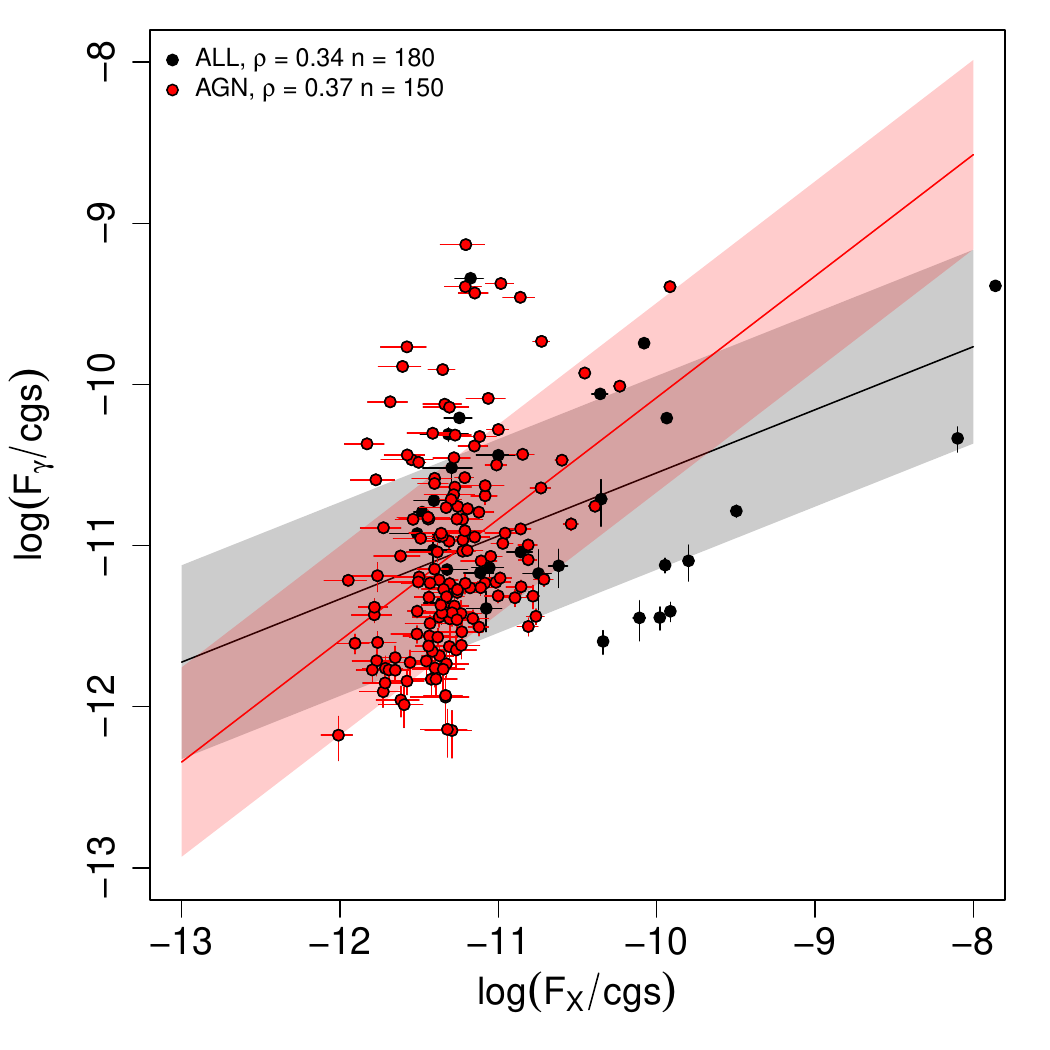}
		\caption{Double logarithmic plot of \(100 \text{ MeV} - 100 \text{ GeV}\) \(\gamma\)-ray flux against \(15-150 \text{ keV}\) hard X-ray flux (left panel) for the 4FGL-DR4 4PBC cross-match sample, and against \(4-12 \text{ keV}\) hard X-ray flux (right panel) for the 4FGL-DR4 SRG/ART-XC cross-match sample (see Sect \ref{sec:sample_selection}). The full samples are represented with black circles, while sources associated to AGNs are represented with red circles. The full black and red lines represent the best linear fit, while light black and red areas represent \(1-\sigma\) scatter around such best fit. Spearman rank correlation coefficients \(\rho\) and sample size (\(n\)) are indicated in the legend.}\label{fig:hardx_vs_gamma}
	\end{figure*}

	We also investigated the connection between emission in \(\gamma\)-rays  and hard X-rays in associated/identified sources in 4FGL-DR4. We note that 2 sources in the 4FGL-DR4 4PBC cross-match sample do not have a reliable hard X-ray flux estimate and were therefore excluded from this analysis. In Fig. \ref{fig:hardx_vs_gamma} we present the \(100 \text{ MeV} - 100 \text{ GeV}\) \(\gamma\)-ray flux (\textsc{Energy\_Flux100} column in 4FGL-DR4) versus the \(15-150 \text{ keV}\) flux for the 4FGL-DR4 4PBC cross-match sample (left panel) and versus the \(4-12 \text{ keV}\) flux for the 4FGL-DR4 SRG/ART-XC cross-match sample (right panel). In particular we see that hard X-ray and \(\gamma\)-ray fluxes appear to be correlated with a Spearman rank correlation coefficient of 0.34 both for 4FGL-DR4 4PBC and 4FGL-DR4 SRG/ART-XC cross-match samples. When considering only sources associated to AGNs, the correlation coefficients are 0.43 and 0.37 for the 4FGL-DR4 4PBC and 4FGL-DR4 SRG/ART-XC cross-match sample, respectively. Similar correlation coefficient have been found for the connection between radio flux density and \(\gamma\)-ray flux \citep{2011ApJ...741...30A}.
	
	We then proceeded to look for hard X-ray counterparts to unidentified \(\gamma\)-ray sources. As mentioned in Sect. \ref{sec:introduction}, 2423 sources in 4FGL-DR4 lack an identified or likely associated lower-energy counterpart, and are therefore classified as UFOs. We then cross-matched the UFOs in 4FGL-DR4 with sources in 4PCB and SRG/ART-XC catalogs to find hard X-ray sources overlapping (considering their \(99\%\) positional uncertainties) with the UFO \(99\%\) uncertainty ellipses.
	
	The result of this cross-matching is presented in Tables \ref{tab:bat} and \ref{tab:artxc}. In total, we find 17 UFOs with at least one overlapping hard X-ray source. In particular, we find 7 UFOs with one overlapping 4PCB source, 6 UFOs with one overlapping SRG/ART-XC source, 3 UFOs with both one overlapping 4PCB source and one overlapping SRG/ART-XC source, and one UFO with two overlapping 4PCB sources and one overlapping SRG/ART-XC source.

	\section{X-ray Data}\label{sec:xray}
	
	In this section we aim at finding soft X-ray sources lying in the overlapping regions of UFOs and hard X-ray sources uncertainty regions. To exploit the most recent X-ray data available, we analyze archival observations from \textit{Swift}-XRT, \textit{Chandra}-ACIS and \textit{XMM-Newton}-EPIC covering the UFO uncertainty ellipses.
	
	\subsection{X-ray Data Reduction}\label{sec:xray_reduction}
	We briefly describe here the data reduction procedures adopted for this work. More information can be found in \citet{2020A&A...641A..62P}.
	
	Photon counting (PC) mode data for the X-ray Telescope \citep[XRT,][]{2005SSRv..120..165B} on board of the Neil Gehrels \textit{Swift} Observatory were downloaded from HEASARC\footnote{\href{https://heasarc.gsfc.nasa.gov/}{https://heasarc.gsfc.nasa.gov/}} data archive. Observations were processed and reduced with the \textsc{XRTDAS} software \citep{capalbi2005} included in the HEAsoft package (v. 6.33.2). The XRT CCD edges were monitored to excluded time intervals with count-rate \(>40 \text{ counts/s}\) \citep{2011A&A...528A.122P}. In addition, we selected time intervals for which the CCD temperatures was \(<-50\degree\text{ C}\) \citep{2013A&A...551A.142D}. Using the \textsc{xrtpipeline} task (ver. 0.13.7), we produced calibrated and cleaned \(0.3-10 \text{ keV}\) PC mode event files and exposure maps for each observation.
	
	Pointing observations for the Advanced CCD Imaging Spectrometer \citep[ACIS,][]{1987ApL&C..26...35N} on board of the \textit{Chandra} X-ray Observatory were retrieved from the \textit{Chandra} Data Archive\footnote{\href{http://cda.harvard.edu/chaser}{http://cda.harvard.edu/chaser}}. Data were processed and reduced with the \textit{Chandra} Interactive Analysis of Observations \citep[CIAO,][]{2006SPIE.6270E..1VF} data analysis system (v. 4.16) and the \textit{Chandra} calibration database CALDB version 4.11.1, adopting standard procedures. Up-to-date calibrations were applied through \textsc{chandra\_repro} task, and time intervals of high background (i.e., exceeding \(3\) \(\sigma\) from the average level) were excluded with the \textsc{deflare} task. For each observation, we produced \(0.3-7 \text{ keV}\) full-band images, as well as exposure, psf, and pileup maps.
	
	Data for the European Photon Imaging Camera \citep[EPIC,][]{2001A&A...365L..18S, 2001A&A...365L..27T} on board of the \textit{XMM-Newton} space telescope were retrieved from the \textit{XMM-Newton} Science Archive\footnote{\href{http://nxsa.esac.esa.int/nxsa-web}{http://nxsa.esac.esa.int/nxsa-web}}. Data were reduced with the Science Analysis System (\textsc{SAS}\footnote{\href{http://www.cosmos.esa.int/web/xmm-newton/sas}{http://www.cosmos.esa.int/web/xmm-newton/sas}}) software (v. 21.0.0). Following \citet{2005ApJ...629..172N}, we excluded time intervals where both the high-energy count-rate (i.e., \(9.5-12 \text{ keV}\) for MOS1 and MOS2, and \(10-12 \text{ keV}\) for PN) evaluated on the whole EPIC detector, and the soft-energy count-rate (i.e., \(1-5 \text{ keV}\)) evaluated at the detector edge (that is, between \(12\arcmin\) and \(14\arcmin\) from the detector center) were more than 3\(\sigma\) away from their averages. We then produced full-band \(0.3-10 \text{ keV}\) images and exposure maps for all available detector in each observation.

	\subsection{Source Detection and Flux estimates}\label{sec:src}
	
	When multiple observations were available for one UFO, we merged them to detect the fainter sources that would not be detected otherwise, using the \textsc{xselect}, \textsc{flux\_obs} and \textsc{edetect\_stack} tasks for \textit{Swift}-XRT, \textit{Chandra}-ACIS and \textit{XMM-Newton}-EPIC data, respectively.
	
	To detect X-ray sources in the \(0.3-10 \text{ keV}\) \textit{Swift}-XRT images, we made use of the \textsc{ximage} detection algorithm \textsc{detect}. For the \(0.3-7 \text{ keV}\) \textit{Chandra}-ACIS images we run the \textsc{wavdetect} task with a \(\sqrt{2}\) sequence of wavelet scales (i.e., 1, 1.41, 2, 2.83, 4, 5.66, 8, 11.31, 16 pixels) and a false-positive probability threshold of \({10}^{-6}\). For the \(0.3-10 \text{ keV}\) \textit{XMM-Newton}-EPIC images we used the standard SAS sliding box task \textsc{edetect\_chain}. Along the source detection processes we also evaluated the \(99\%\) positional uncertainty (including telescope systematic astrometric uncertainty) for each detected source.
	
	We evaluated source and background counts for each detected source using circular  and annular regions centered on the source coordinates (excluding other detected X-ray sources). We then produced appropriate Auxiliary Response Files (ARFs) and Redistribution Matrix Files (RMFs) at each source location for each observation, making use of the \textsc{xrtproducts}, \textsc{specextract}, \textsc{arfgen} and \textsc{rmfgen} for \textit{Swift}-XRT, \textit{Chandra}-ACIS and \textit{XMM-Newton}-EPIC data, respectively.
	
	In order to get a first estimate of fluxes for X-ray sources, we converted the observed net count-rates to \(0.3-10 \text{ keV}\) intrinsic (i.e. unabsorbed) fluxes assuming a standard power-law model with a slope of \(1.8\) (as appropriate for AGNs, see for example \citealt{2006A&A...451..457T}) and an absorption component fixed to the Galactic value \citep{2005A&A...440..775K} using the appropriate ARF and RMF files generated before. For \textit{XMM-Newton}-EPIC data, the source flux was evaluated as the mean of the fluxes of each available detector weighted by its uncertainty \citep{2020A&A...641A.136W}. When a source is not detected in an observation with a minimum significance of \(3\sigma\) level, we evaluate its \(3\sigma\) upper limit.

	\subsection{Spectral Extraction and Fitting}\label{sec:spectra}

	To obtain better estimates on the X-ray source fluxes, we extracted spectra for the X-ray sources for each available observation.
	
	To avoid pile-up contamination, we excluded the inner part of spectra extraction regions as follows. For \textit{Swift}-XRT data, we excluded regions of high pile-up by comparing the observed profiles with the updated XRT PSF analytical model\footnote{\href{https://www.swift.ac.uk/analysis/xrt/SWIFT-XRT-CALDB-10_v01.pdf}{https://www.swift.ac.uk/analysis/xrt/SWIFT-XRT-CALDB-10\_v01.pdf}}. For \textit{Chandra}-ACIS data, we excluded inner pixels with pileup larger than \(5\%\) as estimated from the pileup maps. For \textit{XMM-Newton}-EPIC data, regions of high pileup were estimated and excluded by means of the \textit{epatplot} task following the procedure explained in the SAS Data Analysis Threads\footnote{\href{https://www.cosmos.esa.int/web/xmm-newton/sas-thread-epatplot}{https://www.cosmos.esa.int/web/xmm-newton/sas-thread-epatplot}}.

	Spectral fitting was performed with the Sherpa\footnote{\href{http://cxc.harvard.edu/sherpa}{http://cxc.harvard.edu/sherpa}} modeling and fitting application \citep{2001SPIE.4477...76F} in the \(0.3-10\text{ keV}\) energy range for \textit{Swift}-XRT and \textit{XMM-Newton}-EPIC spectra, and in the \(0.3-7\text{ keV}\) energy range for \textit{Chandra}-ACIS spectra, adopting Gehrels weighting \citep{1986ApJ...303..336G}. Source spectra were binned to a minimum of 20 counts/bin to ensure the validity of \(\chi^2\) statistics. For the EPIC spectra we fitted simultaneously MOS1, MOS2 and PN spectra, if available.

	For the spectral fitting we used three different models:
	\begin{enumerate}
		\item a model comprising an absorption component fixed to the Galactic value and a power-law with slope \(a\)
		\item a model comprising an absorption component fixed to the Galactic value and a log-parabola with slope \(a\) and curvature \(b\)
		\item a model comprising an absorption component fixed to the Galactic value and two power-laws with slopes \(a\) and \(a_2\)        
	\end{enumerate}
	We then considered only models that provided a fit with a reduced \(\chi^2<1.3\) and constrained parameters. If more than one model satisfied these conditions, we selected the best fit model making use of the Akaike information criterion \citep[AIC, see for example][]{2007MNRAS.377L..74L}.

	\subsection{eROSITA Catalog}\label{sec:erass}
	
	In addition to data available from observations of \textit{Swift}-XRT, \textit{Chandra}-ACIS and \textit{XMM-Newton}-EPIC, we looked for X-ray counterparts to UFO sources in the 
	extended ROentgen Survey with an Imaging Telescope Array \citep[eROSITA,][]{2021A&A...647A...1P} main catalog \citep[eRASS1,][]{2024A&A...682A..34M}.

	The fluxes reported in eRASS1 are derived from the count-rates assuming a model comprising a power-law with slope 2.0 and a Galactic absorption of \(3 \times {10}^{20} \text{ cm}^{-2}\) \citep{2024A&A...682A..34M}. We therefore converted the \(0.2-5 \text{ keV}\) count-rates reported in the catalog to \(0.3-10 \text{ keV}\) intrinsic fluxes using PIMMS, again assuming a model comprising an absorption component fixed to the Galactic value and a power-law with a 1.8 slope. For each source we also evaluated its \(99\%\) positional uncertainty.

	\section{Results}\label{sec:results}

	\subsection{Counterpart Selection}\label{sec:sources}
	
\begin{figure*}
		\centering
		\includegraphics[scale=0.35]{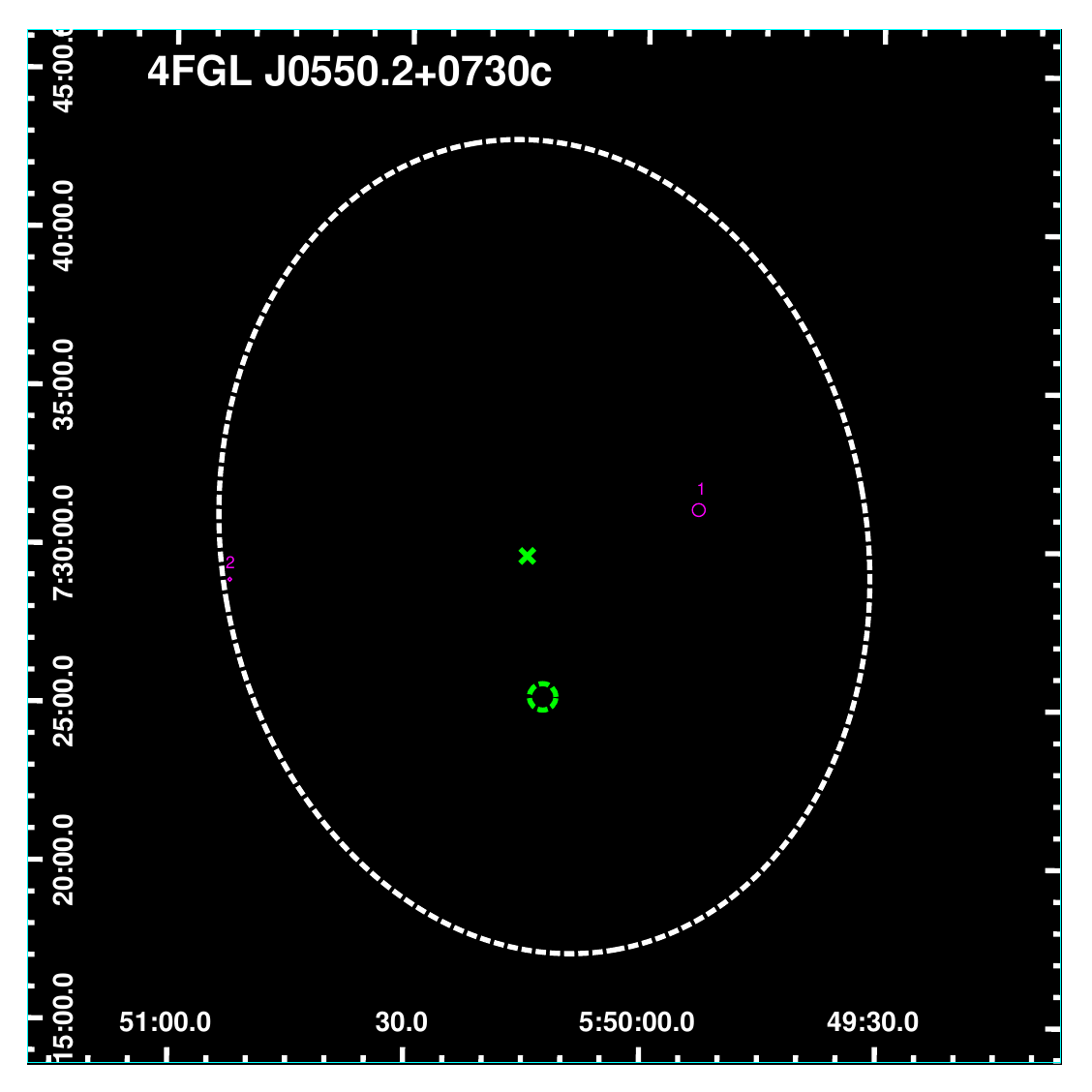}\\
		\includegraphics[scale=0.35]{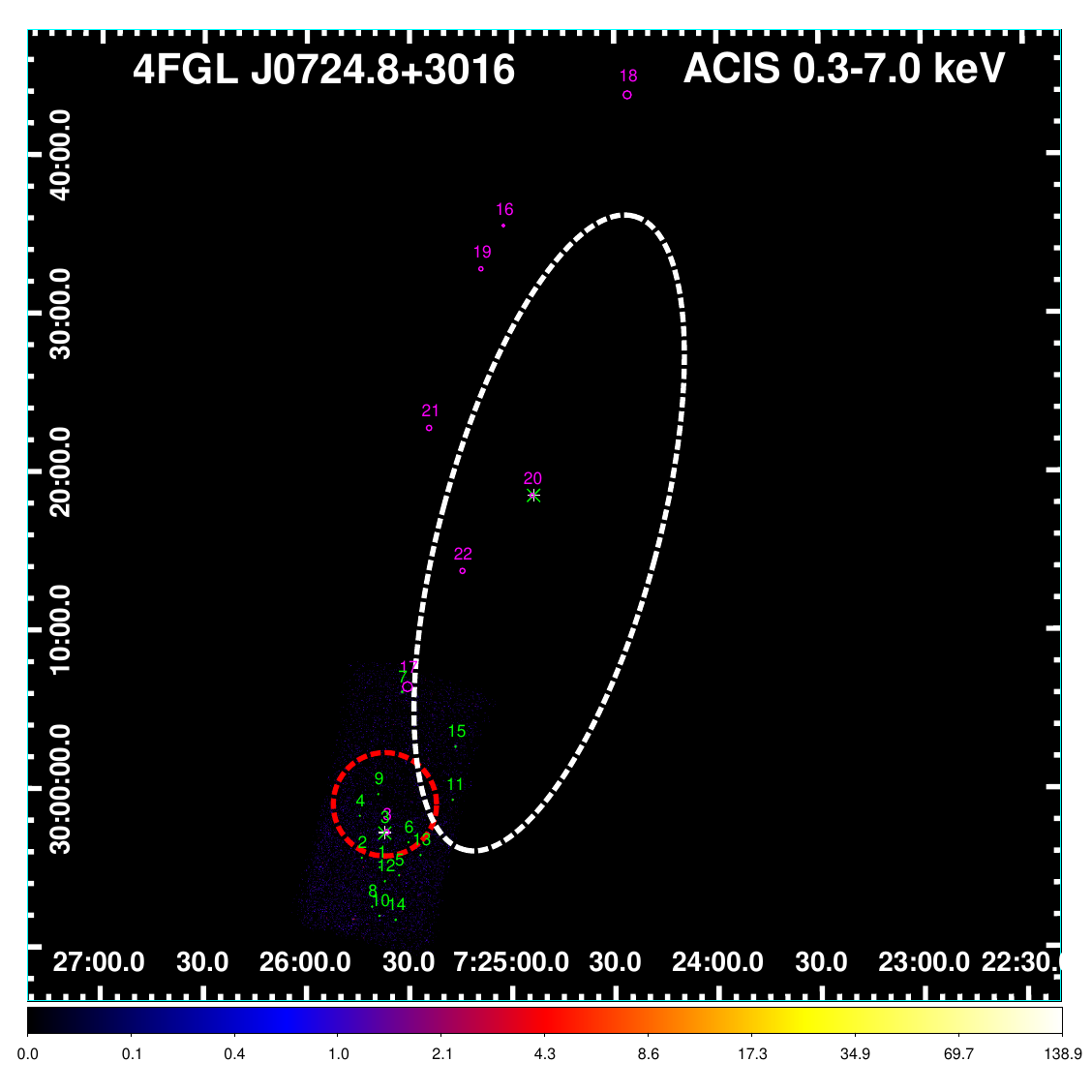}
		\includegraphics[scale=0.35]{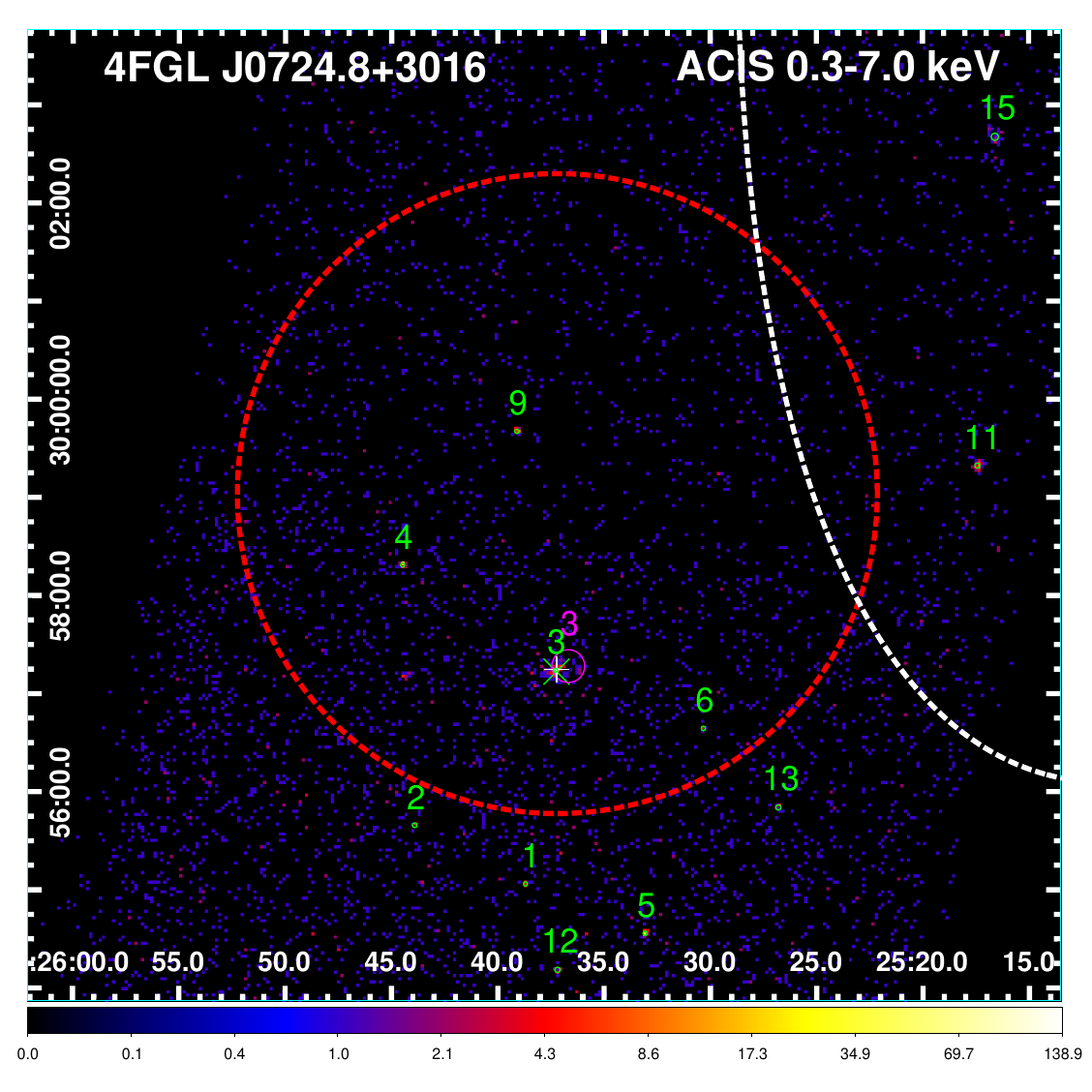}\\
		\includegraphics[scale=0.35]{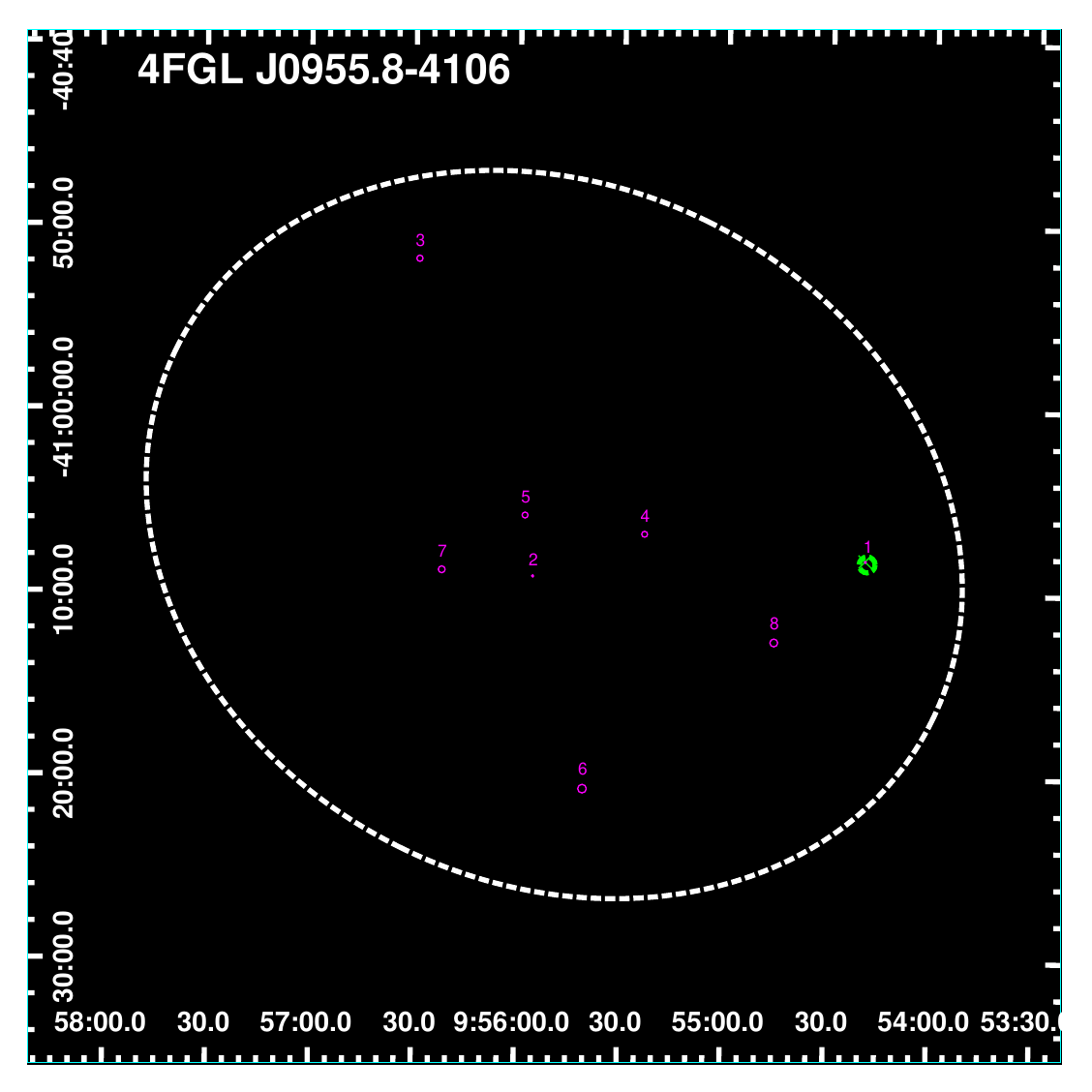}
		\includegraphics[scale=0.35]{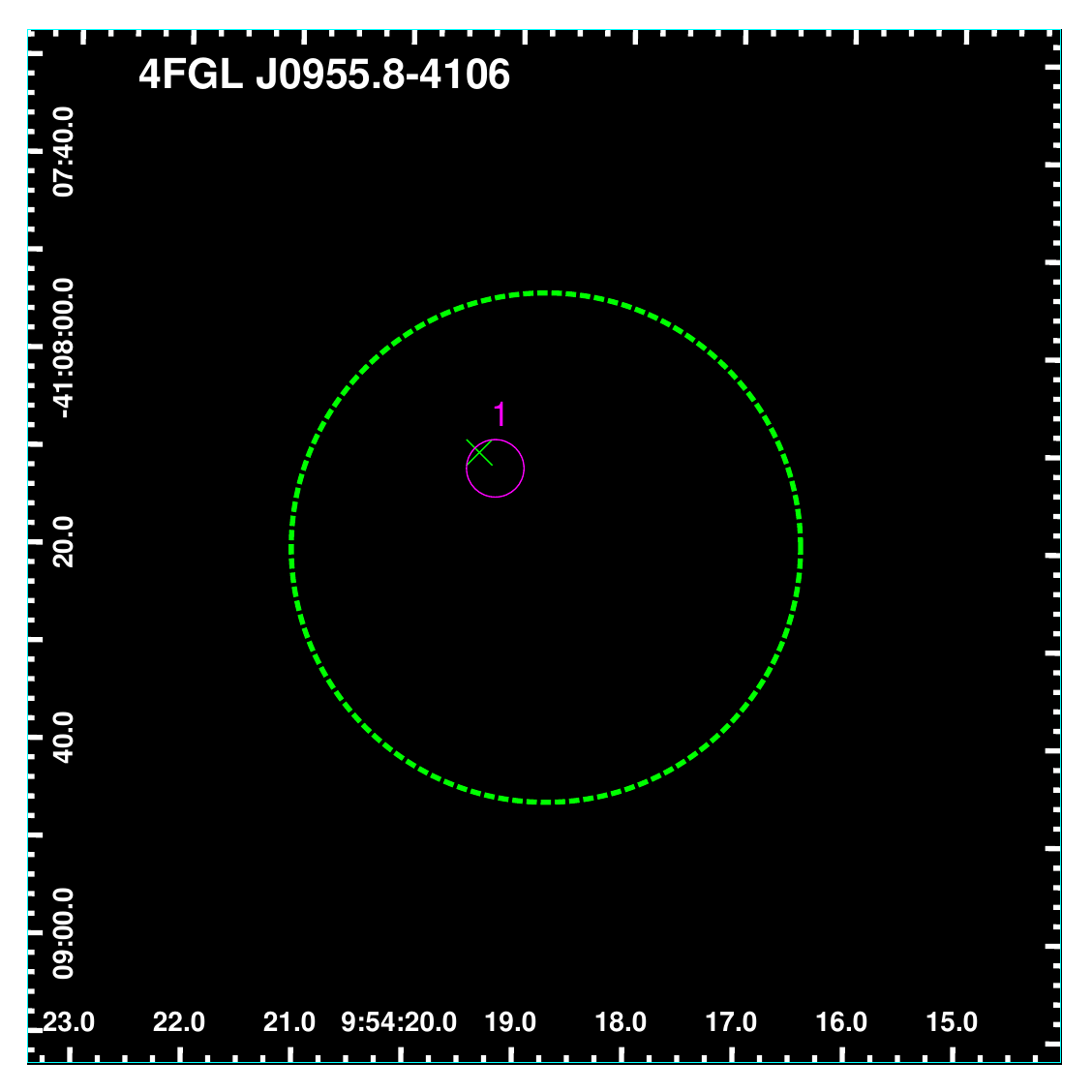}\\        
		\caption{Soft X-ray maps of the UFOs listed in Tables \ref{tab:bat} and \ref{tab:artxc}. The leftmost panel for each row shows the whole LAT uncertainty region, while right panels show a zoom-in of the hard X-ray uncertainty regions. Dashed white ellipses represent \(99\%\) LAT uncertainty regions, red and green dashed circles represent \(99\%\) BAT and SRG/ART-XC uncertainty regions, respectively. \textit{Swift}-XRT, \textit{Chandra}-ACIS, \textit{XMM-Newton}-EPIC and eRASS1 catalog sources are indicated with white, green, red and magenta circles, respectively, with radii indicating the respective \(99\%\) positional uncertainty. Thick green xs indicate the sources selected by \citet{2024A&A...684A.208M} as possible pulsar and/or blazar type X-ray counterparts to the UFO. Note that this zoom-in panel is not shown for 4FGL J0550.2+0730c, since for this source there are no soft X-ray sources compatible with the hard X-ray source position. Radio counterparts to soft X-ray sources are marked with thin white crosses and green Xs for VLASS and RACS sources, respectively. For sources 4FGL J0550.2+0730c and 4FGL J0955.8-4106 no colorbar is shown since we have no available soft X-ray data.}\label{fig:xraymap}
	\end{figure*}

	Detected soft X-ray sources are presented in Fig. \ref{fig:xraymap}. In this figure we present for each selected UFO its soft X-ray map with superimposed as small circles the soft X-ray sources detected in the field, with radii indicating the respective \(99\%\) positional uncertainty. When sources detected by different telescopes or found in eRASS1 catalog are positionally compatible within their \(99\%\) positional uncertainty, they are marked with the same number.
	
	As shown in Fig. \ref{fig:xraymap}, in 10 cases we find only one soft X-ray source positionally compatible with the hard X-ray source, and we therefore select this as the soft X-ray counterpart to the hard X-ray source. For 4FGL J1407.7-3017 and 4FGL J2109.3+3531 we have several spurious \textit{Swift}-XRT detections likely due to the bright wings of the detector PSF, and therefore in this cases we ignore these spurious sources. For 4FGL J0724.8+3016, 4FGL J1616.6-5009 and 4FGL J1817.7-2517 we have more than one soft X-ray source positionally compatible with the hard X-ray source, and in these cases we select the source detected by more than one instrument as the soft X-ray counterpart to the hard X-ray source. For source 4FGL J1110.3-6501 we have two sources detected by more than one instrument positionally compatible with the hard X-ray source, namely source 1 (detected by \textit{Swift}-XRT, \textit{Chandra}-ACIS, \textit{XMM-Newton}-EPIC and eROSITA) and source 3 (detected by \textit{Swift}-XRT and \textit{Chandra}-ACIS) as show in Fig. \ref{fig:xraymap}. We note that source 3 positionally compatible with source 4eRASS J110947.3-645931 listed in \citet{2024A&A...684A.208M} as possible pulsar and/or blazar type X-ray counterparts to the UFO. However the combined posterior Bayesian probabilities for this source to be the counterpart of the \(\gamma\)-ray source and to be a pulsar or a blazar are \(\sim 0.026\) and 0.020, respectively, and its combined posterior probability for this source to be matched with the UFO is \(\sim 0.046\). We therefore select source 1 as the soft X-ray counterpart to the hard X-ray source. Finally, for 4FGL J0550.2+0730c we have no soft X-ray sources positionally compatible with the hard X-ray source, so in this case we select no soft X-ray counterpart.

	In Table \ref{tab:softx} we summarize the soft X-ray sources that we tentatively associate to the hard X-ray sources. In particular, column \(p_{\theta}\) of this Table indicates the confidence level corresponding to the angular separation \(\theta\) between the soft X-ray counterpart and the UFO coordinates:
	\begin{equation}
		p_{\theta}=1-{\left({1-\frac{95}{100}}\right)}^{
			\theta^2 \left[{
				\frac{\cos^2{\left({PA-{PA}_{95}}\right)}}{a_{95}^2} +
				\frac{\sin^2{\left({PA-{PA}_{95}}\right)}}{b_{95}^2}
				}\right]
			}\,,
	\end{equation}
	where \(a_{95}\), \(b_{95}\) and \(PA_{95}\) are the \(95\%\) UFO LAT uncertainty ellipse semi-major axis, semi-minor axis, and position angle, respectively, and \(PA\) is the soft X-ray counterpart position angle.
	
	The X-ray fluxes and fit results for the potential soft X-ray counterparts to the UFOs are presented in Tables. \ref{tab:swift}, \ref{tab:chandra}, \ref{tab:xmm} and \ref{tab:erass} for \textit{Swift}-XRT, \textit{Chandra}-ACIS, \textit{XMM-Newton}-EPIC and eROSITA, respectively. In particular, in the column \(F_{0.3-10.0 \text{ keV}}\) of tables \ref{tab:swift}, \ref{tab:chandra} and \ref{tab:xmm} we report the flux evaluated from the best fit model (see Sect. \ref{sec:spectra}) or, if no model provided a good enough fit, we report the flux evaluated from the net count-rate assuming a power-law model with a slope of \(1.8\) and an absorption component fixed to the Galactic value (see Sect. \ref{sec:src}).

	\subsection{Multi-Wavelength Data}\label{sec:multilambda}
	
	In order to obtain a better characterization of the selected soft X-ray sources we collected multi-wavelength data for the soft x-ray sources listed in Table \ref{tab:softx}.
	
	First of all, using the coordinates and positional uncertainties listed in Table \ref{tab:softx} we looked into the SIMBAD Astronomical Database\footnote{\href{https://simbad.cds.unistra.fr/simbad/}{https://simbad.cds.unistra.fr/simbad/}} for known lower-energy counterparts, and listed them along with their source type in Table \ref{tab:softx}.
	
	Then we collected available radio data making use of the SSDC SED Builder\footnote{\href{https://tools.ssdc.asi.it/SED/}{https://tools.ssdc.asi.it/SED/}}.
	
\begin{figure*}
		\centering
		\includegraphics[scale=0.37]{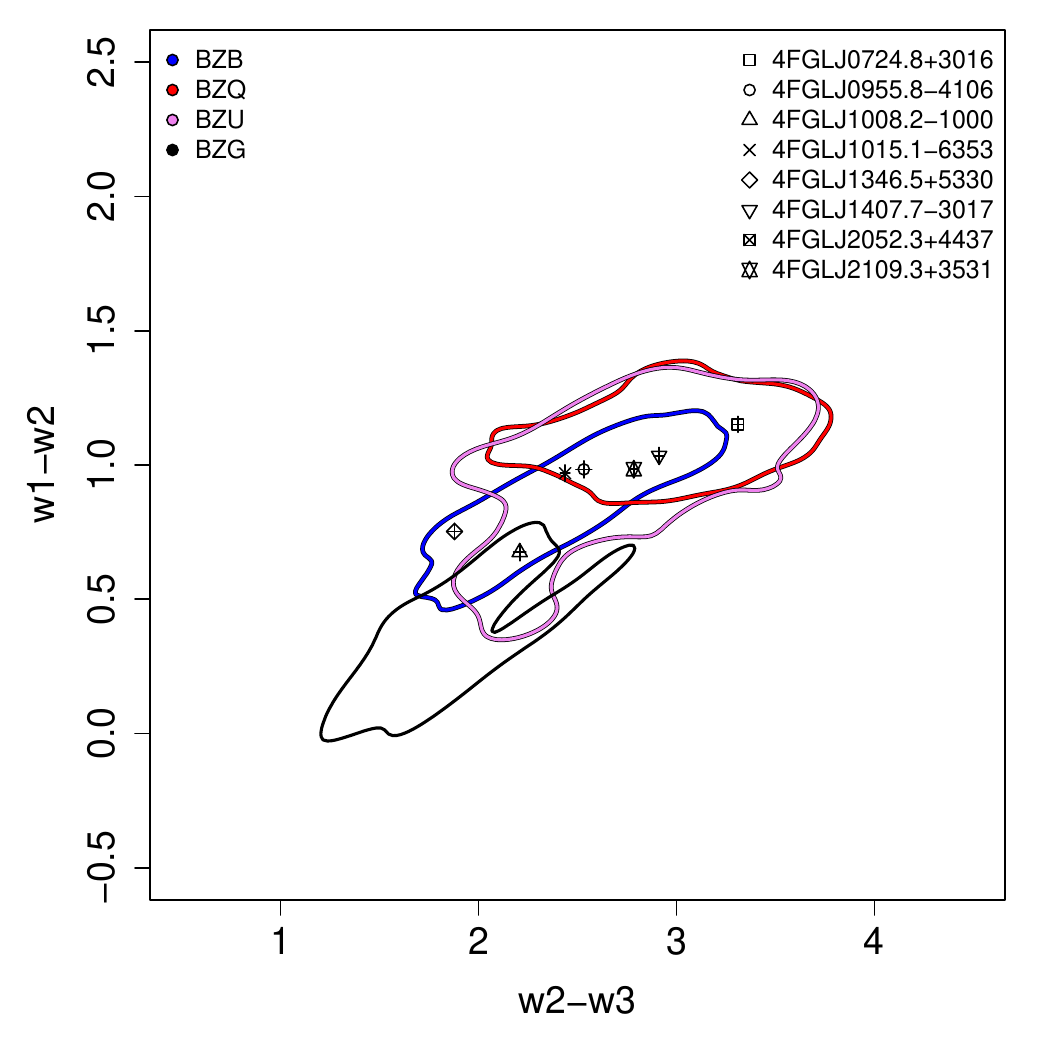}
		\includegraphics[scale=0.37]{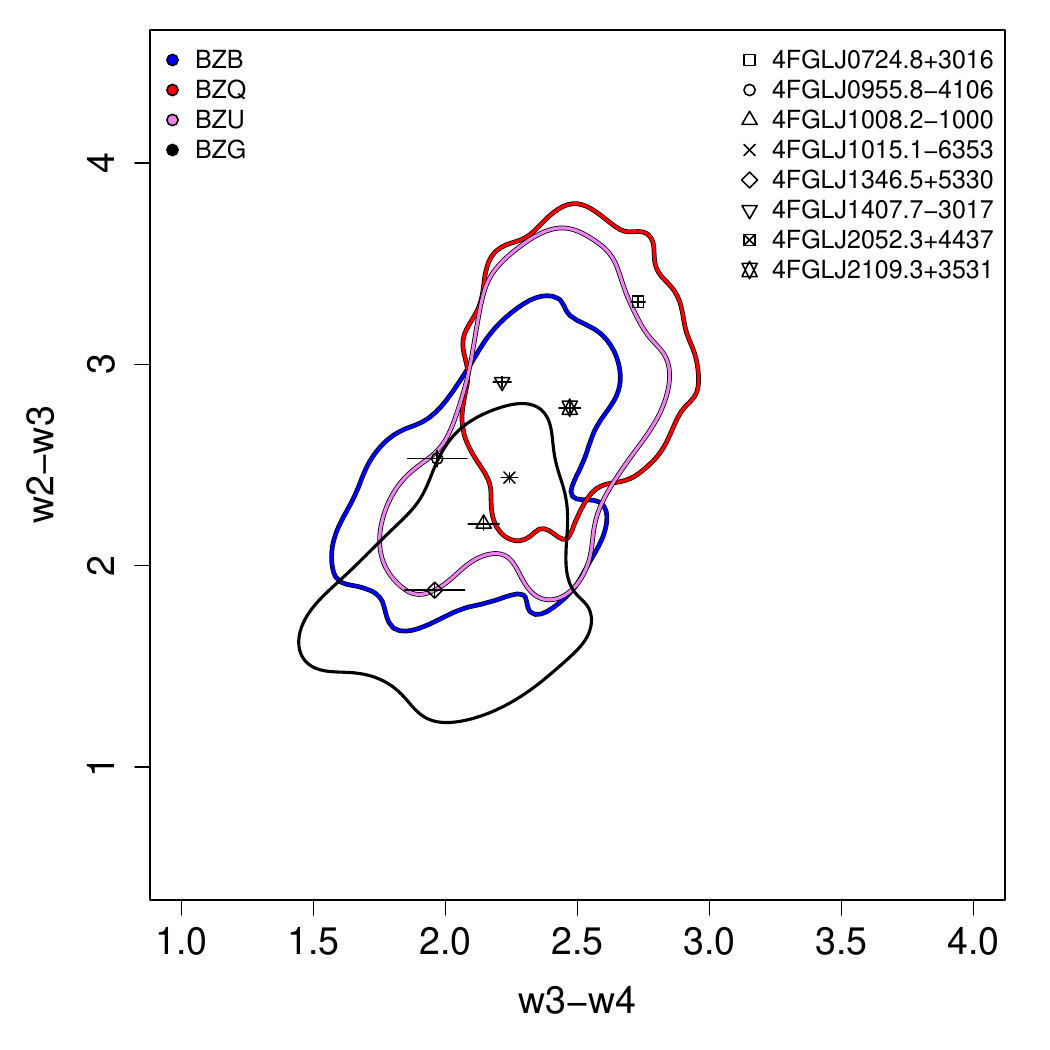}
		\caption{(Left panel) Soft X-ray sources listed in Table \ref{tab:softx} represented in the W1-W2 vs. W2-W3 WISE color-color plane. KDE isodensity contours containing \(90\%\) of known \(\gamma\)-ray blazars are presented with lines of different colors for the different blazar subclasses, as indicated in the legend (BZB=BL Lac, BZQ=FSRQ), BZU=blazar of uncertain type, BZG=BL Lacs with dominating host galactic emission). (Right panel) Same as the left panel, but for the W2-W3 vs. W3-W4 projection.}\label{fig:wise_colors}
	\end{figure*}

	We also looked for mid-IR counterparts to the soft x-ray sources in the Wide-Field Infrared Survey Explorer (WISE) all-sky data \citep[AllWISE,][]{2010AJ....140.1868W}, and present the result of this cross matching in Table \ref{tab:wise}, where all magnitudes have been then corrected for Galactic absorption using reddening estimates from \citet{2011ApJ...737..103S} and the extinction model from \citet{2007ApJ...663..320F}. Using the same procedure presented in \citet{2020A&A...641A..62P}, we use these data to select \(\gamma\)-ray blazar candidates by comparing their position in the two WISE color-color planes with the \(90\%\) KDE isodensity contours of known \(\gamma\)-ray blazars of different classes, as shown in Figure \ref{fig:wise_colors}.
	
	We then collected ultraviolet-optical data available through the \textit{Swift} Ultraviolet and Optical Telescope (UVOT) \citep[UVOT,][]{2005SSRv..120...95R}. The UVOT data were downloaded from HEASARC\footnote{\href{https://heasarc.gsfc.nasa.gov/}{https://heasarc.gsfc.nasa.gov/}} data archive and reduced using standard procedures\footnote{\href{http://www.swift.ac.uk/analysis/uvot/image.php}{http://www.swift.ac.uk/analysis/uvot/image.php}}. After checking the correct World Coordinate System alignment of UVOT images with USNO-B Catalog \citep{2003AJ....125..984M}, all extension in each image were summed using \textsc{uvotimsum}; the same procedure was applied to produce merged exposure maps. We performed source photometry using the \textsc{uvotsource} task using the appropriate exposure map. We adopted an aperture radius of \(5\arcsec\) for all filters, and used as background region an annulus centered at the source coordinates with inner and outer radii of \(10\arcsec\) and \(15\arcsec\), respectively. All magnitudes have been then corrected for Galactic absorption, with exception of source 1 in the field of UFO 4FGL J1424.2-6111c and source 1 in the field of UFO 4FGL J1652.2-4516, since these two sources are very close to the Galactic plane (\(|b| < 1 \degree\)), yielding very high reddening at their location. The result of this analysis is presented in Table \ref{tab:uvot}.

\begin{figure*}
		\centering
		\includegraphics[scale=0.37]{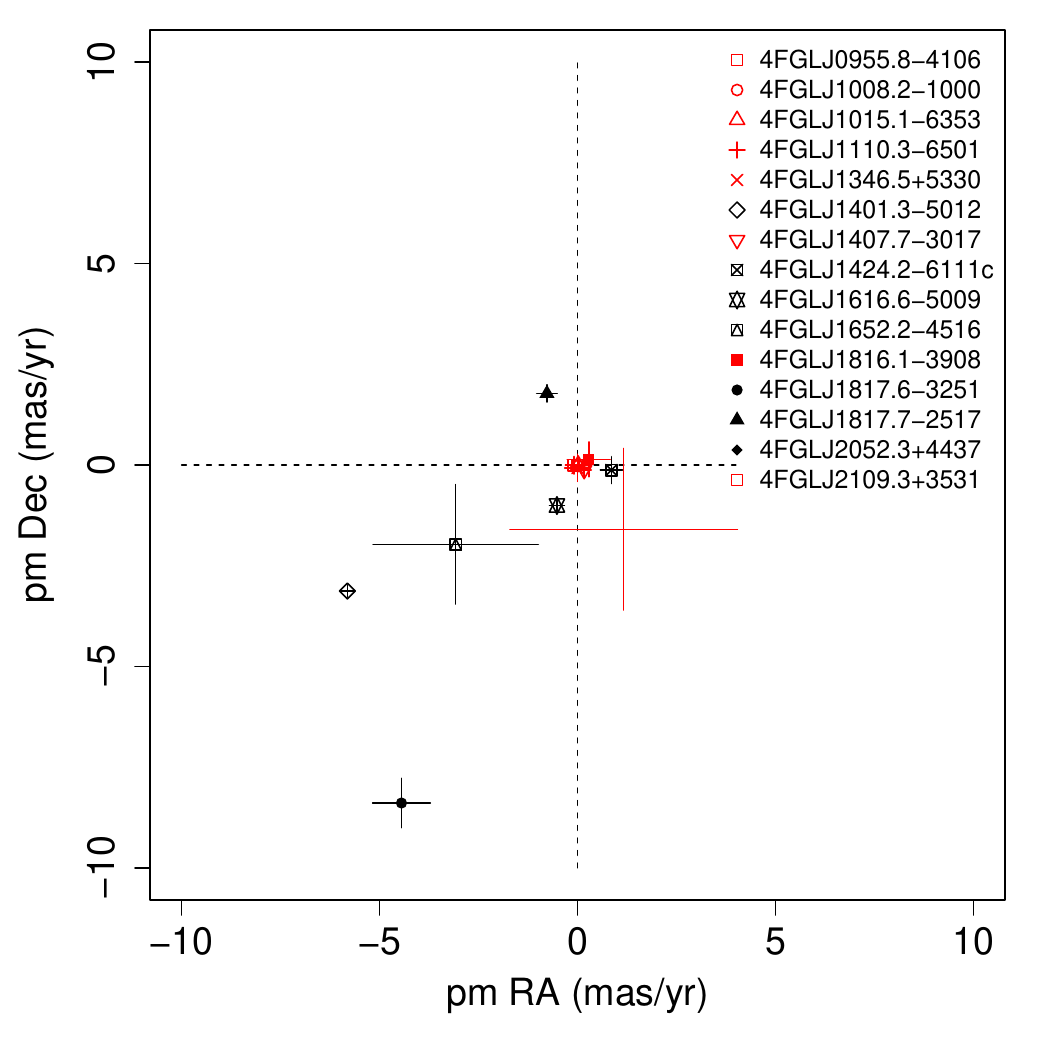}
		\caption{Distribution of the soft X-ray sources presented in Table \ref{tab:softx} in the GAIA motion plane. Sources shown in red have proper motions compatible with zero at a \(3\sigma\) level.}\label{fig:gaia}
	\end{figure*}

	We also collected Gaia \citep{2016A&A...595A...1G} parallax and proper motion data from the Gaia DR3 database \citep{2023A&A...674A...1G} and present them in Table \ref{tab:gaia}. In Figure \ref{fig:gaia} we show the Gaia counterparts to the soft X-ray sources present in Table \ref{tab:softx} in a proper motion Dec vs. RA plot, highlighting in red the sources that have proper motions compatible with zero at a \(3\sigma\) level.
	
\begin{figure*}
		\centering
		\includegraphics[scale=0.16]{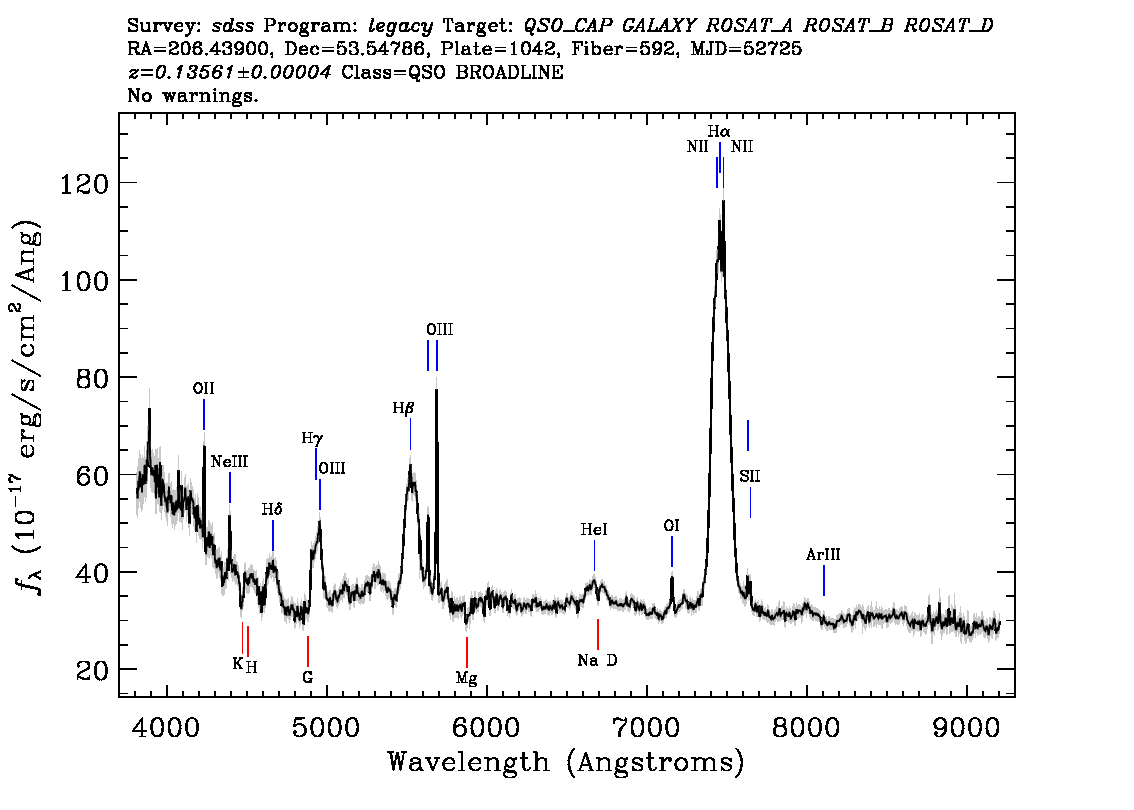}
		\includegraphics[scale=0.22]{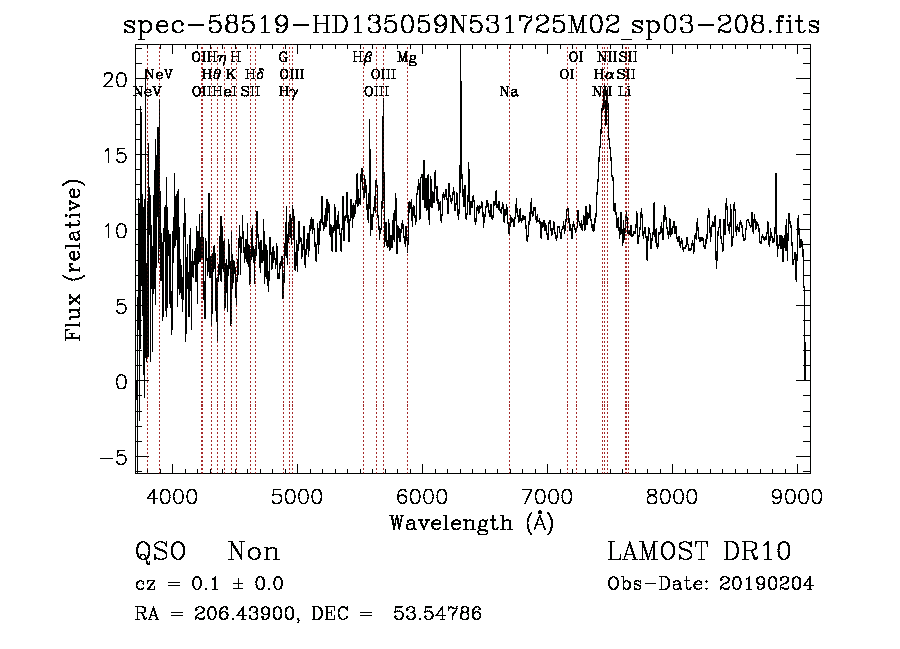}    
		\caption{SDSS DR18 (left panel) and LAMOST DR10 (right panel) optical spectra of the selected soft X-ray source lying in the field of UFO 4FGL J1346.5+5330.}\label{fig:sdss}
	\end{figure*}

	Finally, we looked in the SDSS DR18 \citep{2023ApJS..267...44A} and LAMOST \citep{2012RAA....12.1197C} DR10 for available optical spectra for the soft X-ray sources listed in Table \ref{tab:softx}. We only found optical spectra (both in SDSS DR18 and LAMOST DR10) for the selected soft X-ray source in the field of 4FGL J1346.5+5330 (presented in Figure \ref{fig:sdss}), showing broad, quasar-like lines.
	
	These multi-wavelength data were used to build broad-band spectral energy distributions (SEDs) of these soft X-ray sources, augmented by the X-ray data collected in this work and assuming that their \(\gamma\)-ray emission is that of UFO as reported in 4FGL, with the goal of modeling them with the JetSeT fitting tool \citep{2020ascl.soft09001T}. However, we were only able to collect enough multi-wavelength data to build reasonable SEDs for two sources, namely the soft X-ray sources in the field of 4FGL J1346.5+5330 and 4FGL J2109.3+3531, presented in Figure \ref{fig:seds}. In the same figure we show the best fit model to the SEDs, whose parameters are presented in Table \ref{tab:sed}.

	\subsection{Discussion}\label{sec:discussion}

	We found 22 hard X-ray sources from 4PBC and SRG/ART-XC all-sky X-ray survey Catalog compatible with the \(99\%\) LAT uncertainty ellipses of 17 \textit{Fermi} UFOs. In particular we have 13 UFOs whose uncertainty region is compatible with one hard X-ray source, 3 UFOs whose uncertainty region is compatible with two hard X-ray sources, and 1 UFO whose uncertainty region is compatible with three hard X-ray sources.

	We were able to collect soft X-ray data for these 17 UFOs, and identify 16 unique soft X-ray counterparts to the hard X-ray sources. The list of the soft X-ray counterparts is presented in Table \ref{tab:softx}.

	Since most gamma-ray emitting AGNs are also found to be radio-loud objects, correlation in AGNs between emission in gamma-rays and at radio frequencies is expected and observed \citep{2011ApJ...741...30A}. Being most of associated/identified \(\gamma\)-ray sources in 4FLG-DR4 classified as AGNs, we expect a significant fraction of UFOs to be AGNs as well. It is therefore instructive to investigate radio emission from the selected soft X-ray counterparts to the UFOs. To this end we make use of the Epoch 2 Quick Look Catalogue for the Very Large Array Sky Survey \citep[VLASS,][]{2020PASP..132c5001L}, covering the whole sky at \(\delta > -40\degree\), with a synthesized beam size of \(\sim 2\farcs 5\), a median rms of  \(120 \,\upmu \text{Jy/beam}\), listing \(\sim 3\) millions sources with coordinates, \(3 \text{ GHz}\) fluxes, etc. We also exploit the Rapid Australian SKA Pathfinder (ASKAP) Continuum Survey \citep[RACS,][]{2024PASA...41....3D} covering the whole sky at \(\delta < 49\degree\), with a resolution of \(\sim 10''\), a median sensitivity of  \(200 \,\upmu \text{Jy/beam}\), listing more than \(3\) millions sources with coordinates, \(1367.5 \text{ MHz}\) fluxes, etc. As a first test we look for radio counterparts for the whole sample of 393 soft X-ray sources that we find in the field of the UFOs considered in this work. To do this we cross-matched the soft X-ray source coordinates with the radio source coordinates listed in the VLASS and RACS catalog considering their \(99\%\) positional uncertainties. For 323 soft X-ray sources at \(\delta > -40\degree\) we found 15 VLASS counterparts, while for 389 soft X-ray sources at \(\delta < 49\degree\) we found 19 RACS counterparts. In total we found 20 radio counterparts for 393 soft X-ray source (\(\sim 5\%\)). Radio counterparts to soft X-ray sources are marked in Fig. \ref{fig:xraymap} with thin white crosses and green xs for VLASS and RACS sources, respectively. If we consider the soft X-ray sources that we select as counterparts to the hard X-ray sources, we have 9 sources at \(\delta > -40\degree\) for which we found 5 VLASS counterparts, and 15 sources at \(\delta < 49\degree\) for which we found 7 RACS counterparts. In total we found 8 radio counterparts for 16 soft X-ray counterparts to the hard X-ray sources (\(50\%\)). The radio fluxes for the selected soft X-ray counterpart to the hard X-ray sources are listed in Table \ref{tab:softx}.

	In addition, we were able to associate 15 of these sources with known astronomical objects. For 14 of them, we confirm the previous hard X-ray association, while for the hard X-ray source SRGA J181636.0-391251 in the field of 4FGL J1816.1-3908 we update the association from the \textit{INTEGRAL} source IGR J18165-3912 to the Mira variable star V\({}^*\) V371 CrA.
	
	These 15 associated soft X-ray sources are classified as: 1 blazar, 1 blazar candidate, 2 Seyfert 2 galaxies, 3 Seyfert 1 galaxies, 2 X-ray binaries, 1 generic X-ray source, 2 cataclysmic variables (CVs), and 3 variable stars.
	
\begin{figure*}
		\centering
		\includegraphics[scale=0.38]{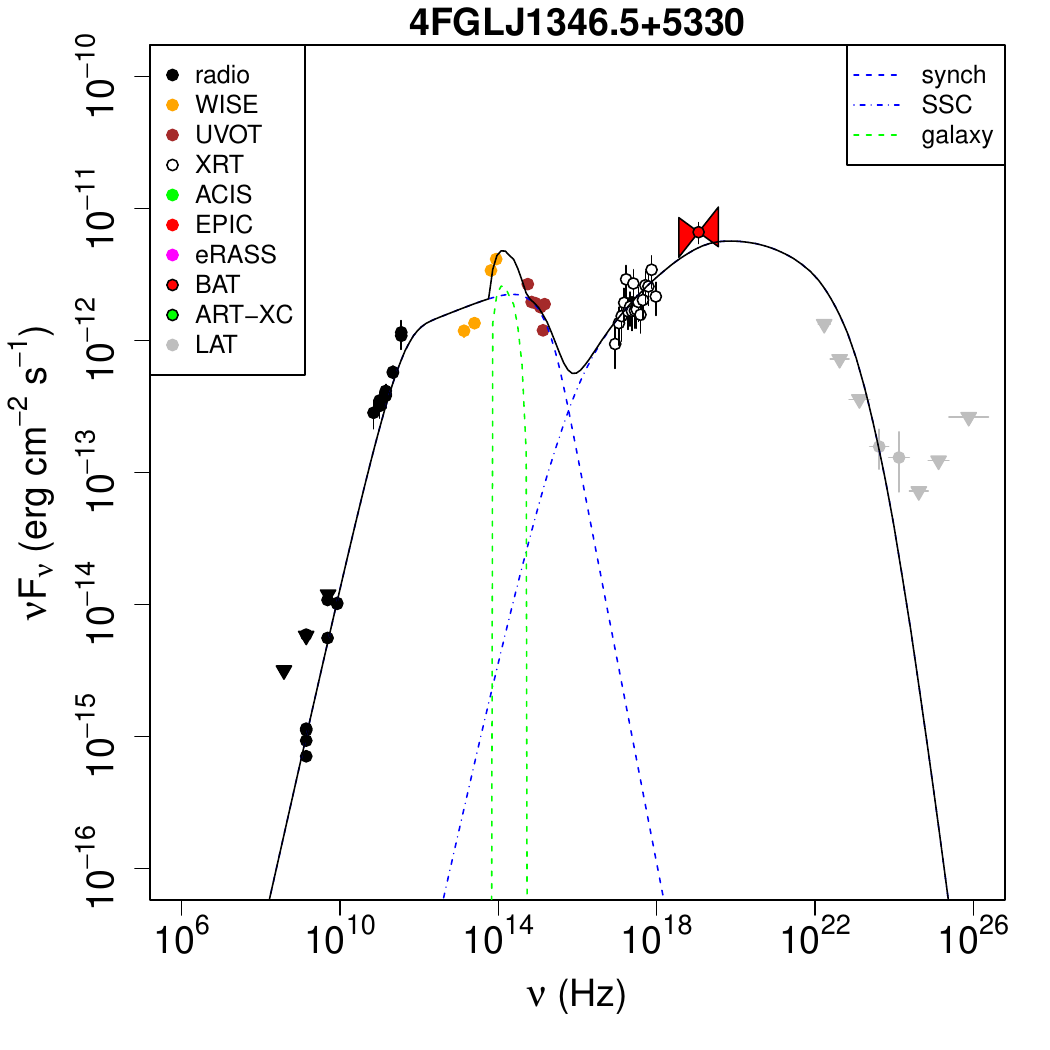}
		\includegraphics[scale=0.37]{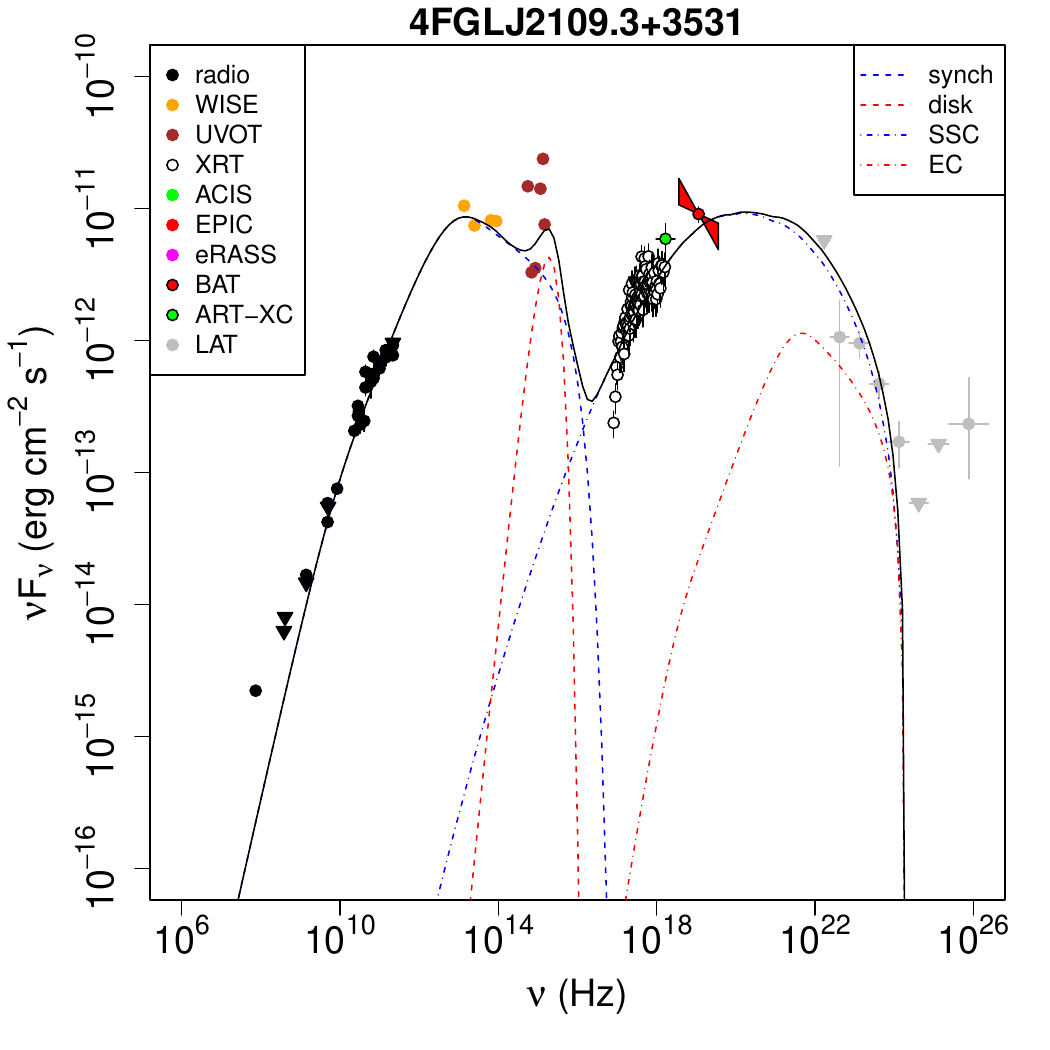}
		\caption{Broad-band SEDs of the selected soft X-ray sources lying in the field of UFOs 4FGL J1346.5+5330 and 4FGL J2109.3+3531. Data points from different bands/instruments are shown as colored circles (shown in the left legend). The colored dashed lines represent the model components (shown in the right legend: synch = synchrotron, SSC = synchrotron self-Compton, disk = accretion disk, EC = external Compton, galaxy = host galaxy emission) while the full black line represent the total best fit model.}\label{fig:seds}
	\end{figure*}

	The blazar source is BZU J1345+5332 (source 2 in the field of 4FGL J1346.5+5330), that has been classified as a blazar of unknown type \citep{2015Ap&SS.357...75M}. We note that, while its WISE colors (see Table \ref{tab:wise}) and SED (see Fig. \ref{fig:seds}, left panel) seem to favor a BL Lac classification, its optical spectra (presented in Fig. \ref{fig:sdss}) show broad quasar-like emission lines, favoring a FSRQ classification. 
	
	The blazar candidate source is PMN J0954-4108 (source 1 in the field of 4FGL J0955.8-4106). This radio source has been classified as a \(\gamma\)-ray BL Lac candidate by \citet{2014ApJS..215...14D} on the basis of its mid-IR WISE, and we confirm this classification (see Table \ref{tab:wise}).
	
	The 2 Seyfert 2 galaxies are source 3 in the field of 4FGL J0724.8+3016 (associated with NVSS J072537+295714) and source 5 in the field of 4FGL J1008.2-1000 (associated with 2MASX J10084862-0954510). The 3 Seyfert 1 galaxies are source 2 in the field of 4FGL J1015.1-6353 (associated with IGR J10147-6354), source 2 in the field of 4FGL J1407.7-3017 (associated with 2MASX J14080674-3023537), and source 3 in the field of 4FGL J2109.3+3531 (associated with ICRF J210931.8+353257). Source NVSS J072537+295714 has been classified as a Seyfert 2 galaxy \citep{2000A&A...355...89D}, although an alternate FR-I radio galaxy classification has been proposed \citep{2009MNRAS.396.1522E}. In addition, we note that 2MASX J10084862-0954510 has been classified as a Seyfert 2 galaxy \citep{2013ApJS..207...19B}, a Seyfert 1 galaxy \citep{2022ApJS..261....2K}, or a LINER \citep{2022ApJS..258...29C}. Sources IGR J10147-6354, 2MASX J14080674-3023537 and ICRF J210931.8+353257 has been classified as Seyfert 1 galaxies \citep{2010A&A...518A..10V, 2012MNRAS.426.1750M, 2022ApJS..261....2K}. Sources NVSS J072537+295714, 2MASX J10084862-0954510, IGR J10147-6354, 2MASX J10084862-0954510 and ICRF J210931.8+353257 show mid-IR WISE colors compatible with those of known \(\gamma\)-ray emitting blazars, indicating a possible jetted non-thermal emission. In addition, the latter source ICRF J210931.8+353257 shows a broad-band SED adequately represented by a leptonic jet model (see Fig. \ref{fig:seds}, right panel). \(\gamma\)-ray emission from Seyfert 1 galaxies has been first reported in narrow-line Seyfert 1 galaxies (NLSy1) by \citet{2009ApJ...707L.142A} and subsequently by other studies \citep[e.g.,][]{2018ApJ...853L...2P, 2024MNRAS.533.1281Y}, while \(\gamma\)-ray emission from Seyfert 2 galaxies has been so far observed in two sources (NGC 1068 and NGC 4945, \citealt{2010A&A...524A..72L}).
	
	The 2 X-ray binaries are source 1 in the field of 4FGL J1652.2-4516 (associated with XTE J1652-453) and source 1 in the field of 4FGL J1817.6-3251 (associated with XTE J1817-330). Source J1817-330 is classified as a low-mass X-ray binary \citep[LMXB,][]{2010A&A...523A..61K}, while source XTE J1652-453 is classified as a generic X-ray binary \citep{2011MNRAS.411..137H}. There are so far 8 confirmed \(\gamma\)-ray high-mass X-ray binaries \citep[HMXBs,][]{2013A&ARv..21...64D, 2019frap.confE..44P, 2020mbhe.confE..45C, 2023hxga.book..143F}, while in LMXBs only a few \(\gamma\)-ray emitting microquasars have been proposed \citep{2013A&A...550A..89D, 2016ApJ...831...89S, 2025ApJ...979L..40M}.
	
	The generic X-ray source is source 1 in the field of 4FGL J1110.3-6501 (associated with CXOU J110926.4-650224). We note that CXOU J110926.4-650224 has been proposed as a binary transitional pulsar candidate \citep{2019A&A...622A.211C}, a class of known \(\gamma\)-ray emitters \citep[e.g.,][]{2015ApJ...803L..27B}.
	
	The 2 CVs are source 1 in the field of 4FGL J1424.2-6111c (associated with IGR J14257-6117, \citealt{2013A&A...556A.120M}) and source 1 the field of 4FGL J1616.6-5009 (associated with IGR J16167-4957, \citealt{2007ApJS..170..175B}). This class of sources has been considered as potential \(\gamma\)-ray emitters \citep{1995ApJ...439..322S, 2011MNRAS.411.1701B, 2013EAS....61..255P, 2024heas.confE..20M}, however \textit{Fermi}-LAT data so far only provided upper limits \citep{2024AAS...24346105S}.
	
	The 3 variable stars are source 1 in the field of 4FGL J1816.1-3908 (associated with V\({}^*\) V371 CrA), source 1 in the field of 4FGL J1817.7-2517 (associated with OGLE BLG529.32 1342), and source 1 in the field of 4FGL J2052.3+4437 (associated with V\({}^*\) V1794 Cyg). In particular V\({}^*\) V371 CrA is classified as a Mira \citep{2002AcA....52..397P}, OGLE BLG529.32 1342 is classified as a delta Sct \citep{2021AcA....71..189S}, and V\({}^*\) V1794 Cyg is classified as a generic rotating variable \citep{2017ARep...61...80S}. Although stars are expected to emit \(\gamma\)-rays, the \(\gamma\)-ray flux from stars other than the Sun would of course be too low to be detected by \textit{Fermi}-LAT. There is one reported \(\gamma\)-ray detection of a symbiotic binary system V407 Cygni, involving a Mira accreting a white dwarf through stellar winds and triggering a nova explosion \citep{2010Sci...329..817A}. Based on the Mira variable stars listed in the fourth phase of the Optical Gravitational Lensing Experiment \citep[OGLE-IV,][]{2015AcA....65....1U} catalog, the probability of a random coincidence of a Mira star with an UFO is \(\sim 3.2\%\), therefore an association of V\({}^*\) V371 CrA with the UFO 4FGL J1816.1-3908 is possible, although unlikely. On the other hand, delta Sct and rotating variable stars have not been detected in \(\gamma\)-rays so far. Based on the delta Sct stars listed in OGLE-IV catalog, the probability of a random coincidence of a delta Sct star with an UFO is \(\sim 2.0\%\), so again although unlikely, an association of OGLE BLG529.32 1342 with the UFO 4FGL J1817.7-2517. An association of these sources with the UFOs would represent the first case of delta Sct or rotating variable star to be detected in \(\gamma\)-rays, and so we find unlikely for these sources to be the low-energy counterparts of the UFOs.
	
	As expected, all selected soft X-ray counterparts to hard X-ray sources classified as AGNs (namely NVSS J072537+295714, PMN J0954-4108, 2MASX J10084862-0954510, IGR J10147-6354, BZU J1345+5332, 2MASX J14080674-3023537 and ICRF J210931.8+353257) are also detected at radio frequencies in the VLASS and RACS surveys (see Table \ref{tab:softx}). In particular we notice that PMN J0954-4108, IGR J10147-6354 and BZU J1345+5332 are the only soft X-ray sources in the UFO field with a radio counterpart, while NVSS J072537+295714 and ICRF J210931.8+353257 are the soft X-ray sources in the UFO field with the brightest radio counterpart. The sources 2MASX J10084862-0954510 and 2MASX J14080674-3023537 are not the soft X-ray sources in the UFO field with the brightest radio counterpart. On the other hand, the selected soft X-ray counterparts to hard X-ray sources not classified as AGNs (CXOU J110926.4-650224, IGR J14257-6117, IGR J16167-4957, XTE J1652-453, V\({}^*\) V371 CrA, XTE J1817-330 and OGLE BLG529.32 1342) are not detected at radio frequencies in the VLASS and RACS surveys. In particular, in the field of UFOs 4FGL J1110.3-6501 and 4FGL J1816.1-3908 we have only one soft X-ray source with a radio counterpart that is not the one that we select as soft X-ray counterpart to the hard X-ray source. Finally we note that the rotating variable V\({}^*\) V1794 Cyg is instead detected  in both the VLASS and RACS surveys, and is the soft X-ray source in the UFO field with the brightest radio counterpart.

\begin{figure*}
		\centering
		\includegraphics[scale=0.5]{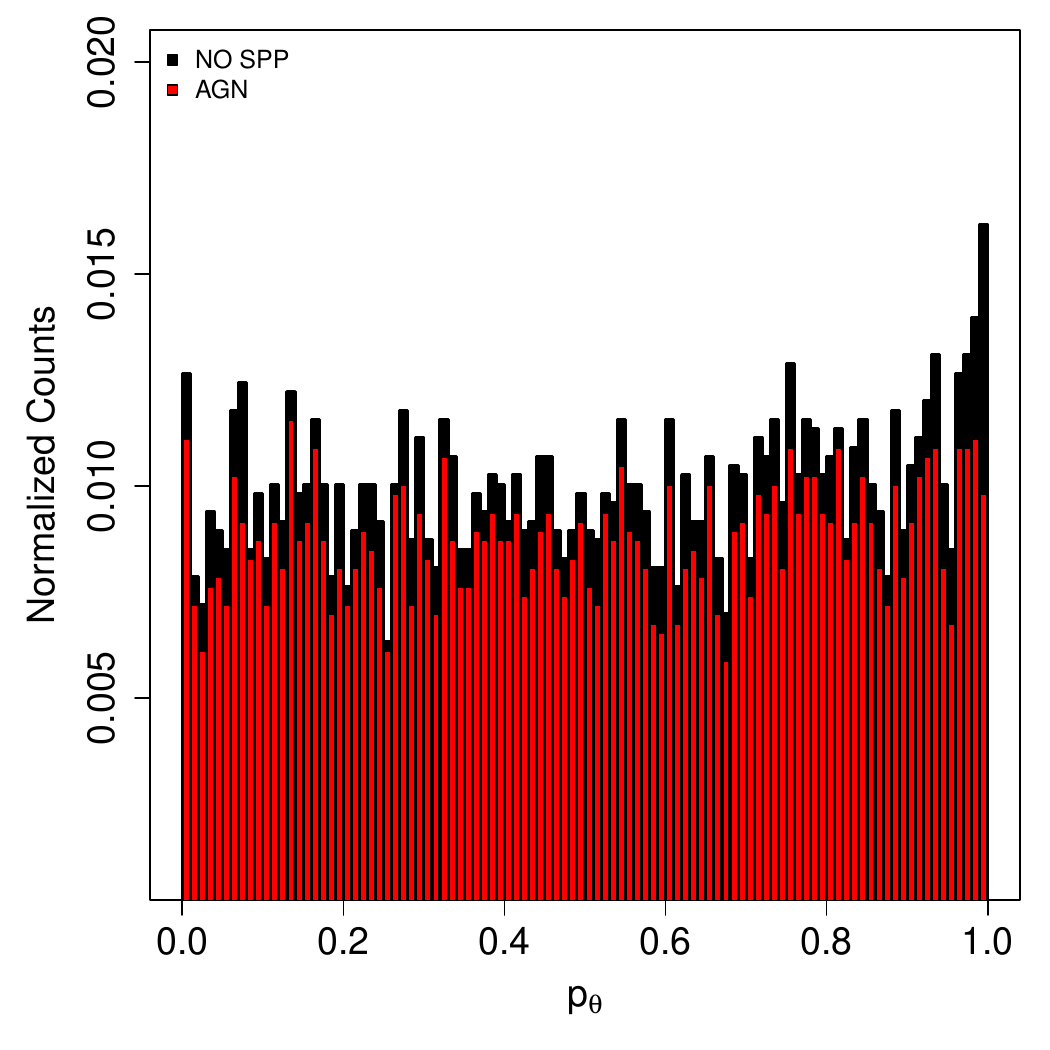}		
		\caption{Distribution of \(p_{\theta}\) for the associated sources in 4FGL-DR4 excluding SPPs (see Sect. \ref{sec:discussion}). Black and red bars represent the distribution of \(p_{\theta}\) for all sources (excluding SPPs) and AGNs, respectively.}\label{fig:sep_hist}
	\end{figure*}

	We note that for 4 out of 15 tentatively associated sources (namely source 3 in the field of 4FGL J0724.8+3016, source 2 in the field of 4FGL J1015.1-6353, source 2 in the field of 4FGL J1346.5+5330, and source 1 in the field of 4FGL J1817.6-3251) the soft X-ray counterparts lie outside the \(99\%\) LAT uncertainty regions of the \textit{Fermi} UFO, that is, with \(p_{\theta}>0.99\). In Fig. \ref{fig:sep_hist} we show the distribution of \(p_{\theta}\) for the associated sources in 4FGL-DR4. In particular, we excluded from this plot 4FGL-DR4 sources of unknown nature but overlapping with known supernova remnants or pulsar wind nebulae and thus candidates to these classes (SPPs), since the majority of these sources (\(\sim 75\%\)) lie at \(p_{\theta}>0.99\). In this figure we show with black and red bars the distribution of \(p_{\theta}\) for all sources (excluding SPPs) and AGNs, respectively. Regarding AGNs we don't see a clear trend of associated sources decreasing with increasing \(p_{\theta}\). We instead note an increased number of associations in the no SPPs sample at \(p_{\theta}>0.96\), mainly due to an increased number of low galactic latitude pulsars and supernova remnants association at large separations. In any case we do not see a decreasing trend of associated sources with increasing \(p_{\theta}\), that is, with increasing separation between the coordinates of the \(\gamma\)-ray source  and those of its associated counterpart. In particular the last bin of the histograms shown in Fig. \ref{fig:sep_hist} (at \(p_{\theta}>0.99\)) does not appear significantly less populated than other bins. Therefore we still consider as potential counterparts to the UFOs the four soft X-ray sources lying outside the \(99\%\) LAT uncertainty regions of the \textit{Fermi} UFOs. In addition we note that LAT uncertainty ellipses changed and improved between \textit{Fermi}-LAT catalog releases \citep[see for example][]{2023IAUS..375...71M}, and therefore these soft X-ray sources may lay inside LAT uncertainty regions in catalog future releases.

	Despite the small sample size of 16 soft X-ray counterparts, we can compare these results with the expectations from the composition of the \(\gamma\)-ray sky. Considering the 16 soft X-ray counterparts we found, we note that 4 are located at high Galactic latitude (\(|b| > 15 \degree\)) and 12 at low Galactic latitude (\(|b| < 15 \degree\)). For associated sources in 4FGL-DR4, we see that high Galactic latitude sources are dominated by blazars (\(\sim 93.2\%\)) with a small contribution from pulsars (\(\sim 3.2\%\)). For low Galactic latitude associated sources in 4FGL-DR4 we still have a major contribution from blazars (\(\sim 53.8\%\)), followed by pulsars (\(\sim 17.4\%\)) and sources of unknown nature (\(\sim 10.5\%\)). Assuming that the same source class composition holds also for UFOs, in our sample of 16 UFOs we would expect to find \(\sim 10\) blazars and \(\sim 2\) pulsars. To investigate if the hard X-ray selection used in this analysis can disfavor blazar of pulsar selection, we cross matched the associated sources from 4FGL-DR4 with both 4PBC and SRG/ART-XC catalogs. We find however a representation in the matched sources of blazars and pulsars that is similar to that of associated 4FGL-DR4 sources, since in the high and low Galactic latitude sources blazars represent \(\sim 88.7\%\) and \(\sim 54.9\%\) of the matched hard X-ray sources, respectively, while pulsars represent \(\sim 1.1\%\) and \(\sim 15.9\%\) of the matched hard X-ray sources, respectively. It is therefore possible that UFOs represent a sample of sources significantly different from 4FGL-DR4 associated sources.

	We note that for the UFOs 4FGL J0550.2+0730c, 4FGL J1110.3-6501, 4FGL J1424.2-6111c, 4FGL J1616.6-5009, and 4FGL J1652.2-4516, \citet{2024A&A...684A.208M} report 4eRASS sources inside the \(95\%\) LAT uncertainty ellipses as possible pulsar and/or blazar type X-ray counterparts to the UFO (see Fig. \ref{fig:xraymap}). We note however that these sources are always outside the hard X-ray source uncertainty regions (with the exception of source 3 in the field of 4FGL J1110.3-6501, see Sect. \ref{sec:sources}), and that their combined posterior Bayesian probability to be the counterpart of the \(\gamma\)-ray source and to be a pulsar or a blazar are \(\lesssim 0.06\), with the exception of source 4eRASS J165154.2-451450 in the field of 4FGL J1652.2-4516, with a combined posterior Bayesian probability for the X-ray source to be the counterpart of the \(\gamma\)-ray source and to be a pulsar \(\sim 0.12\), and source 4eRASS J055014.8+072944 in the field of 4FGL J0550.2+0730c, with a combined posterior Bayesian probability for the X-ray source to be the counterpart of the \(\gamma\)-ray source and to be a blazar \(\sim 0.17\).
	
	We compare our results with those of \citet{2021ApJ...916...28T}. These authors cross-matched  the 105 month \textit{Swift}-BAT catalog \citep{2018ApJS..235....4O} - listing 1632 \(14–195 \text{ keV}\) sources at detected at \(>4.8 \sigma\) significance level - and a previous release of 4FGL, namely 4FGL-DR2 \citep{2020arXiv200511208B} listing 5788 \(\gamma\)-ray sources. The authors find 132 matches for point-like sources by means of spatial cross-match (with a separation threshold of \(0.08\degree\) between the \(\gamma\)-ray source and the hard X-ray source) and 24 matches for point-like sources by means of identifier match, for a total of 156 point-like source matches. In particular, we note the UFOs 4FGL J1110.3-6501, 4FGL J1424.2-6111c and 4FGL J2109.3+3531 are not present in 4FGL-DR2, so they are not included in \citet{2021ApJ...916...28T} analysis. For 4FGL-DR2 UFO 4FGL J1808.5-3701 the authors report a cross match with a \textit{Swift}-LAT source, but this \(\gamma\)-ray source has been associated with the LMXB SAX J1808.4-3658 in 4FGL-DR4, so it not considered a UFO anymore. For UFOs 4FGL J0552.0+2656c, 4FGL J0955.8-4106, 4FGL J1015.1-6353, 4FGL J1401.3-5012, 4FGL J1816.1-3908 and 4FGL J2052.3+4437 we find an overlapping hard X-ray source only in the SRG/ART-XC catalog (not in the 4PBC), so for these sources \citet{2021ApJ...916...28T} do not report any cross-match. For UFOs 4FGL J1008.2-1000, 4FGL J1346.5+5330, 4FGL J1616.6-5009 and 4FGL J1817.7-2517 the authors do not report any cross-match, but these associations were likely excluded because the separation between the \(\gamma\)-ray source and the \textit{Swift}-LAT source is \(>0.15\degree\). Finally, for UFOs 4FGL J0724.8+3016, 4FGL J1407.7-3017, 4FGL J1652.2-4516 and 4FGL J1817.6-3251 the authors report a cross-match between the \(\gamma\)-ray source and the \textit{Swift}-LAT source, identifying the same lower-energy counterparts reported in our analysis.

	To conclude this section we briefly mention the soft X-ray source for which we did not find any lower-energy counterpart, namely source 1 in the field of UFO 4FGL J1401.3-5012 (see Fig. \ref{fig:xraymap}). This source lies in the uncertainty region of SRG/ART-XC source SRGA J140146.3-501333 and is detected both by \textit{Swift}-XRT and eROSITA. It appears quite bright in soft X-rays with fluxes of \(5.6\) and \(4.5 \times{10}^{-12} \text{ erg} \text{ cm}^{-2} \text{ s}\) in \textit{Swift}-XRT and eROSITA observations, respectively. The source is also detected in AllWISE but only in the first three WISE bands, and in GAIA DR3 with significant parallax and proper motion, indicating that this source is likely Galactic.

	\section{Conclusions}\label{sec:conclusion}
	
	We have looked in the 100-month Palermo BAT Catalog and in SRG/ART-XC all-sky X-ray survey Catalog for hard X-ray sources compatible with the uncertainty regions of unidentified \(\gamma\)-ray sources in the 4FGL catalog, finding 22 hard X-ray sources compatible with the \(99\%\) LAT uncertainty ellipses of 17 \textit{Fermi} UFOs.
	
	We collected \textit{Swift}-XRT, \textit{Chandra}-ACIS, \textit{XMM-Newton}-EPIC and eRASS soft X-ray data for these sources, identifying 17 soft X-ray counterparts for the hard X-ray sources, and associating 16 of these sources with known astronomical objects. In addition, we collected available multi-wavelength data for these sources.
	
	\begin{itemize}
		\item 1 of these these associated sources (namely BZU J1345+5332) has been classified as a blazar of unknown type. This source shows mid-IR WISE colors compatible with known \(\gamma\)-ray blazars, a blazar-like broad band SED, and quasar-like optical spectra. We suggest that this sources is the lower-energy counterpart to the UFO 4FGL J1346.5+5330.
		
		\item 1 of these these associated sources (namely PMN J0954-4108) has been classified as a \(\gamma\)-ray BL Lac candidate on the basis of its mid-IR WISE. We confirm this classification, and suggest that this sources is likely the lower-energy counterpart to the UFO 4FGL J0955.8-4106.
		
		\item 5 of these associated sources are classified as Seyfert galaxies (namely, NVSS J072537+295714, 2MASX J10084862-0954510, IGR J10147-6354, 2MASX J14080674-3023537, ICRF J210931.8+353257). In addition these sources show mid-IR WISE colors compatible with known \(\gamma\)-ray blazars, and one (ICRF J210931.8+353257) shows a blazar-like broad band SED. Since \(\gamma\)-ray emission has been observed in NLSy1 and Seyfert 2 galaxies, we suggest that these sources can be the lower-energy counterpart to the UFOs 4FGL J0724.8+3016, 4FGL J1008.2-1000, 4FGL J1015.1-6353, 4FGL J1407.7-3017, and 4FGL J2109.3+3531
		
		\item 2 of these associated sources (namely XTE J1652-453, and XTE J1817-330) are classified as X-ray binaries, a class of sources that has been already observed (although rarely) to emit in \(\gamma\)-ray. We therefore suggest that these sources may be the the lower-energy counterpart to the UFOs 4FGL J1652.2-4516 and 4FGL J1817.6-3251.
		
		\item 1 of these these associated sources (namely CXOU J110926.4-650224) is classified as a generic X-ray source, but it has has been proposed as a binary transitional pulsar candidate. We therefore suggest that this sources can be the lower-energy counterpart to the UFO 4FGL J1110.3-6501. 
		
		\item 2 of these associated sources (namely IGR J14257-6117 and IGR J16167-4957) are classified as CVs, which have been suggested as potential \(\gamma\)-ray sources. However, \(\gamma\)-ray emission from these source has not been yet observed, so, if the association with UFOs 4FGL J1424.2-6111c and 4FGL J1616.6-5009 will be confirmed, these sources would represent the first \(\gamma\)-ray emitting CVs.
		
		\item 2 of these associated sources (namely V\({}^*\) V371 CrA, OGLE BLG529.32 1342 and V\({}^*\) V1794 Cyg) are classified as Mira, delta Sct and rotating variable stars. We consider the association of these sources with the UFOs 4FGL J1816.1-3908, 4FGL J1817.7-2517 and 4FGL J2052.3+4437 possible but unlikely.
	\end{itemize}
	
	Correlated multi-wavelength variability is the key to firmly identify the sources listed in this work as low-energy counterparts to the UFOs. Monitoring programs like those currently performed at radio \citep[e.g.,][]{2018ApJS..234...12L, 2022ApJS..260...12W}, optical \citep[e.g.,][]{2003PASP..115..844L, 2020AAS...23530508L} and X-ray frequencies \citep[e.g.,][]{2013ApJS..207...28S} would represent the ideal tool to correlate the UFO \(\gamma\)-ray variability with that of the candidate counterpart at lower energies, therefore allowing a firm identification.

	\begin{acknowledgements}
		We thank the anonymous referee for their comments and suggestions which helped us to considerably improve the manuscript.
		A. P. was funded by the European Union ERC-2022-STG - BOOTES - 101076343. Views and opinions expressed are however those of the author(s) only and do not necessarily reflect those of the European Union or the European Research Council Executive Agency. Neither the European Union nor the granting authority can be held responsible for them. I.L. received support from the ESA Archival Research Visitor Programme. This research has made use of the Palermo BAT Catalog and database operated at INAF - IASF Palermo. This work is based on observations with the Mikhail Pavlinsky ART-XC telescope, hard X-ray instrument on board the SRG observatory. The SRG observatory was created by Roskosmos in the interests of the Russian Academy of Sciences represented by its Space Research Institute (IKI) in the framework of the Russian Federal Space Program, with the participation of Germany. This work is based on data from eROSITA, the soft X-ray instrument aboard SRG, a joint Russian-German science mission supported by the Russian Space Agency (Roskosmos), in the interests of the Russian Academy of Sciences represented by its Space Research Institute (IKI), and the Deutsches Zentrum für Luft- und Raumfahrt (DLR). The SRG spacecraft was built by Lavochkin Association (NPOL) and its subcontractors, and is operated by NPOL with support from the Max Planck Institute for Extraterrestrial Physics (MPE). The development and construction of the eROSITA X-ray instrument was led by MPE, with contributions from the Dr. Karl Remeis Observatory Bamberg \& ECAP (FAU Erlangen-Nuernberg), the University of Hamburg Observatory, the Leibniz Institute for Astrophysics Potsdam (AIP), and the Institute for Astronomy and Astrophysics of the University of Tübingen, with the support of DLR and the Max Planck Society. The Argelander Institute for Astronomy of the University of Bonn and the Ludwig Maximilians Universität Munich also participated in the science preparation for eROSITA. This work has made use of data from the European Space Agency (ESA) mission \textit{Gaia} (\href{https://www.cosmos.esa.int/gaia}{https://www.cosmos.esa.int/gaia}), processed by the \textit{Gaia} Data Processing and Analysis Consortium (DPAC, \href{https://www.cosmos.esa.int/web/gaia/dpac/consortium}{https://www.cosmos.esa.int/web/gaia/dpac/consortium}). Funding for the DPAC has been provided by national institutions, in particular the institutions participating in the \textit{Gaia} Multilateral Agreement. Funding for the Sloan Digital Sky Survey IV has been provided by the Alfred P. Sloan Foundation, the U.S. Department of Energy Office of Science, and the Participating Institutions. SDSS-IV acknowledges support and resources from the Center for High-Performance Computing at the University of Utah. The SDSS web site is \href{www.sdss.org}{www.sdss.org}. SDSS-IV is managed by the Astrophysical Research Consortium for the Participating Institutions of the SDSS Collaboration including the Brazilian Participation Group, the Carnegie Institution for Science, Carnegie Mellon University, the Chilean Participation Group, the French Participation Group, Harvard-Smithsonian Center for Astrophysics, Instituto de Astrof\'isica de Canarias, The Johns Hopkins University, Kavli Institute for the Physics and Mathematics of the Universe (IPMU) / University of Tokyo, the Korean Participation Group, Lawrence Berkeley National Laboratory, Leibniz Institut f\"ur Astrophysik Potsdam (AIP), Max-Planck-Institut f\"ur Astronomie (MPIA Heidelberg), Max-Planck-Institut f\"ur Astrophysik (MPA Garching), Max-Planck-Institut f\"ur Extraterrestrische Physik (MPE), National Astronomical Observatories of China, New Mexico State University, New York University, University of Notre Dame, Observat\'ario Nacional / MCTI, The Ohio State University, Pennsylvania State University, Shanghai Astronomical Observatory, United Kingdom Participation Group, Universidad Nacional Aut\'onoma de M\'exico, University of Arizona, University of Colorado Boulder, University of Oxford, University of Portsmouth, University of Utah, University of Virginia, University of Washington, University of Wisconsin, Vanderbilt University, and Yale University. This publication makes use of data products from the Wide-field Infrared Survey Explorer, which is a joint project of the University of California, Los Angeles, and the Jet Propulsion Laboratory/California Institute of Technology, funded by the National Aeronautics and Space Administration. Guoshoujing Telescope (the Large Sky Area Multi-Object Fiber Spectroscopic Telescope LAMOST) is a National Major Scientific Project built by the Chinese Academy of Sciences. Funding for the project has been provided by the National Development and Reform Commission. LAMOST is operated and managed by the National Astronomical Observatories, Chinese Academy of Sciences. We acknowledge the use of public data from the \textit{Swift} data archive. This research has made use of data obtained from the \textit{Chandra} Data Archive. This research has made use of observations obtained with \textit{XMM-Newton}, an ESA science mission with instruments and contributions directly funded by ESA Member States and NASA. This research has made use of data, software and/or web tools obtained from the High Energy Astrophysics Science Archive Research Center (HEASARC), a service of the Astrophysics Science Division at NASA/GSFC and of the Smithsonian Astrophysical Observatory's High Energy Astrophysics Division. This research has made use of software provided by the \textit{Chandra} X-ray Center (CXC) in the application packages CIAO, ChIPS, and Sherpa. Part of this work is based on archival data, software or online services provided by the Space Science Data Center - ASI. This research has made use of the SIMBAD database, operated at CDS, Strasbourg, France. This research has made use of the TOPCAT software \citep{2005ASPC..347...29T}.
	\end{acknowledgements}

	\software{HEAsoft (\href{https://heasarc.gsfc.nasa.gov/docs/software/lheasoft/}{https://heasarc.gsfc.nasa.gov/docs/software/lheasoft/}), CIAO \citep{2006SPIE.6270E..1VF}, Sherpa \citep{2001SPIE.4477...76F}, TOPCAT \citep{2005ASPC..347...29T}.}

	\begin{table}
		\caption{List of 4FGL UFOs with at least one 4PBC source overlapping the UFO uncertainty ellipse. Note that UFO 4FGL J1616.6-5009 has two overlapping 4PBC source, namely 4PBC J1616.7-4959 and 4PBC J1617.4-5019.}\label{tab:bat}
		\resizebox{\textwidth}{!}{
			\begin{tabular}{c|c|c|c|c|c|c|c|c|c|c|c|c}
				\hline\hline
				4FGL Name & 4FGL RA & 4FGL Dec & 4FGL semimaj & 4FGL semimin & 4FGL PA & 4PBC Name & 4PBC association & 4PBC RA & 4PBC Dec & 4PBC err & \(F_{15-150 \text{ keV}}\) & \(\Gamma_{15-150 \text{ keV}}\) \\
				& HH:MM:SS.SSS & DD:MM:SS.SS & deg & deg & deg & & & HH:MM:SS.SSS & DD:MM:SS.SS & arcmin & \({10}^{-12} \text{ erg} \text{ cm}^{-2} \text{ s}^{-1}\) &  \\
				\hline
				4FGL J0724.8+3016    & 07:24:49.200   & +30:16:09.12   & 0.34                      & 0.11                      & -15.06    & 4PBC J0725.6+2959 & NVSS J072537+295714       & 07:25:37.200   & +29:59:02.40    & 3.26              & \({10.7}\pm{1.6}\)                & \(2.51\pm 0.30\)  \\
				\hline
				4FGL J1008.2-1000    & 10:08:16.056  & -10:00:59.76  & 1.03                      & 0.76                      & -52.04    & 4PBC J1008.7-0955 & 2MASX J10084862-0954510   & 10:08:46.560   & -09:55:19.20    & 2.73              & \({19.3}\pm{2.0}\)                & \(1.78\pm 0.10\)  \\
				\hline
				4FGL J1110.3-6501    & 11:10:22.776  & -65:01:06.25  & 0.13                      & 0.10                      & -38.42    & 4PBC J1109.3-6501 & CXOU J110926.4-650224      & 11:09:18.479   & -65:01:30.01   & 4.22              & \({8.4}\pm{1.7}\)                 & \(1.88\pm 0.30\)  \\
				\hline
				4FGL J1346.5+5330    & 13:46:35.424  & +53:30:06.12   & 0.14                      & 0.12                      & 1.03      & 4PBC J1345.4+5331 & BZU J1345+5332             & 13:45:27.601   & +53:31:37.20    & 4.12              & \({6.63}\pm{1.2}\)                & \(1.92\pm 0.30\)  \\
				\hline
				4FGL J1407.7-3017    & 14:07:44.976  & -30:17:08.88  & 0.21                      & 0.16                      & -32.58    & 4PBC J1408.1-3023 & 2MASX J14080674-3023537   & 14:08:11.521   & -30:23:49.20   & 2.62              & \({14.8}\pm{1.5}\)                & \(2.04\pm 0.20\)  \\
				\hline
				4FGL J1424.2-6111c   & 14:24:15.384  & -61:11:51.72  & 0.18                      & 0.16                      & 68.63     & 4PBC J1425.1-6118 & IGR J14257-6117           & 14:25:10.800   & -61:18:50.40   & 2.74              & \({10.8}\pm{1.3}\)                & \(2.50\pm 0.20\)  \\
				\hline
				4FGL J1616.6-5009    & 16:16:37.800  & -50:09:02.52  & 0.21                      & 0.16                      & -61.92    & 4PBC J1616.7-4959 & IGR J16167-4957           & 16:16:44.161   & -49:59:16.80   & 1.29              & \({19.2}\pm{1.1}\)                & \(2.80\pm 0.10\)  \\
				&   &   & & &    & 4PBC J1617.4-5019 & 1RXS J161728.1-502238     & 16:17:25.441   & -50:19:08.40 & 2.95              & \({6.6}\pm{0.9}\)                 & \(3.38\pm 0.30\)  \\
				\hline
				4FGL J1652.2-4516    & 16:52:15.840   & -45:16:56.28  & 0.12                      & 0.09                      & -4.11     & 4PBC J1652.3-4520 & XTE J1652-453             & 16:52:21.841   & -45:20:52.80   & 0.47              & \({420.0}\pm{6.4}\)               & \(2.06\pm 0.02\)  \\
				\hline
				4FGL J1817.6-3251    & 18:17:38.352  & -32:51:20.53  & 0.16                      & 0.15                      & 15.76     & 4PBC J1817.7-3300 & XTE J1817-330             & 18:17:43.682   & -33:00:57.60 & 0.40              & \({751.0}\pm{9.1}\)               & \(2.32\pm 0.02\)  \\
				\hline
				4FGL J1817.7-2517    & 18:17:46.344  & -25:17:16.80   & 0.22                      & 0.13                      & -65.83    & 4PBC J1817.3-2509 & OGLE BLG529.32 1342           & 18:17:22.800   & -25:09:46.80   & 1.79              & \({11.4}\pm{0.9}\)                & \(3.23\pm 0.20\)  \\
				\hline
				4FGL J2109.3+3531    & 21:09:19.584  & +35:31:27.12   & 0.16                      & 0.13                      & 88.99     & 4PBC J2109.4+3532 & ICRF J210931.8+353257      & 21:09:28.799    & +35:32:09.60    & 3.05              & \({9.1}\pm{1.2}\)                 & \(2.34\pm 0.20\)  \\
				\hline\hline
			\end{tabular}
		}
		\tablecomments{For each UFO we show the 4FGL Name, the 4FGL coordinates (4FGL RA and 4FGL Dec), the 99\% 4FGL uncertainty ellipse semimajor axis, semiminor axis and position angle (4FGL semimaj, 4FGL semimin and 4FGL PA), the name of the overlapping 4PBC source (4PBC Name), the association of the 4PBC source as presented in the 4PBC catalog (4PBC association), the 4PBC coordinates (4PBC RA and 4PBC Dec) the 4PBC 99\% uncertainty radius (4PBC err), the 4PBC \(15-150 \text{ keV}\) flux (\(F_{15-150 \text{ keV}}\)) and the 4PBC power-law slope (\(\Gamma_{15-150 \text{ keV}}\)).}
	\end{table}

	\begin{table}
		\caption{List of 4FGL UFOs with at least one SRG/ART-XC source overlapping the UFO uncertainty ellipse.}\label{tab:artxc}
		\resizebox{\textwidth}{!}{
			\begin{tabular}{c|c|c|c|c|c|c|c|c|c|c|c}
				\hline\hline
				4FGL Name & 4FGL RA & 4FGL Dec & 4FGL semimaj & 4FGL semimin & 4FGL PA & ART-XC Name & ART-XC association & ART-XC RA & ART-XC Dec & ART-XC err & \(F_{4-12 \text{ keV}}\) \\
				& HH:MM:SS.SSS & DD:MM:SS.SS & deg & deg & deg & & & HH:MM:SS.SSS & DD:MM:SS.SS & arcsec & \({10}^{-12} \text{ erg} \text{ cm}^{-2} \text{ s}^{-1}\) \\
				\hline
				4FGL J0550.2+0730c & 05:50:12.697 & +07:30:02.88 & 0.22 & 0.17 & 10.49 & SRGA J054958.5+072518 & & 05:49:58.485 & +07:25:18.40 & 25.05 & \({4.4}\pm{1.5}\) \\
				\hline
				4FGL J0955.8-4106 & 09:55:49.391 & -41:06:51.84 & 0.38 & 0.32 & 64.45 & SRGA J095418.7-410820 & PMN J0954-4108 & 09:54:18.748 & -41:08:19.89 & 26.07 & \({3.9}\pm{1.3}\) \\
				\hline
				4FGL J1015.1-6353 & 10:15:06.262 & -63:53:00.24 & 0.10 & 0.08 & -40.16 & SRGA J101416.2-635142 & IGR J10147-6354 & 10:14:16.192 & -63:51:41.82 & 24.63 & \({1.5}\pm{0.5}\) \\
				\hline
				4FGL J1401.3-5012 & 14:01:20.928 &  -50:12:10.80 & 0.11 & 0.09 & -28.69 & SRGA J140146.3-501333 &  & 14:01:46.318   & -50:13:32.66 & 15.67 & \({4.5}\pm{1.1}\) \\
				\hline
				4FGL J1424.2-6111c & 14:24:15.384 & -61:11:51.72 & 0.18 & 0.16 & 68.63 & SRGA J142508.7-611903 & IGR J14257-6117 & 14:25:08.722 & -61:19:02.76 & 19.22 & \({4.4}\pm{1.2}\) \\
				\hline
				4FGL J1616.6-5009 & 16:16:37.800 & -50:09:02.52 & 0.21 & 0.16 & -61.92 & SRGA J161637.8-495847 & IGR J16167-4957 & 16:16:37.848 & -49:58:46.85 & 8.54 & \({16.5}\pm{2.3}\) \\
				\hline
				4FGL J1816.1-3908 & 18:16:10.104 & -39:08:18.96 & 0.25  & 0.15 & 38.06 & SRGA J181636.0-391251 & IGR J18165-3912 & 18:16:36.045 & -39:12:51.34 & 17.30 & \({6.7}\pm{1.9}\) \\
				\hline
				4FGL J1817.7-2517 & 18:17:46.344 & -25:17:16.80 & 0.22 & 0.13 & -65.83 & SRGA J181721.6-250839 & IGR J18170-2511 & 18:17:21.595 & -25:08:38.68 & 19.58 & \({11.4}\pm{0.9}\)  \\
				\hline
				4FGL J2052.3+4437 & 20:52:19.680 & +44:37:59.88 & 0.48 & 0.21 & -41.27 & SRGA J205353.8+442304 & V1794 Cyg & 20:53:53.758 & +44:23:04.41 & 21.15 & \({5.2}\pm{1.5}\)  \\
				\hline
				4FGL J2109.3+3531 & 21:09:19.584 & +35:31:27.12 & 0.16  & 0.13 & 88.99 & SRGA J210932.5+353309 & ICRF J210931.8+353257 & 21:09:32.511   & +35:33:08.75 & 24.19 & \({5.9}\pm{1.8}\) \\
				\hline\hline
			\end{tabular}
		}
		\tablecomments{For each UFO we show the 4FGL Name, the 4FGL coordinates (4FGL RA and 4FGL Dec), the 99\% 4FGL uncertainty ellipse semimajor axis, semiminor axis and position angle (4FGL semimaj, 4FGL semimin and 4FGL PA), the name of the overlapping SRG/ART-XC source (ART-XC Name), the association of the SRG/ART-XC source as presented in the SRG/ART-XC catalog (ART-XC association), the SRG/ART-XC coordinates (ART-XC RA and ART-XC Dec) the SRG/ART-XC 99\% uncertainty radius (ART-XC err), and the SRG/ART-XC \(4-12 \text{ keV}\) flux (\(F_{4-12 \text{ keV}}\)).}
	\end{table}

	\begin{table}
		\caption{List of the selected soft X-ray counterparts to the hard X-ray sources overlapping with UFO uncertainty ellipses.}\label{tab:softx}
		\resizebox{\textwidth}{!}{
			\begin{tabular}{c|c|c|c|c|c|c|c|c|c|c}
				\hline\hline
				4FGL Name & Source \# & \(p_{\theta}\) & Insts. & RA & Dec & err. & Assoc. & Type & \(F_{3 \text{ GHz}}\) & \(F_{1.4 \text{ GHz}}\)  \\
				& & & & HH:MM:SS.SSS & DD:MM:SS.SS & arcsec & & & mJy &mJy \\
				\hline
				4FGL J0724.8+3016 & \({3}^{*}\) & \(0.999\) & \textbf{C} E & 07:25:37.293 & +29:57:14.35 & 1.247 & NVSS J072537+295714 & Seyfert 2 & \(98.4\pm1.8\) & \(150.2\pm 9.1\) \\
				\hline
				4FGL J0955.8-4106 & 1 & \(0.943\) & \textbf{E} & 09:54:19.217 & -41:08:11.84 & 2.942 & PMN J0954-4108 & Blazar cand. & & \(154.9\pm 9.4\) \\
				\hline
				4FGL J1008.2-1000 & 5 & \(0.117\) & S \textbf{E} & 10:08:48.687 & -9:54:50.91 & 2.233 & 2MASX J10084862-0954510 & Seyfert 2 & \(147.2\pm 1.3\) & \(628.0\pm 38.0\) \\
				\hline
				4FGL J1015.1-6353 & \({2}^{*}\) & \(0.996\) & S \textbf{E} & 10:14:15.109 & -63:51:50.05 & 3.593 & IGR J10147-6354 & Seyfert 1 & & \(1.3\pm 0.3\) \\
				\hline
				4FGL J1110.3-6501 & 1 & \(0.973\) & S \textbf{C} X E & 11:09:26.436 & -65:02:24.77 & 1.302 & CXOU J110926.4-650224 & X-ray & & \\
				\hline
				4FGL J1346.5+5330 & \({2}^{*}\)  & \(0.994\) & \textbf{S} & 13:45:45.598 & +53:32:49.78 & 5.035 & BZU J1345+5332 & Blazar & \(119.3\pm 0.4\) & \\
				\hline
				4FGL J1401.3-5012 & 1 & \(0.920\) & S \textbf{E} & 14:01:46.631 & -50:13:21.79 & 2.178 & & & & \\
				\hline
				4FGL J1407.7-3017 & 2 & \(0.870\) & S \textbf{C} E & 14:08:06.831 & -30:23:54.68 & 1.244 & 2MASX J14080674-3023537 & Seyfert 1 & & \(1.3\pm 0.3\) \\
				\hline
				4FGL J1424.2-6111c & 1 & \(0.989\) & S \textbf{X} E & 14:25:07.702 & -61:18:57.82 & 2.410 & IGR J14257-6117 & CV & & \\
				\hline
				4FGL J1616.6-5009 & 1 & \(0.990\) & S \textbf{C} X E & 16:16:37.752 & -49:58:44.59 & 1.247 & IGR J16167-4957 & CV & & \\
				\hline
				4FGL J1652.2-4516 & 1 & \(0.723\) & S \textbf{X} & 16:52:20.421 & -45:20:40.09 & 2.404 & XTE J1652-453 & XB & &  \\
				\hline
				4FGL J1816.1-3908 & 1 & \(0.811\) & \textbf{S} E & 18:16:35.849 & -39:12:44.52 & 7.566 & V\({}^*\) V371 CrA & Mira & & \\
				\hline
				4FGL J1817.6-3251 & \({1}^{*}\) & \(0.994\) & \textbf{S} & 18:17:43.649 & -33:01:08.60 & 4.978 & XTE J1817-330 & LMXB & & \\
				\hline
				4FGL J1817.7-2517 & 3 & \(0.983\) & S \textbf{C} X & 18:17:22.185 & -25:08:42.55 & 1.245 & OGLE BLG529.32 1342 & delta Sct & & \\
				\hline
				4FGL J2052.3+4437 & 1 & \(0.981\) & \textbf{C} X & 20:53:53.687 & +44:23:10.66 & 1.243 & V\({}^*\) V1794 Cyg & Ro\({}^{*}\) & \(30.4\pm 0.3\) & \(5.0\pm 0.6\) \\
				\hline
				4FGL J2109.3+3531 & 3 & \(0.400\) & \textbf{S} & 21:09:31.893 & +35:32:56.85 & 4.992 & ICRF J210931.8+353257 & Seyfert 1 & \(2360.4\pm 7.7\) & \(1082\pm 65\) \\
				\hline\hline
			\end{tabular}
		}
		\tablecomments{For each UFO we show the 4FGL Name, the soft X-ray source number as shown in Figure \ref{fig:xraymap} (Source \#), the confidence level corresponding to the separation between the soft X-ray source and the \(\gamma\)-ray source (\(p_{\theta}\)), the instruments that detected the soft X-ray counterpart (Insts., S=\textit{Swift}-XRT, C=\textit{Chandra}-ACIS, X=\textit{XMM-Newton}-EPIC, E=eROSITA), the soft X-ray counterpart coordinates and 99\% positional uncertainty (RA, Dec, and err.), the lower-energy association to the  soft X-ray counterpart and its type (Assoc. and Type), and the \(3 \text { GHz}\) and \(1.4 \text{ GHz}\) radio fluxes from VLASS and RACS surveys (\(F_{3 \text{ GHz}}\) and \(F_{1.4 \text{ GHz}}\), see Sect. \ref{sec:discussion}). When a soft X-ray counterpart is detected by more than one instrument, the coordinates and positional error refer to the instrument which provides the smaller positional uncertainty (marked in boldface in the column Insts.). Sources marked with an asterisk in column Source \#  fall outside the UFO \(99\%\) uncertainty ellipses.}
	\end{table}

	\begin{table}
		\caption{\textit{Swift}-XRT fluxes evaluated from the net count-rates and spectral fit results for the sources listed in Table \ref{tab:softx}.}\label{tab:swift}
		\resizebox*{!}{\dimexpr\textheight-5\baselineskip\relax}{
			\begin{tabular}{c|c|c|c|c|c|c|c|c|c|c|c}
				\hline\hline
				4FGL Name & Source \# & XRT RA & XRT Dec & OBSID & Exposure & MJD & \(a\)& \(b\) & \(a_2\) & \(\chi^2\)(d.o.f.) & \(F_{0.3-10.0 \text{ keV}}\)\\
				& & HH:MM:SS.SSS & DD:MM:SS.SS & & ks & d &  &  & & & \({10}^{-12} \text{ erg} \text{ cm}^{-2} \text{ s}^{-1}\) \\
				\hline
				4FGL J1008.2-1000 & 5 & 10:08:48.616 & -09:54:50.05 & 00045369001 & 0.64 & 55627.1 & \({2.00}_{-0.23}^{+0.24}\)  & - & - & 0.63 (6) & \({12.86}_{-1.83}^{+2.72}\) \\
				& &  &  & 00045369002 & 0.25 & 55628.3 & - & - & - & - & \({15.90}_{-1.88}^{+1.88}\) \\
				& &  &  & 00045369003 & 0.25 & 55643.8 & \({1.85}_{-0.31}^{+0.34}\) & - & - & 0.02 (2) & \({20.65}_{-5.45}^{+8.16}\) \\
				& &  &  & 00045369004 & 0.55 & 55645.0 & \({1.64}_{-0.14}^{+0.14}\) & - & - & 0.49 (10) & \({22.41}_{-3.33}^{+3.60}\) \\
				& &  &  & 00045369005 & 6.83 & 55645.1 & \({2.10}_{-0.09}^{+0.16}\) & - & \({0.08}_{-1.15}^{+0.89}\) & 0.55 (102) & \({22.25}_{-4.80}^{+38.67}\) \\
				& &  &  & 00074442010 & 0.01 & 58779.3 & - & - & - & - & \(<58.91\) \\
				& &  &  & merged & 8.60 & & - & - & - & - & \({35.02}_{-0.63}^{+0.63}\) \\
				\hline
				4FGLJ1015.1-6353 & 2 &10:14:15.098 & -63:51:50.85 & 00037048001 & 4.50 & 54405.0 & \({0.78}_{-0.59}^{+0.58}\) & - & - & 0.84 (1) & \({2.36}_{-1.44}^{+5.00}\) \\
				& & & & 03110146001 & 0.57 & 58681.5 & - & - & - & - & \({1.95}_{-0.54}^{+0.54}\) \\
				& & & & 03110146002 & 1.27 & 58689.2 & - & - & - & - & \({1.44}_{-0.31}^{+0.31}\) \\
				& & & & 03110146003 & 0.81 & 58745.8 & - & - & - & - & \({2.59}_{-0.77}^{+0.77}\) \\
				& & & & 03110146004 & 0.68 & 58747.2 & - & - & - & - & \({2.09}_{-0.52}^{+0.52}\) \\
				& &  &  & merged & 7.83 &  & \({0.84}_{-0.24}^{+0.24}\) & - & - & 0.35 (5) & \({2.59}_{-1.04}^{+1.54}\) \\
				\hline
				4FGL J1110.3-6501 & 1 & 11:09:26.321 & -65:02:24.40 & 00011274001 & 2.57 & 58586.1 & \({2.17}_{-0.43}^{+0.49}\) & - & - & 0.20 (1) & \({2.34}_{-0.91}^{+1.07}\) \\
				& &  &  & 00011274002 & 3.55 & 58732.9 & \({1.62}_{-0.24}^{+0.25}\) & - & - & 0.95 (6) & \({3.85}_{-1.07}^{+1.46}\) \\
				& &  &  & 00011274003 & 3.39 & 58912.8 & \({1.59}_{-0.16}^{+0.16}\) & - & - & 0.86 (8) & \({4.73}_{-0.97}^{+1.07}\) \\
				& &  &  & 00011274004 & 3.71 & 59655.0 & \({1.50}_{-0.23}^{+0.23}\) & - & - & 0.37 (6) & \({3.79}_{-1.03}^{+1.43}\) \\
				& &  &  & 00037050001 & 3.46 & 54487.0 & \({1.11}_{-0.20}^{+0.20}\) & - & - & 0.47 (6) & \({4.60}_{-1.40}^{+1.90}\) \\
				& &  &  & 00037050002 & 0.73 & 54507.8 & - & - & - & - & \({2.45}_{-0.67}^{+0.67}\) \\
				& &  &  & 00037050003 & 1.36 & 54887.2 & \({1.20}_{-0.34}^{+0.34}\) & - & - & 0.84 (2) & \({6.75}_{-2.81}^{+5.09}\) \\
				& &  &  & 00037050004 & 1.73 & 58172.1 & \({1.39}_{-0.49}^{+0.49}\) & - & - & 0.80 (1) & \({4.66}_{-2.13}^{+4.58}\) \\
				& &  &  & 00037050005 & 2.99 & 58177.2 & \({1.27}_{-0.29}^{+0.29}\) & - & - & 0.27 (4) & \({4.50}_{-1.64}^{+2.29}\) \\
				& &  &  & merged & 23.49 & & - & - & - & - & \({4.49}_{-0.15}^{+0.15}\) \\
				\hline
				4FGL J1346.5+5330 & 2 & 13:45:45.598 & +53:32:49.78 & 00016681001 & 1.49 & 60486.9 & \({1.73}_{-0.14}^{+0.15}\) & - & - & 0.65 (11) & \({7.22}_{-0.97}^{+1.29}\) \\
				& &  &  & 00016681002 & 0.44 & 60488.0 & \({1.93}_{-0.26}^{+0.29}\) & - & - & 0.09 (2) & \({8.19}_{-1.61}^{+2.33}\) \\
				& &  &  & 00016681003 & 1.35 & 60490.1 & \({1.67}_{-0.22}^{+0.22}\) & - & - & 0.41 (5) & \({6.45}_{-1.28}^{+1.55}\) \\
				& &  &  & 00016681004 & 1.94 & 60499.1 & \({1.75}_{-0.12}^{+0.12}\) & - & - & 0.54 (14) & \({7.57}_{-0.83}^{+1.02}\) \\
				& &  &  & 00016681006 & 2.04 & 60503.4 & \({1.78}_{-0.11}^{+0.11}\) & - & - & 0.58 (16) & \({7.41}_{-0.77}^{+0.96}\) \\
				& &  &  & 00016681007 & 1.23 & 60507.6 & \({1.85}_{-0.18}^{+0.18}\) & - & - & 0.43 (8) & \({6.20}_{-0.89}^{+1.18}\) \\
				& &  &  & 03110482001 & 0.21 & 58652.5 & - & - & - & - & \({3.25}_{-0.87}^{+0.87}\) \\
				& &  &  & 03110482002 & 0.74 & 58741.5 & - & - & - & - & \({3.98}_{-0.55}^{+0.55}\) \\
				& &  &  & 03110482003 & 0.86 & 58754.7 & \({1.18}_{-0.45}^{+0.47}\) & - & - & 0.93 (1) & \({6.04}_{-2.73}^{+6.32}\) \\
				& &  &  & 03110482004 & 1.35 & 58766.2 & \({1.42}_{-0.34}^{+0.34}\) & - & - & 0.39 (1) & \({3.03}_{-1.00}^{+1.76}\) \\
				& &  &  & 03110482005 & 0.77 & 58769.7 & - & - & - & - & \({2.75}_{-0.49}^{+0.49}\) \\
				& &  &  & merged & 12.41 & & - & - & - & - & \({7.15}_{-0.19}^{+0.19}\) \\
				\hline
				4FGL J1401.3-5012 & 1 & 14:01:47.273 & -50:13:21.16 & 00089673001 & 1.67 & 60197.2 & \({1.06}_{-0.26}^{+0.27}\) & - & - & 0.50 (3) & \({6.71}_{-2.41}^{+3.79}\) \\
				& &  &  & 00089673002 & 1.85 & 60465.2 & \({0.98}_{-0.42}^{+0.44}\) & - & - & 0.99 (2) & \({6.23}_{-3.05}^{+5.40}\) \\
				& &  &  & 00089673003 & 1.90 & 60466.3 & - & - & - & - & \({7.13}_{-0.57}^{+0.57}\) \\
				& &  &  & merged & 5.42 & & - & - & - & - & \({5.62}_{-0.29}^{+0.29}\) \\
				\hline
				4FGL J1407.7-3017 & 2 & 14:08:06.824 & -30:23:54.84 & 00037384002 & 7.30 & 54726.1 & \({1.17}_{-0.12}^{+0.11}\) & - & \({4.35}_{-0.51}^{+0.60}\) & 0.90 (41) & \({10.60}_{-2.07}^{+2.42}\) \\
				& &  &  & 00037384003 & 1.28 & 54816.9 & \({1.25}_{-0.45}^{+0.29}\) & - & \({4.46}_{-1.27}^{+2.01}\) & 0.82 (9) & \({19.85}_{-8.21}^{+21.00}\) \\
				& &  &  & 00037384004 & 9.31 & 54817.0 & \({1.13}_{-0.13}^{+0.11}\) & - & \({3.88}_{-0.40}^{+0.46}\) & 0.73 (71) & \({14.27}_{-2.65}^{+3.33}\) \\
				& &  &  & 00080127001 & 6.02 & 60128.1 & \({3.61}_{-0.69}^{+0.91}\) & - & \({0.61}_{-0.39}^{+0.28}\) & 0.71 (13) & \({6.20}_{-2.71}^{+4.26}\) \\
				& &  &  & merged & 23.90 & & - & - & - & - & \({9.61}_{-0.17}^{+0.17}\) \\
				\hline
				4FGL J1424.2-6111c & 1 & 14:25:07.567 & -61:18:59.59 & 00033091001 & 0.45 & 56666.8 & - & - & - & - & \({6.95}_{-1.83}^{+1.83}\) \\
				& &  &  & 00040976001 & 0.98 & 55548.3 & - & - & - & - & \({12.24}_{-1.60}^{+1.60}\) \\
				& &  &  & 00040976002 & 0.48 & 55642.6 & - & - & - & - & \({6.25}_{-2.21}^{+2.21}\) \\
				& &  &  & 00040976003 & 1.07 & 55736.7 & - & - & - & - & \({7.18}_{-1.34}^{+1.34}\) \\
				& &  &  & 00040976004 & 0.47 & 55774.5 & - & - & - & - & \({8.28}_{-2.27}^{+2.27}\) \\
				& &  &  & 00040976005 & 0.47 & 55788.5 & - & - & - & - & \({6.01}_{-1.76}^{+1.76}\) \\
				& &  &  & 00040976006 & 3.13 & 55789.3 & \({0.92}_{-0.44}^{+0.42}\) & - & - & 0.38 (3) & \({6.20}_{-3.56}^{+7.40}\) \\
				& &  &  & 00040976007 & 0.52 & 55817.1 & - & - & - & - & \({2.36}_{-1.19}^{+1.19}\) \\
				& &  &  & 00040976009 & 1.01 & 55830.3 & - & - & - & - & \({7.72}_{-1.40}^{+1.40}\) \\
				& &  &  & 00040976010 & 1.27 & 55831.0 & - & - & - & - & \({4.38}_{-1.24}^{+1.24}\) \\
				& &  &  & 00042378001 & 0.67 & 56001.8 & - & - & - & - & \(<676.90\) \\
				& &  &  & merged & 11.50 & & \({1.05}_{-0.20}^{+0.19}\) & - & - & 0.68 (18) & \({5.63}_{-1.67}^{+2.53}\) \\
				\hline
				4FGL J1616.6-5009 & 1 & 16:16:37.711 & -49:58:45.25 & 00035084001 & 2.22 & 53622.3 & - & - & - & - & \({36.43}_{-1.83}^{+1.83}\) \\
				& &  &  & 00035084002 & 3.07 & 53761.4 & - & - & - & - & \({38.00}_{-1.59}^{+1.59}\) \\
				& &  &  & 00035084003 & 5.17 & 54146.6 & - & - & - & - & \({37.52}_{-1.20}^{+1.20}\) \\
				& &  &  & 00042851001 & 0.58 & 56078.6 & \({2.37}_{-0.48}^{+0.49}\) & - & - & 0.01 (1) & \({33.32}_{-15.19}^{+16.30}\) \\
				& &  &  & 00042852001 & 0.49 & 56077.7 & - & - & - & - & \({69.95}_{-8.83}^{+8.83}\) \\
				& &  &  & 00087348001 & 5.15 & 57871.8 & - & - & - & - & \({123.73}_{-8.15}^{+8.15}\) \\
				& &  &  & 03110581001 & 1.01 & 58863.0 & \({1.85}_{-0.35}^{+0.35}\) & - & - & 0.16 (3) & \({31.90}_{-13.12}^{+14.59}\) \\
				& &  &  & 03110581002 & 0.26 & 60103.8 & - & - & - & - & \({85.62}_{-14.32}^{+14.32}\) \\
				& &  &  & 03110581003 & 0.19 & 60152.2 & - & - & - & - & \({79.33}_{-12.35}^{+12.35}\) \\
				& &  &  & 03110581004 & 0.62 & 60172.3 & - & - & - & - & \({60.75}_{-8.05}^{+8.05}\) \\
				& &  &  & merged & 23.63 & & - & - & - & - & \({61.24}_{-1.21}^{+1.21}\) \\
				\hline
				4FGL J1652.2-4516 & 1 & 16:52:20.411 & -45:20:39.97 & 00031440001 & 1.36 & 55015.6 & - & - & - & - & \({4044.50}_{-22.69}^{+22.69}\) \\
				& &  &  & 00031440003 & 1.98 & 55097.6 & - & - & - & - & \({335.76}_{-5.39}^{+5.39}\) \\
				& &  &  & 00031440004 & 1.96 & 55107.2 & - & - & - & - & \({408.94}_{-6.07}^{+6.07}\) \\
				& &  &  & 03110091001 & 0.14 & 60204.6 & - & - & - & - & \({5.23}_{-3.01}^{+3.01}\) \\
				& &  &  & merged & 5.43 & & - & - & - & - & \({1332.21}_{-6.64}^{+6.64}\) \\
				\hline
				4FGL J1816.1-3908 & 1 & 18:16:35.849 & -39:12:44.52 & 00011480001 & 1.79 & 58680.8 & - & - & - & - & \({1.06}_{-0.21}^{+0.21}\) \\
				\hline
				4FGL J1817.6-3251 & 1 & 18:17:43.649 & -33:01:08.60 & 00030367020 & 0.87 & 53924.4 & \({1.06}_{-0.41}^{+0.33}\) & - & \({3.39}_{-0.30}^{+0.40}\) & 0.73 (45) & \({149.23}_{-45.53}^{+75.11}\) \\
				& &  &  & 00030367021 & 0.71 & 53932.7 & \({2.08}_{-0.25}^{+0.26}\) & - & - & 0.78 (6) & \({55.26}_{-10.32}^{+11.69}\) \\
				& &  &  & 00030367022 & 1.16 & 53939.5 & \({3.41}_{-0.15}^{+0.18}\) & - & \({0.97}_{-0.39}^{+0.34}\) & 0.78 (75) & \({238.98}_{-44.95}^{+75.95}\) \\
				& &  &  & 03110554001 & 0.04 & 58793.7 & - & - & - & - & \(<46.24\) \\
				& &  &  & 03110554002 & 0.40 & 59349.9 & - & - & - & - & \(<7.53\) \\
				& &  &  & 03110554003 & 0.22 & 59885.7 & - & - & - & - & \(<10.91\) \\
				& &  &  & 03110554004 & 1.27 & 60256.6 & - & - & - & - & \(<13.60\) \\
				& &  &  & merged & 4.68 & & - & - & - & - & \({143.37}_{-2.03}^{+2.03}\) \\
				\hline
				4FGL J1817.7-2517 & 3 & 18:17:22.067 & -25:08:41.49 & 00031148001 & 5.78 & 54530.5 & \({-0.07}_{-0.11}^{+0.11}\) & - & - & 1.06 (21) & \({14.79}_{-3.34}^{+3.75}\) \\
				& &  &  & 00031148002 & 7.79 & 54533.0 & \({-0.16}_{-0.09}^{+0.09}\) & - & - & 1.03 (31) & \({15.93}_{-2.96}^{+3.83}\) \\
				& &  &  & 00031148003 & 1.79 & 55770.0 & \({0.01}_{-0.43}^{+0.42}\) & - & - & 0.91 (3) & \({11.86}_{-7.73}^{+20.14}\) \\
				& &  &  & 00031148004 & 0.44 & 55772.7 & - & - & - & - & \({8.99}_{-1.36}^{+1.36}\) \\
				& &  &  & merged & 15.80 & & \({-0.11}_{-0.07}^{+0.06}\) & - & - & 0.96 (61) & \({15.35}_{-2.04}^{+2.31}\) \\
				\hline
				4FGL J2109.3+3531 & 3 & 21:09:31.893 & +35:32:56.85 & 00011521001 & 1.53 & 58716.9 & \({2.40}_{-0.28}^{+0.31}\) & - & - & 1.02 (4) & \({7.68}_{-1.33}^{+1.39}\) \\
				& &  &  & 00041114001 & 11.97 & 55246.0 & \({2.01}_{-0.06}^{+0.06}\) & - & - & 0.68 (62) & \({8.80}_{-0.42}^{+0.45}\) \\
				& &  &  & 00081255001 & 5.56 & 58783.4 & \({1.83}_{-0.08}^{+0.09}\) & - & - & 0.77 (32) & \({9.45}_{-0.74}^{+0.87}\) \\
				& &  &  & 00081255002 & 2.98 & 58786.2 & \({1.77}_{-0.13}^{+0.13}\) & - & - & 0.73 (16) & \({8.88}_{-1.07}^{+1.49}\) \\
				& &  &  & merged & 22.05 & & \({1.92}_{-0.04}^{+0.04}\) & - & - & 0.73 (110) & \({8.96}_{-0.37}^{+0.36}\) \\
				\hline\hline
			\end{tabular}
		}
		\tablecomments{For each UFO we show the 4FGL Name, the soft X-ray source number as shown in Figure \ref{fig:xraymap} (Source \#), the \textit{Swift}-XRT coordinates of the soft X-ray counterpart to the hard X-ray source (XRT RA and XRT Dec), the \textit{Swift}-XRT OBSID, exposure and time of observation (OBSID, Exposure and MJD), the best fit power-law slope, log-parabola curvature and second power-law slope (\(a\), \(b\) and \(a_2\)), the fit reduced \(\chi^2\) and degrees of freedom (\(\chi^2\) and d.o.f.), and the flux evaluated from the best fit model or, if it was not possible to get a good fit, the flux evaluated from the net count-rate assuming a power-law model with a slope of \(1.8\) and an absorption component fixed to the Galactic value (\(F_{0.3-10.0 \text{ keV}}\)). Errors correspond to the \(1\)-\(\sigma\) confidence level for one interesting parameter.}
	\end{table}
	
	\begin{table}
		\caption{\textit{Chandra}-ACIS fluxes evaluated from the net count-rates and spectral fit results for the sources listed in Table \ref{tab:softx}.}\label{tab:chandra}
		\resizebox{\textwidth}{!}{
			\begin{tabular}{c|c|c|c|c|c|c|c|c|c|c|c}
				\hline\hline
				4FGL Name & Source \# & ACIS RA & ACIS Dec & OBSID & Exposure & MJD & \(a\)& \(b\) & \(a_2\) & \(\chi^2\)(d.o.f.) & \(F_{0.3-10.0 \text{ keV}}\)\\
				& & HH:MM:SS.SSS & DD:MM:SS.SS & & ks & d &  &  & & & \({10}^{-12} \text{ erg} \text{ cm}^{-2} \text{ s}^{-1}\) \\
				\hline
				4FGL J0724.8+3016 & 3 & 07:25:37.293 & +29:57:14.35 & 28160 & 9.26 & 60297.6 & - & - & - & - & \({0.39}_{-0.03}^{+0.03}\) \\
				& &  &  & 29132 & 9.26 & 60297.9 & - & - & - & - & \({0.34}_{-0.03}^{+0.03}\) \\
				& &  &  & merged & 18.52 &  & - & - & - & - & \({0.36}_{-0.02}^{+0.02}\) \\
				\hline
				4FGL J1110.3-6501 & 1 & 11:09:26.436 & -65:02:24.77 & 09066 & 5.11 & 54721.6 & \({1.43}_{-0.09}^{+0.09}\) & - & - & 0.48 (36) & \({4.71}_{-0.58}^{+0.63}\) \\
				\hline
				4FGL J1407.7-3017 & 2 & 14:08:06.831 & -30:23:54.68 & 23706 & 1.95 & 59312.8 & - & - & - & - & \({6.00}_{-0.25}^{+0.25}\) \\
				\hline
				4FGL J1616.6-5009 & 1 & 16:16:37.752 & -49:58:44.59 & 05473 & 4.98 & 53534.8 & \({1.93}_{-0.51}^{+0.50}\) & - & - & 0.85 (2) & \({44.10}_{-23.56}^{+31.90}\) \\
				\hline
				4FGL J1817.7-2517 & 3 & 18:17:22.185 & -25:08:42.55 & 09067 & 4.90 & 54592.0  & - & - & - & - & \({2.65}_{-0.16}^{+0.16}\) \\
				\hline
				4FGLJ2052.3+4437 & 1 & 20:53:53.687 & +44:23:10.66 & 23418 & 24.65 & 59476.5 & - & - & - & - & \({50.50}_{-0.33}^{+0.33}\) \\
				& & &  & 26126 & 16.74 & 59492.8 & - & - & - & - & \({42.72}_{-0.36}^{+0.36}\) \\
				& &  &  & merged & 41.39 &  & - & - & - & - & \({47.21}_{-0.24}^{+0.24}\) \\
				\hline\hline
			\end{tabular}
		}
		\tablecomments{For each UFO we show the 4FGL Name, the soft X-ray source number as shown in Figure \ref{fig:xraymap} (Source \#), the \textit{Chandra}-ACIS coordinates of the soft X-ray counterpart to the hard X-ray source (ACIS RA and ACIS Dec), the \textit{Chandra}-ACIS OBSID, exposure and time of observation (OBSID, Exposure and MJD), the best fit power-law slope, log-parabola curvature and second power-law slope (\(a\), \(b\) and \(a_2\)), the fit reduced \(\chi^2\) and degrees of freedom (\(\chi^2\) and d.o.f.), and the flux evaluated from the best fit model or, if it was not possible to get a good fit, the flux evaluated from the net count-rate assuming a power-law model with a slope of \(1.8\) and an absorption component fixed to the Galactic value (\(F_{0.3-10.0 \text{ keV}}\)). Errors correspond to the \(1\)-\(\sigma\) confidence level for one interesting parameter.}
	\end{table}

	\begin{table}
		\caption{\textit{XMM-Newton}-EPIC fluxes evaluated from the net count-rates and spectral fit results for the sources listed in Table \ref{tab:softx}.}\label{tab:xmm}
		\resizebox{\textwidth}{!}{
			\begin{tabular}{c|c|c|c|c|c|c|c|c|c|c|c}
				\hline\hline
				4FGL Name & Source \# & EPIC RA & EPIC Dec & OBSID & Exposure & MJD & \(a\)& \(b\) & \(a_2\) & \(\chi^2\)(d.o.f.) & \(F_{0.3-10.0 \text{ keV}}\)\\
				& & HH:MM:SS.SSS & DD:MM:SS.SS & & ks & d &  &  & & & \({10}^{-12} \text{ erg} \text{ cm}^{-2} \text{ s}^{-1}\) \\
				\hline
				4FGL J1110.3-6501 & 1 & 11:09:26.620 & -65:02:25.08 & 0764344301 & 57.05 & 58289.8 & \({1.78}_{-0.08}^{+0.08}\) & - & - & 0.21 (200) & \({4.39}_{-0.39}^{+0.43}\) \\
				& &  &  & 0851180201 & 14.15 & 58648.8 & \({1.49}_{-0.41}^{+0.42}\) & - & - & 0.03 (233) & \({3.95}_{-1.73}^{+3.65}\) \\
				& &  &  & 0851180301 & 33.49 & 58649.6 & - & - & - & - & \({4.72}_{-0.05}^{+0.05}\) \\
				& &  &  & 0864190201 & 20.66 & 59641.0 & \({1.73}_{-0.46}^{+0.47}\) & - & - & 0.03 (166) & \({3.79}_{-1.63}^{+2.72}\) \\
				& &  &  & merged & 125.35 &  & \({1.68}_{-0.02}^{+0.02}\) & - & - & 0.69 (843) & \({4.04}_{-0.09}^{+0.10}\) \\
				\hline
				4FGL J1424.2-6111c & 1 & 14:25:07.702 & -61:18:57.82 & 0780700101 & 20.47 & 57773.3 & \({1.05}_{-0.04}^{+0.04}\) & - & - & 0.95 (459) & \({5.61}_{-0.42}^{+0.45}\) \\
				\hline
				4FGL J1616.6-5009 & 1 & 16:16:37.781 & -49:58:45.92 & 0402920101 & 19.04 & 53964.7 & - & - & - & - & \({36.19}_{-0.16}^{+0.16}\) \\
				\hline
				4FGL J1652.2-4516 & 1 & 16:52:20.421 & -45:20:40.09 & 0610000701 & 33.30 & 55065.1 & - & - & - & - & \({1130.28}_{-1.39}^{+1.39}\) \\
				\hline
				4FGL J1743.9-3539 & 1 & 17:43:01.340 & -36:22:22.63 & 0675040101 & 33.59 & 55986.8 & - & - & - & - & \({4.00}_{-0.03}^{+0.03}\) \\
				\hline
				4FGL J1817.7-2517 & 3 & 18:17:22.218 & -25:08:42.64 & 0601270301 & 29.60 & 55081.1 & \({0.56}_{-0.18}^{+0.17}\) & \({1.11}_{-0.23}^{+0.24}\) & - & 0.72 (100) & \({0.28}_{-0.07}^{+0.10}\) \\
				\hline
				4FGLJ2052.3+4437 & 1 & 20:53:53.584 & +44:23:11.29 & 0100241101 & 5.32 & 52590.0 & - & - & - & - & \({93.21}_{-0.38}^{+0.38}\) \\
				4FGLJ2052.3+4437 & & & & 0100241201 & 4.41 & 52602.0 & - & - & - & - & \({137.88}_{-0.51}^{+0.51}\) \\
				4FGLJ2052.3+4437 & & & & 0556050101 & 10.11 & 54588.5 & - & - & - & - & \({89.60}_{-0.26}^{+0.26}\) \\
				4FGLJ2052.3+4437 & & & & 0679580201 & 5.91 & 55889.4 & - & - & - & - & \({596.39}_{-1.63}^{+1.63}\) \\
				4FGLJ2052.3+4437 & & &  & merged & 25.75 & & - & - & - & - & \({177.99}_{-0.28}^{+0.28}\) \\
				\hline\hline
			\end{tabular}
		}
		\tablecomments{For each UFO we show the 4FGL Name, the soft X-ray source number as shown in Figure \ref{fig:xraymap} (Source \#), the \textit{XMM-Newton}-EPIC coordinates of the soft X-ray counterpart to the hard X-ray source (EPIC RA and EPIC Dec), the \textit{XMM-Newton}-EPIC OBSID, exposure and time of observation (OBSID, Exposure and MJD), the best fit power-law slope, log-parabola curvature and second power-law slope (\(a\), \(b\) and \(a_2\)), the fit reduced \(\chi^2\) and degrees of freedom (\(\chi^2\) and d.o.f.), and the flux evaluated from the best fit model or, if it was not possible to get a good fit, the flux evaluated from the net count-rate assuming a power-law model with a slope of \(1.8\) and an absorption component fixed to the Galactic value (\(F_{0.3-10.0 \text{ keV}}\)). Errors correspond to the \(1\)-\(\sigma\) confidence level for one interesting parameter.}
	\end{table}

	\begin{table}
		\caption{eROSITA fluxes evaluated from the net count-rates for the sources listed in Table \ref{tab:softx}.}\label{tab:erass}
		\resizebox{\textwidth}{!}{
			\begin{tabular}{c|c|c|c|c|c|c|c}
				\hline\hline
				4FGL Name & Source \# & eRASS Name & eRASS RA & eRASS Dec & \({\text{MJD}}_{\text{start}}\) &  \({\text{MJD}}_{\text{stop}}\) & \(F_{0.3-10 \text{ keV}}\)  \\
				& & & HH:MM:SS.SSS & DD:MM:SS.SS & d & d & \({10}^{-12} \text{ erg} \text{ cm}^{-2} \text{ s}\) \\
				\hline
				4FGL J0724.8+3016  & 3 & 1eRASS J072536.6+295716 & 07:25:36.683 & +29:57:16.94 & 58954.4 & 58954.9 & \(0.21\pm0.12\) \\
				\hline
				4FGL J0955.8-4106 & 1 & 1eRASS J095419.2-410811 & 09:54:19.217 & -41:08:11.84 & 59001.1 & 59002.1 & \(3.56\pm0.31\) \\
				\hline
				4FGL J1008.2-1000 & 5 & 1eRASS J100848.6-095450 & 10:08:48.687 & -09:54:50.91 & 58991.6 & 58992.2& \(8.10\pm0.47\)  \\
				\hline
				4FGLJ1015.1-6353 & 2 & 1eRASS J101415.1-635150 & 10:14:15.109 & -63:51:50.05 & 58852.97 & 58855.97 & \(1.02\pm0.10\) \\
				\hline
				4FGL J1110.3-6501   & 1 & 1eRASS J110926.1-650224 & 11:09:26.186 & -65:02:24.70 & 58866.5 & 58869.3 & \(5.50\pm0.29\) \\
				\hline
				4FGL J1401.3-5012  & 1 & 1eRASS J140146.6-501321 & 14:01:46.631 & -50:13:21.79 & 58888.2 & 58894.4 & \(4.47\pm0.23\) \\
				\hline
				4FGL J1407.7-3017   & 2 & 1eRASS J140806.8-302354 & 14:08:06.880 & -30:23:54.78 & 58877.3 & 58879.0 & \(8.90\pm0.38\) \\
				\hline
				4FGL J1424.2-6111c   & 1 & 1eRASS J142507.9-611859 & 14:25:07.908 & -61:18:59.17 & 58904.9 & 58906.0 & \(4.90\pm0.61\) \\
				\hline
				4FGL J1616.6-5009   & 1 & 1eRASS J161637.9-495844 & 16:16:37.907 & -49:58:44.86 & 58920.3 & 58921.5 & \(32.17\pm1.51\) \\
				\hline
				4FGL J1816.1-3908  & 1 & 1eRASS J181635.8-391236 & 18:16:35.834 & -39:12:36.19 & 58943.176 & 58943.84 & \(0.43\pm0.12\) \\
				\hline\hline
			\end{tabular}
		}
		\tablecomments{For each UFO we show the 4FGL Name, the soft X-ray source number as shown in Figure \ref{fig:xraymap} (Source \#), the eRASS1 name and coordinates of the soft X-ray counterpart to the hard X-ray source (eRASS Name, eRASS RA and eRASS Dec), the start and stop time of observation (\({\text{MJD}}_{\text{start}}\) and  \({\text{MJD}}_{\text{stop}}\)), and the flux evaluated from the net count-rate assuming a power-law model with a slope of \(1.8\) and an absorption component fixed to the Galactic value (\(F_{0.3-10.0 \text{ keV}}\)).}
	\end{table}

	\begin{table}
		\caption{AllWISE counterparts to soft X-ray sources listed in Table \ref{tab:softx}.}\label{tab:wise}
		\resizebox{\textwidth}{!}{
			\begin{tabular}{c|c|c|c|c|c|c|c|c|c}
				\hline\hline
				4FGL Name & Source \# & AllWISE Name & AllWISE RA & AllWISE Dec &W1 & W2 & W3 & W4 & Blazar CAND \\
				& & & HH:MM:SS.SSS & DD:MM:SS.SS & & & & &  \\
				\hline
				4FGL J0724.8+3016 & 3 & WISE J072537.26+295714.9 &  07:25:37.262 & +29:57:14.99 & \(10.577\pm0.023\) & \(9.425\pm 0.019\) & \(6.115\pm 0.014\) & \(3.385 \pm 0.023\) & BZQ \\
				\hline
				4FGLJ0955.8-4106 & 1 & WISE J095419.07-410812.8 & 09:54:19.070 & -41:08:12.84 & \(12.890\pm 0.023\) & \(11.907 \pm 0.022\) & \(9.375\pm 0.034\) & \(7.407\pm 0.107\) & BZB \\
				\hline
				4FGL J1008.2-1000 & 5 & WISE J100848.57-095450.7 & 10:08:48.576 & -09:54:50.70 & \(10.673\pm 0.024\) & \(9.998\pm 0.021\) & \(7.790\pm 0.023\) & \(5.645\pm 0.054\) & BZB \\
				\hline
				4FGLJ1015.1-6353 & 2 & WISE J101415.55-635150.1 & 10:14:15.553 & -63:51:50.18 & \(10.614\pm 0.024\) & \(9.645\pm 0.020\) & \(7.207\pm 0.014\) & \(4.965\pm 0.028\) & MIXED \\
				\hline
				4FGL J1346.5+5330 & 2 & WISE J134545.35+533252.2 & 13:45:45.353 & +53:32:52.29 & \(12.047\pm 0.023\) & \(11.294\pm 0.021\) & \(9.416\pm 0.031\) & \(7.457\pm 0.109\) & BZB \\
				\hline
				4FGL J1401.3-5012 & 1 & WISE J140146.68-501321.9 & 14:01:46.682 & -50:13:21.99 & \(13.715\pm 0.025\) & \(13.426\pm 0.031\) & \(12.273\pm 0.219\) & \(>9.157\) & \\
				\hline
				4FGL J1407.7-3017 & 2 & WISE J140806.78-302353.9 & 14:08:06.785 & -30:23:53.95 & \(11.396\pm 0.023\)  & \(10.361\pm 0.021\) & \(7.449\pm 0.016\) & \(5.234\pm 0.030\) & MIXED \\
				\hline
				4FGLJ2052.3+4437 & 1 & WISE J205353.67+442310.9 & 20:53:53.674 & +44:23:10.99 & \(4.352\pm 0.170\) & \(4.655 \pm 0.038\) & \(4.905\pm 0.016\) & \(4.162 \pm 0.107\) & BZQ \\
				\hline
				4FGL J2109.3+3531 & 3 & WISE J210931.87+353257.6 & 21:09:31.874 & +35:32:57.63 & \(11.325\pm 0.023\) & \(10.341 \pm 0.019\) & \(7.556\pm 0.017\) & \(5.086 \pm 0.037\) & MIXED \\
				\hline\hline
			\end{tabular}
		}
		\tablecomments{For each UFO we show the 4FGL Name, the soft X-ray source number as shown in Figure \ref{fig:xraymap} (Source \#), the AllWISE name and coordinates of the AllWISE counterpart to the soft X-ray source (AllWISE Name, AllWISE RA and AllWISE Dec), the AllWISE magnitudes corrected for Galactic absorption (W1, W2, W3 and W4), and the blazar candidate class (BZB=BL Lac, BZQ=FSRQ, MIXED=blazar of uncertain type, see Figure \ref{fig:wise_colors}).}
	\end{table}

	\begin{table}
		\caption{UVOT magnitudes for the soft X-ray sources listed in Table \ref{tab:softx}.}\label{tab:uvot}
		\resizebox{\textwidth}{!}{
			\begin{tabular}{c|c|c|c|c|c|c|c|c|c|c|c|c|c|c|c|c|c|c}
				\hline\hline
				4FGL Name & Source \# & UVOT RA & UVOT Dec & OBSID & MJD & \(\text{Exposure}_{W2}\) & \(\text{mag}_{W2}\) & \(\text{Exposure}_{M2}\) & \(\text{mag}_{M2}\) & \(\text{Exposure}_{W1}\) & \(\text{mag}_{W1}\) & \(\text{Exposure}_{U}\) & \(\text{mag}_{U}\) & \(\text{Exposure}_{B}\) & \(\text{mag}_{B}\) & \(\text{Exposure}_{V}\) & \(\text{mag}_{V}\) \\
				& & HH:MM:SS.SSS & DD:MM:SS.SS & & d & ks & & ks & & ks & & ks & & ks & & ks & \\
				\hline
				4FGL J1008.2-1000 & 5 & 10:08:48.616 & -09:54:50.05 & 00045369001 & 55627.1 &  &  &  &  & 0.63 & \(13.97(0.02)\) &  &  &  &  &  &  \\
				& & & & 00045369002 & 55628.3 &  &  &  &  &  &  & 0.25 & \(14.04(0.03)\) &  &  &  &  \\
				& & & & 00045369003 & 55643.8 &  &  &  &  & 0.25 & \(13.96(0.03)\) &  &  &  &  &  &  \\
				& & & & 00045369004 & 55645.0 & 0.55 & \(13.93(0.03)\) &  &  &  &  &  &  &  &  &  &  \\
				& & & & 00045369005 & 55646.0 & 5.08 & \(13.94(0.02)\) & 1.69 & \(14.02(0.02)\) &  &  &  &  &  &  &  &  \\
				& & & & merged &  & 5.63 & \(13.90(0.02)\) &  &  & 0.89 & \(14.11(0.02)\) & 0.31 & \(14.26(0.03)\) &  &  &  &  \\
				\hline
				4FGLJ1015.1-6353 & 2 & 10:14:15.098 & -63:51:50.85 & 00037048001 & 54405.0 & 4.68 & \(13.86(0.02)\) &  &  &  &  &  &  &  &  &  &  \\
				& & & & 03110146001 & 58681.5 & 0.56 & \(13.88(0.04)\) &  &  &  &  &  &  &  &  &  &  \\
				& & & & 03110146002 & 58689.2 & 1.26 & \(13.88(0.03)\) &  &  &  &  &  &  &  &  &  &  \\
				& & & & 03110146003 & 58745.8 & 0.80 & \(13.87(0.04)\) &  &  &  &  &  &  &  &  &  &  \\
				& & & & 03110146004 & 58747.2 &  &  &  &  & 0.67 & \(13.91(0.04)\) &  &  &  &  &  &  \\
				& & & & merged &  & 7.30 & \(13.82(0.02)\) &  &  &  &  &  &  &  &  &  &  \\
				\hline
				4FGL J1110.3-6501 & 1 & 11:09:26.321 & -65:02:24.40 & 00011274001 & 58586.1 &  &  & 2.54 & \(>12.63\) &  &  &  &  &  &  &  &  \\
				& & & & 00011274002 & 58732.9 & 0.21 & \(>10.66\) &  &  &  &  & 3.32 & \(11.59(0.11)\) &  &  &  &  \\
				& & & & 00011274003 & 58912.8 &  &  &  &  &  &  & 3.35 & \(11.31(0.07)\) &  &  &  &  \\
				& & & & 00011274004 & 59655.0 &  &  &  &  & 3.66 & \(13.54(0.58)\) &  &  &  &  &  &  \\
				& & & & 00037050004 & 58172.1 &  &  &  &  &  &  &  &  &  &  & 1.67 & \(10.22(0.06)\) \\
				& & & & 00037050005 & 58177.2 &  &  &  &  &  &  &  &  &  &  & 2.94 & \(10.11(0.05)\) \\
				& & & & merged &  &  &  &  &  &  &  & 6.67 & \(15.15(0.06)\) &  &  & 4.60 & \(15.31(0.04)\) \\
				\hline
				4FGL J1346.5+5330 & 2 & 13:45:45.598 & +53:32:49.78 & 00016681001 & 60486.9 & 0.63 & \(16.59(0.04)\) &  &  & 0.41 & \(16.88(0.06)\) & 0.21 & \(17.03(0.06)\) & 0.21 & \(17.79(0.07)\) &  &  \\
				& & & & 00016681002 & 60488.0 & 0.15 & \(16.66(0.08)\) & 0.07 & \(16.76(0.15)\) & 0.07 & \(16.70(0.11)\) & 0.04 & \(16.93(0.12)\) & 0.04 & \(17.83(0.14)\) & 0.04 & \(17.04(0.17)\) \\
				& & & & 00016681003 & 60490.1 & 0.44 & \(16.70(0.05)\) & 0.29 & \(16.82(0.08)\) & 0.22 & \(16.86(0.08)\) & 0.11 & \(16.87(0.07)\) & 0.11 & \(17.90(0.09)\) & 0.11 & \(17.10(0.10)\) \\
				& & & & 00016681004 & 60499.1 & 0.63 & \(16.48(0.04)\) & 0.46 & \(16.76(0.06)\) & 0.31 & \(16.93(0.07)\) & 0.16 & \(16.95(0.07)\) & 0.16 & \(17.89(0.09)\) & 0.16 & \(17.22(0.09)\) \\
				& & & & 00016681007 & 60507.6 & 0.40 & \(16.67(0.06)\) & 0.30 & \(16.88(0.09)\) & 0.20 & \(17.04(0.10)\) & 0.10 & \(16.95(0.10)\) & 0.10 & \(18.15(0.17)\) & 0.10 & \(17.34(0.13)\) \\
				& & & & 03110482001 & 58652.5 &  &  &  &  &  &  & 0.21 & \(17.17(0.07)\) &  &  &  &  \\
				& & & & 03110482002 & 58741.5 & 0.73 & \(17.71(0.07)\) &  &  &  &  &  &  &  &  &  &  \\
				& & & & 03110482003 & 58754.7 &  &  & 0.85 & \(17.78(0.08)\) &  &  &  &  &  &  &  &  \\
				& & & & 03110482004 & 58766.2 &  &  & 1.33 & \(17.78(0.06)\) &  &  &  &  &  &  &  &  \\
				& & & & 03110482005 & 58769.7 & 0.76 & \(17.47(0.06)\) &  &  &  &  &  &  &  &  &  &  \\
				& & & & merged &  & 3.74 & \(16.90(0.03)\) & 3.30 & \(17.34(0.04)\) & 1.22 & \(16.90(0.04)\) & 0.82 & \(17.02(0.04)\) & 0.61 & \(17.89(0.05)\) & 0.40 & \(17.20(0.06)\) \\
				\hline
				4FGL J1401.3-5012 & 1 & 14:01:47.273 & -50:13:21.16 & 00089673001 & 60197.2 & 0.89 & \(15.84(0.06)\) & 0.77 & \(15.70(0.07)\) &  &  &  &  &  &  &  &  \\
				& & & & 00089673002 & 60465.2 & 1.83 & \(15.52(0.04)\) &  &  &  &  &  &  &  &  &  &  \\
				& & & & 00089673003 & 60466.3 &  &  & 1.89 & \(15.25(0.04)\) &  &  &  &  &  &  &  &  \\
				& & & & merged &  & 2.72 & \(15.75(0.03)\) & 2.66 & \(15.45(0.04)\) &  &  &  &  &  &  &  &  \\
				\hline
				4FGL J1407.7-3017 & 2 & 14:08:06.824 & -30:23:54.84 & 00037384002 & 54727.0 &  &  & 6.33 & \(14.52(0.02)\) & 0.88 & \(14.50(0.02)\) &  &  &  &  &  &  \\
				& & & & 00037384003 & 54816.9 &  &  &  &  &  &  & 0.60 & \(14.41(0.03)\) &  &  &  &  \\
				& & & & 00037384004 & 54817.0 & 9.23 & \(14.39(0.02)\) &  &  &  &  &  &  &  &  &  &  \\
				& & & & 00080127001 & 60128.5 & 0.56 & \(14.74(0.03)\) & 0.39 & \(14.61(0.04)\) & 0.28 & \(14.68(0.03)\) & 4.47 & \(14.92(0.02)\) & 0.14 & \(15.96(0.04)\) & 0.14 & \(15.60(0.05)\) \\
				& & & & merged &  & 9.79 & \(14.18(0.02)\) & 6.72 & \(14.32(0.02)\) & 1.16 & \(14.42(0.02)\) & 5.07 & \(14.92(0.02)\) &  &  &  &  \\
				\hline
				4FGL J1424.2-6111c & 1 & 14:25:07.567 & -61:18:59.59 & 00033091001 & 56666.8 &  &  &  &  &  &  & 0.45 & \({18.37}^*(0.11)\) &  &  &  &  \\
				& & & & 00040976001 & 55548.3 &  &  &  &  &  &  & 0.97 & \({18.62}^*(0.11)\) &  &  &  &  \\
				& & & & 00040976002 & 55642.6 &  &  & 0.48 & \(>{19.43}^*\) &  &  &  &  &  &  &  &  \\
				& & & & 00040976003 & 55737.5 & 0.72 & \({20.55}^*(0.66)\) &  &  &  &  & 0.34 & \({18.96}^*(0.23)\) &  &  &  &  \\
				& & & & 00040976004 & 55774.5 &  &  & 0.46 & \({19.86}^*(0.49)\) &  &  &  &  &  &  &  &  \\
				& & & & 00040976005 & 55788.5 &  &  &  &  &  &  & 0.47 & \({18.36}^*(0.14)\) &  &  &  &  \\
				& & & & 00040976006 & 55789.3 & 3.11 & \({19.78}^*(0.18)\) &  &  &  &  &  &  &  &  &  &  \\
				& & & & 00040976009 & 55830.3 &  &  & 1.01 & \(>{19.83}^*\) &  &  &  &  &  &  &  &  \\
				& & & & 00040976010 & 55831.0 &  &  &  &  & 1.26 & \({19.56}^*(0.27)\) &  &  &  &  &  &  \\
				& & & & merged &  & 3.83 & \({19.88}^*(0.18)\) & 3.14 & \({21.05}^*(0.73)\) &  &  & 2.66 & \({18.54}^*(0.07)\) &  &  &  &  \\
				\hline
				4FGL J1652.2-4516 & 1 & 16:52:20.411 & -45:20:39.97 & 00031440001 & 55015.6 &  &  &  &  & 1.34 & \(>{20.57}^*\) &  &  &  &  &  &  \\
				& & & & 00031440003 & 55097.6 & 1.96 & \(>{20.84}^*\) &  &  &  &  &  &  &  &  &  &  \\
				& & & & 00031440004 & 55107.2 &  &  & 1.95 & \(>{20.67}^*\) &  &  &  &  &  &  &  &  \\
				& & & & 03110091001 & 60204.6 &  &  &  &  &  &  & 0.14 & \(>{19.16}^*\) &  &  &  &  \\
				\hline
				4FGL J1816.1-3908 & 1 & 18:16:35.849 & -39:12:44.52 & 00011480001 & 58680.8 &  &  &  &  &  &  & 1.76 & \(18.16(0.08)\) &  &  &  &  \\
				\hline
				4FGL J1817.6-3251 & 1 & 18:17:43.649 & -33:01:08.60 & 00030367020 & 53924.4 &  &  &  &  & 0.87 & \(15.04(0.04)\) &  &  &  &  &  &  \\
				& & & & 00030367021 & 53932.7 &  &  &  &  & 0.85 & \(15.14(0.04)\) &  &  &  &  &  &  \\
				& & & & 00030367022 & 53939.5 &  &  &  &  & 1.22 & \(15.14(0.04)\) &  &  &  &  &  &  \\
				& & & & 03110554004 & 60256.6 &  &  &  &  &  &  & 0.50 & \(>17.38\) &  &  &  &  \\
				& & & & merged &  &  &  &  &  & 2.94 & \(15.11(0.03)\) &  &  &  &  &  &  \\
				\hline
				4FGL J1817.7-2517 & 3 & 18:17:22.067 & -25:08:41.49 & 00031148001 & 54530.5 & 1.96 & \(9.90(0.03)\) & 1.23 & \(9.81(0.04)\) & 0.98 & \(11.57(0.04)\) & 0.49 & \(12.77(0.04)\) & 0.49 & \(14.02(0.04)\) & 0.49 & \(14.59(0.07)\) \\
				& & & & 00031148002 & 54533.0 & 2.51 & \(9.85(0.03)\) & 1.91 & \(9.65(0.03)\) & 1.26 & \(11.47(0.03)\) & 0.63 & \(12.67(0.03)\) & 0.63 & \(14.01(0.03)\) & 0.63 & \(14.58(0.06)\) \\
				& & & & 00031148003 & 55770.0 &  &  & 1.74 & \(10.13(0.04)\) &  &  &  &  &  &  &  &  \\
				& & & & 00031148004 & 55772.7 &  &  &  &  &  &  & 0.43 & \(9.52(0.03)\) &  &  &  &  \\
				& & & & merged &  & 4.48 & \(9.87(0.02)\) & 4.88 & \(9.71(0.03)\) & 2.24 & \(11.52(0.03)\) & 1.55 & \(12.67(0.03)\) & 1.12 & \(14.02(0.03)\) & 1.12 & \(14.59(0.05)\) \\
				\hline
				4FGL J2109.3+3531 & 3 & 21:09:31.893 & +35:32:56.85 & 00011521001 & 58716.9 &  &  &  &  &  &  & 1.52 & \(15.78(0.03)\) &  &  &  &  \\
				& & & & 00041114001 & 55247.4 &  &  & 4.83 & \(14.02(0.02)\) & 7.21 & \(14.31(0.02)\) &  &  &  &  &  &  \\
				& & & & 00081255001 & 58783.6 & 0.53 & \(15.39(0.05)\) & 0.42 & \(15.75(0.08)\) & 4.15 & \(16.03(0.03)\) & 0.13 & \(16.34(0.07)\) & 0.13 & \(17.38(0.08)\) & 0.13 & \(15.33(0.05)\) \\
				& & & & 00081255002 & 58786.2 & 0.54 & \(15.39(0.05)\) & 1.71 & \(15.71(0.04)\) & 0.27 & \(15.96(0.07)\) & 0.14 & \(16.21(0.06)\) & 0.14 & \(17.27(0.08)\) & 0.14 & \(15.36(0.05)\) \\
				& & & & merged &  & 1.07 & \(15.39(0.04)\) & 6.96 & \(14.09(0.02)\) & 11.63 & \(14.66(0.02)\) & 1.78 & \(16.36(0.03)\) & 0.27 & \(17.33(0.06)\) & 0.27 & \(15.34(0.04)\) \\
				\hline\hline
			\end{tabular}
		}
		\tablecomments{For each UFO we show the 4FGL Name, the soft X-ray source number as shown in Figure \ref{fig:xraymap} (Source \#), the coordinates of the UVOT counterpart to the soft X-ray source (UVOT RA and UVOT Dec) the UVOT OBSID and time of observation (OBSID and MJD) and the exposure times and magnitude for each UVOT filter (\(\text{Exposure}_{W2}\), \(\text{mag}_{W2}\), \(\text{Exposure}_{M2}\), \(\text{mag}_{M2}\), \(\text{Exposure}_{W1}\), \(\text{mag}_{W1}\), \(\text{Exposure}_{U}\), \(\text{mag}_{U}\), \(\text{Exposure}_{B}\), \(\text{mag}_{B}\), \(\text{Exposure}_{V}\) and \(\text{mag}_{V}\)). Note that UVOT magnitudes for UFOs 4FGL J1424.2-6111c and 4FGL J1652.2-4516 (marked with an asterisk) are not corrected for Galactic absorption due to the high reddening at their coordinates.}
	\end{table}

	\begin{table}
		\caption{Parallax and proper motion for the GAIA DR3 counterparts to the soft X-ray sources listed in Table \ref{tab:softx}.}\label{tab:gaia}
		\resizebox{\textwidth}{!}{
			\begin{tabular}{c|c|c|c|c|c|c}
				\hline\hline
				4FGL Name & Source \# & GAIA Name & GAIA RA & GAIA Dec & parallax & pm \\
				& & & HH:MM:SS.SSS & DD:MM:SS.SS & \(\text{mas}\) & \(\text{mas} \text{ yr}^{-1}\) \\
				\hline
				4FGL J0955.8-4106 & 1 & Gaia DR3 5419607815353769216 & 09:54:19.061 & -41:08:12.84 & \(0.032 \pm 0.080\) & \(0.095\pm 0.101\) \\
				\hline
				4FGL J1008.2-1000 & 5 & Gaia DR3 3767237487450870912 & 10:08:48.592 & -09:54:50.89 & \(-0.100 \pm 0.047\) & \(0.026\pm 0.095\) \\
				\hline
				4FGL J1015.1-6353 & 2 & Gaia DR3 5252486205464496896 & 10:14:15.544 & -63:51:50.13 & \(-0.005 \pm 0.046\) & \(0.020\pm 0.076\) \\
				\hline
				4FGL J1110.3-6501 & 1 & Gaia DR3 5240167590731178624 & 11:09:26.406 & -65:02:24.82 & \(0.449 \pm 0.637\) &\(1.978\pm 1.168\) \\
				\hline
				4FGL J1346.5+5330 & 2 & Gaia DR3 1561000748426882816 & 13:45:45.355 & +53:32:52.29 & \(-0.052 \pm 0.099\) &\(0.070\pm 0.157\) \\
				\hline
				4FGL J1401.3-5012 & 1 & Gaia DR3 6090855029847914368 & 14:01:46.682 & -50:13:21.89 & \(0.864 \pm 0.058\) & \(6.602\pm 0.076\) \\
				\hline
				4FGL J1407.7-3017 & 2 & Gaia DR3 6172850662806629888 & 14:08:06.782 & -30:23:53.96 & \(-0.029 \pm 0.056\) &\(0.209\pm 0.085\) \\
				\hline
				4FGL J1424.2-6111c & 1 & Gaia DR3 5854580937493175040 & 14:25:07.590 & -61:18:58.05 & \(0.501 \pm 0.111\) &\(0.866 \pm 0.145\) \\
				\hline
				4FGL J1616.6-5009 & 1 & Gaia DR3 5935487817742197120 & 16:16:37.761 & -49:58:44.58 & \(0.642 \pm 0.055\) & \(1.132 \pm 0.079\) \\
				\hline
				4FGL J1652.2-4516 & 1 & Gaia DR3 5964188301032037248 & 16:52:20.517 & -45:20:41.49 & \(0.366 \pm 0.491\) & \(3.654 \pm 0.853\) \\
				\hline
				4FGL J1816.1-3908 & 1 & Gaia DR3 6727363681257990016 & 18:16:35.946 & -39:12:46.43 & \(0.052 \pm 0.183\) & \(0.317 \pm 0.231\) \\
				\hline
				4FGL J1817.6-3251 & 1 & Gaia DR3 4045481137009875712 & 18:17:43.699 & -33:01:07.17 & \(0.226 \pm 0.196\) & \(9.498 \pm 0.318\) \\
				\hline
				4FGL J1817.7-2517 & 3 & Gaia DR3 4065169988601422848 & 18:17:22.174 & -25:08:42.48 & \(0.221 \pm 0.086\) & \(1.938 \pm 0.112\) \\
				\hline
				4FGL J2052.3+4437 & 1 & Gaia DR3 2162964329341318656 & 20:53:53.692 & +44:23:11.07 & \(8.891 \pm 0.015\) &  \(26.465 \pm 0.021\) \\                
				\hline
				4FGL J2109.3+3531 & 3 & Gaia DR3 1868050125092374784 & 21:09:31.879 & +35:32:57.60 & \(0.069 \pm 0.078\) &  \(0.003 \pm 0.096\) \\
				\hline\hline
			\end{tabular}
		}
		\tablecomments{For each UFO we show the 4FGL Name, the soft X-ray source number as shown in Figure \ref{fig:xraymap} (Source \#), the name and coordinates of the GAIA counterpart to the soft X-ray source (GAIA name, GAIA RA and GAIA Dec), the GAIA parallax and proper motion (parallax, pm).}
	\end{table}

	\begin{table}
		\caption{Best fit parameters for the SEDs presented in Fig. \ref{fig:seds}.}\label{tab:sed}
		\resizebox{\textwidth}{!}{
			\begin{tabular}{c|c|c|c|c|c|c|c|c|c|c|c|c|c|}
				\hline\hline
				4FGL Name & Source \# & \(R\) & \(R_H\) & \(T_D\) & \(L_D\) & \(B\) & \(\delta\) & \(\gamma_{\text{min}}\) & \(\gamma_{\text{max}}\) & \(\gamma_{\text{b}}\) & \(N\) & \(p_1\) & \(p_2\) \\
				& & \({10}^{16} \text{ cm}\) & \({10}^{17} \text{ cm}\) & \({10}^4 \text{ k}\) & \({10}^{44} \text{erg}/\text{s}\) & G &  &  &  &  & \({\text{cm}}^{-3}\) & & \\
				\hline
				4FGL J1346.5+5330 & 2 & 0.2 & - & - & - & 1.9 & 7.7 & \(7.9 \times {10}\) & \(3.0 \times {10}^7\) & \(4.3 \times {10}^3\) & \(4.3 \times {10}^3\) & 2.8 & 6.1\\
				4FGL J2109.3+3531 & 3 & 22.6 & 0.2 & \(3.0\) & \(8.4\) & 2.4 & 1.2 & \(2.73 \times {10}\) & \(2.2 \times {10}^4\) & \(9.0 \times {10}^2\) & \(3.7 \times {10}\) & 1.3 & 3.4 \\
				\hline\hline
			\end{tabular}
		}
		\tablecomments{For each UFO we show the 4FGL Name, the soft X-ray source number as shown in Figure \ref{fig:xraymap} (Source \#), the radius of the emitting source (\(R\)), the distance of the emitting region from the central black hole (\(R_H\)), the accretion disk temperature (\(T_D\)), the accretion disk luminosity (\(L_D\)), the magnetic field (\(B\)), the beaming factor (\(\delta\)), the minimum and maximum energy of the electron energy distribution (\(\gamma_{\text{min}}\) and \(\gamma_{\text{max}}\)), the break energy of the electron energy distribution (\(\gamma_{\text{b}}\)), the  electron density in the emitting region (\(N\)), and the slopes of the electron energy distribution (\(p_1\) and \(p_2\)). For source 2 in 4FGL J1346.5+5330 field we find the SED to be adequately represented by a synchrotron self-Compton, with additional emission from the host galaxy. For source 3 in 4FGL J2109.3+3531 field we find the SED to be adequately represented by an external Compton emission.}
	\end{table}

	\begin{figure*}
		\centering
		\includegraphics[scale=0.35]{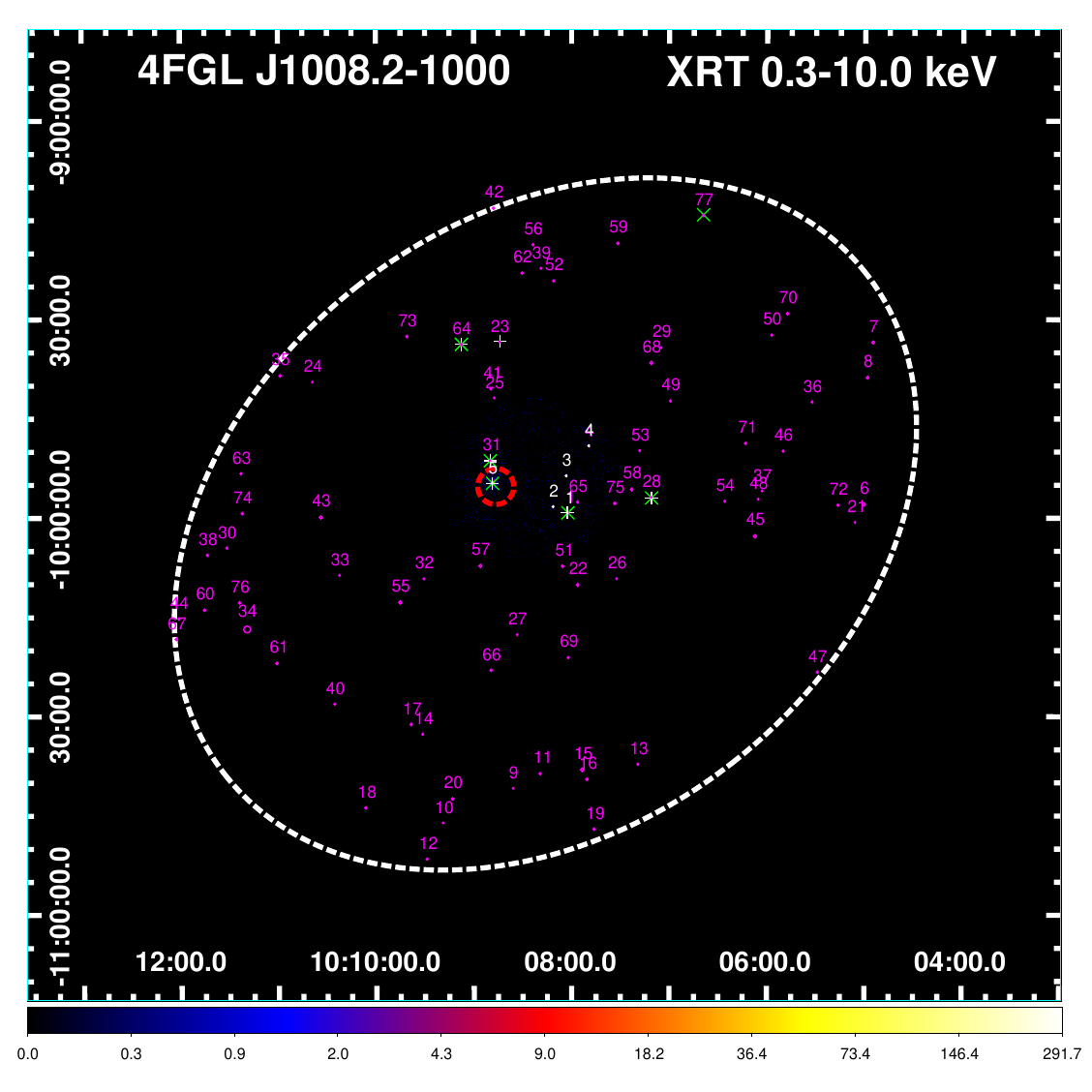}
		\includegraphics[scale=0.35]{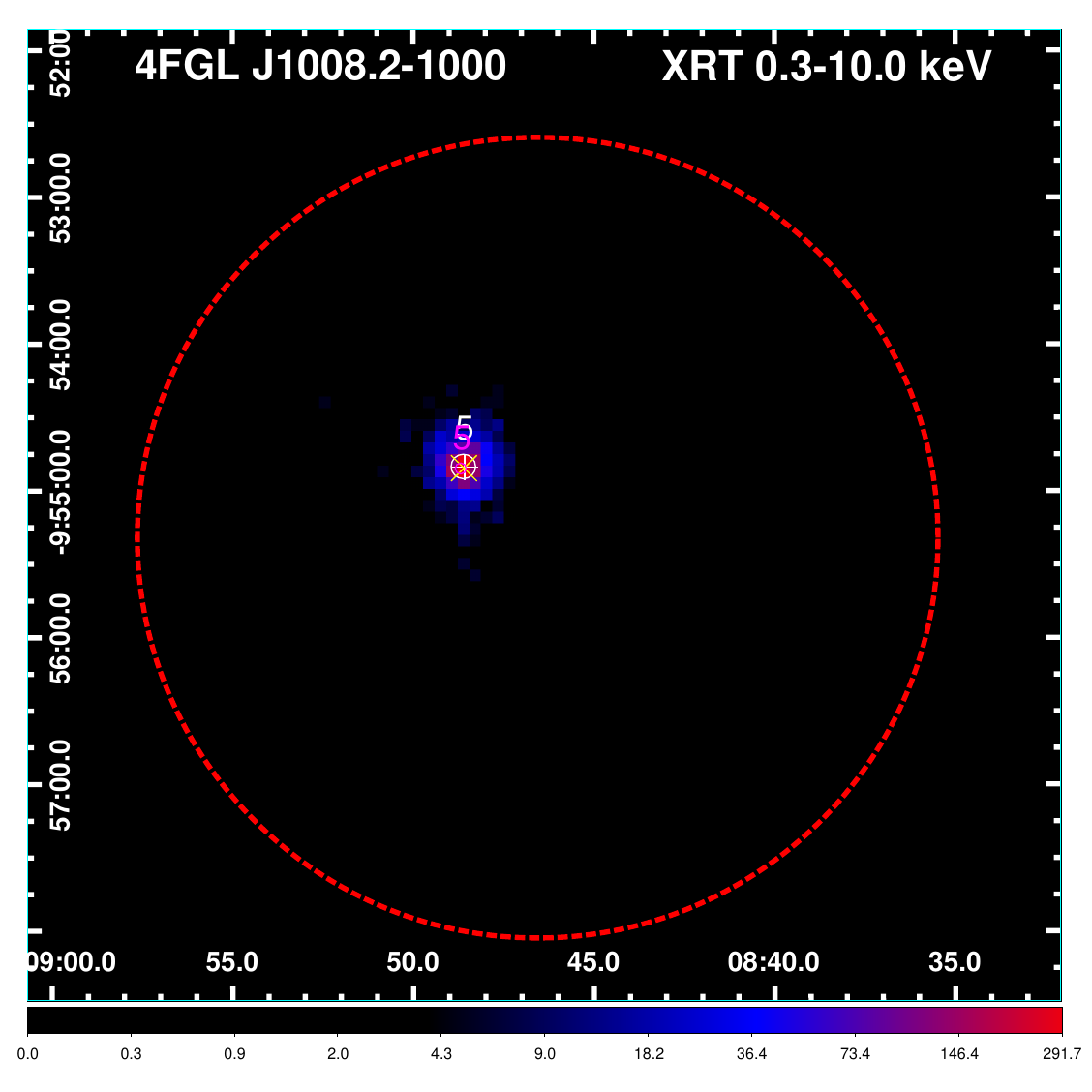}\\
		\includegraphics[scale=0.35]{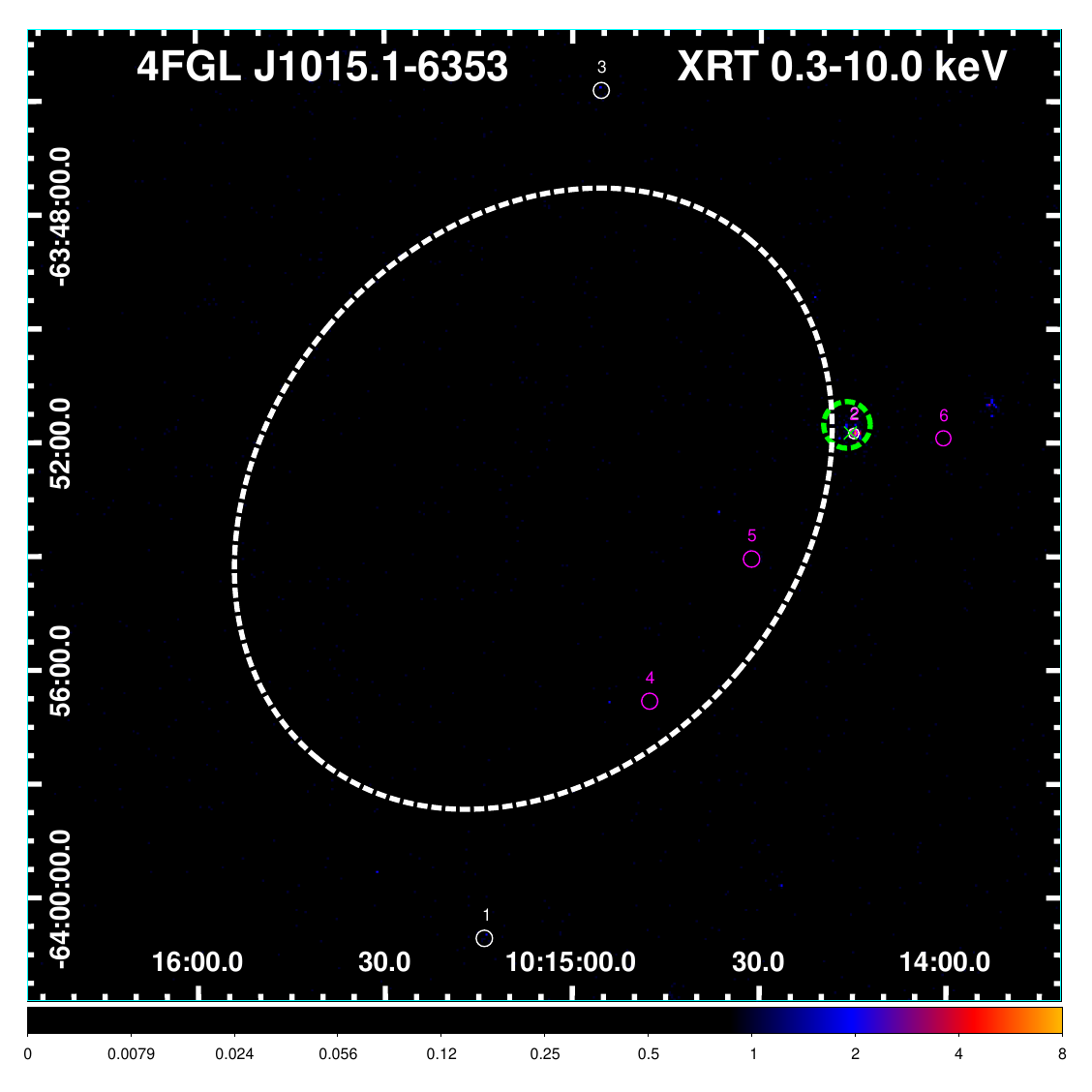}
		\includegraphics[scale=0.35]{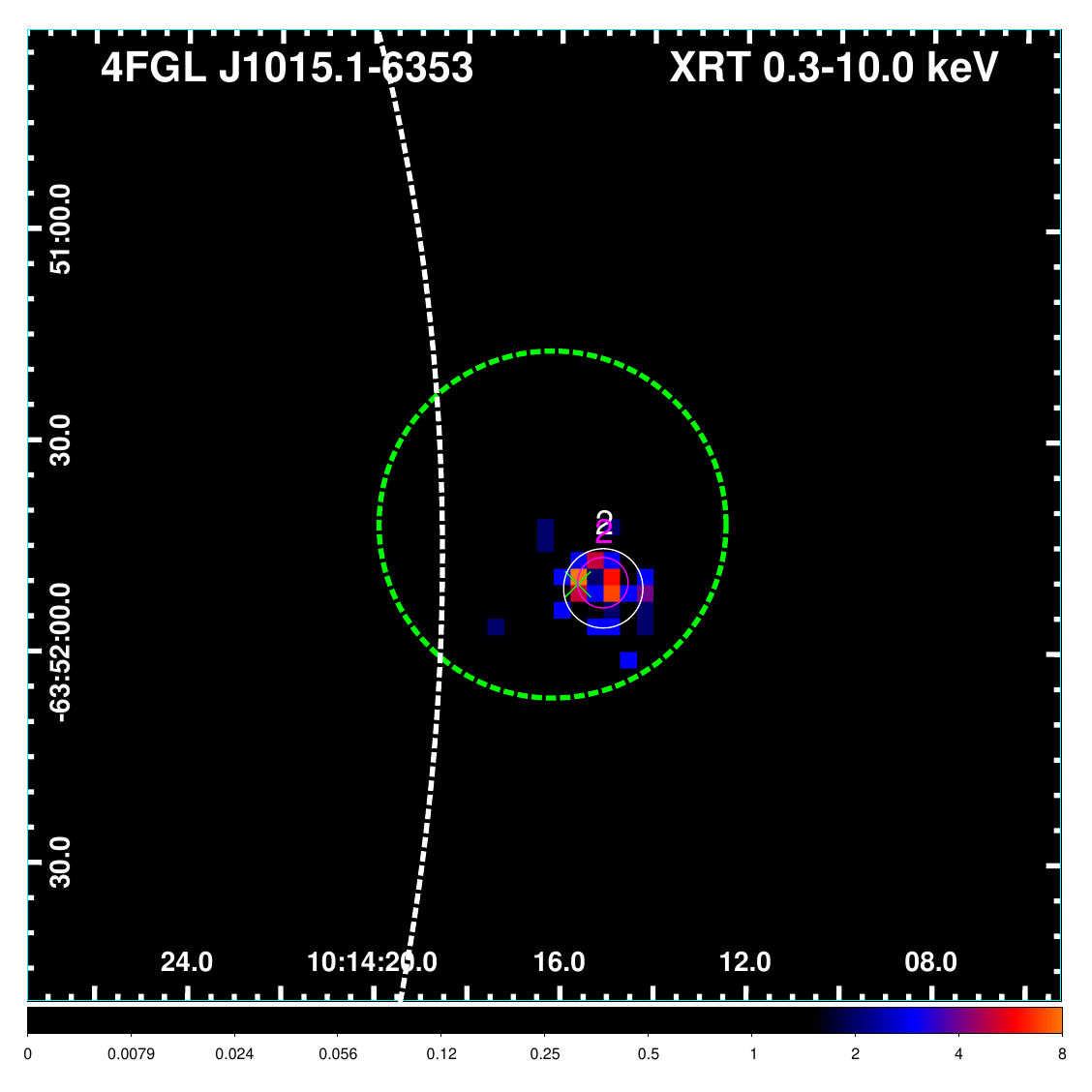}\\
		\includegraphics[scale=0.35]{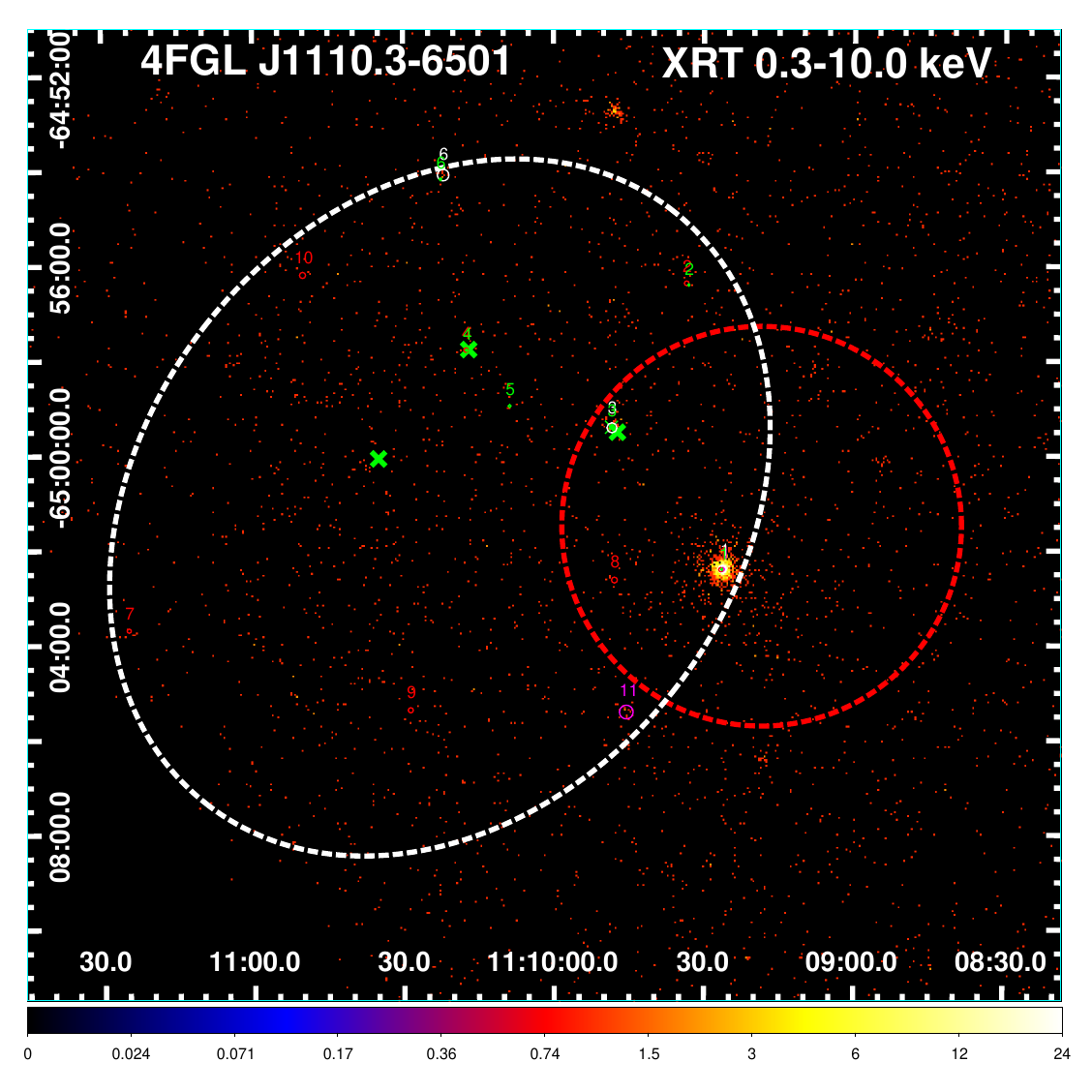}
		\includegraphics[scale=0.35]{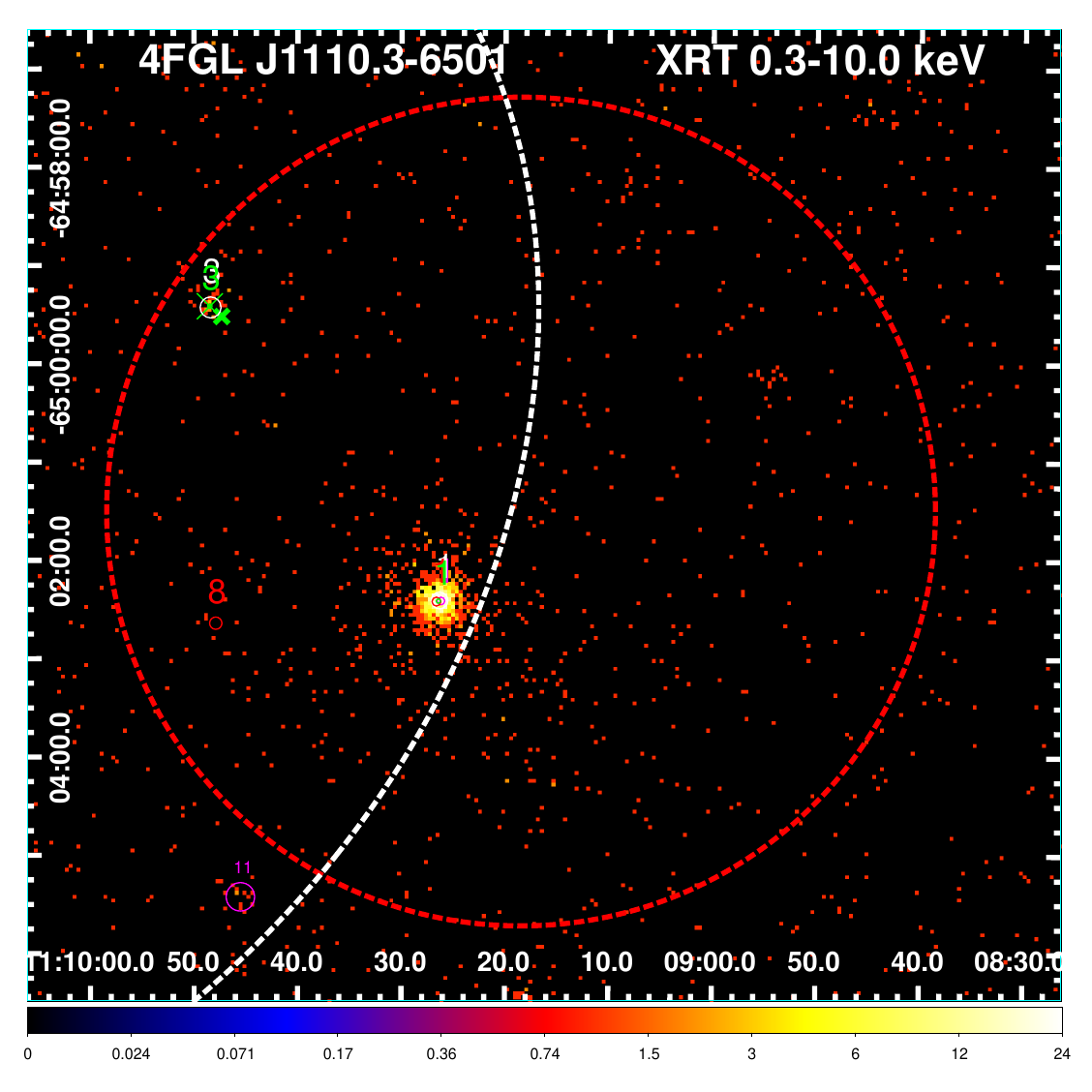}
		\caption*{Fig. 2: Continued.}
	\end{figure*}
	
	\begin{figure*}
		\centering
		\includegraphics[scale=0.35]{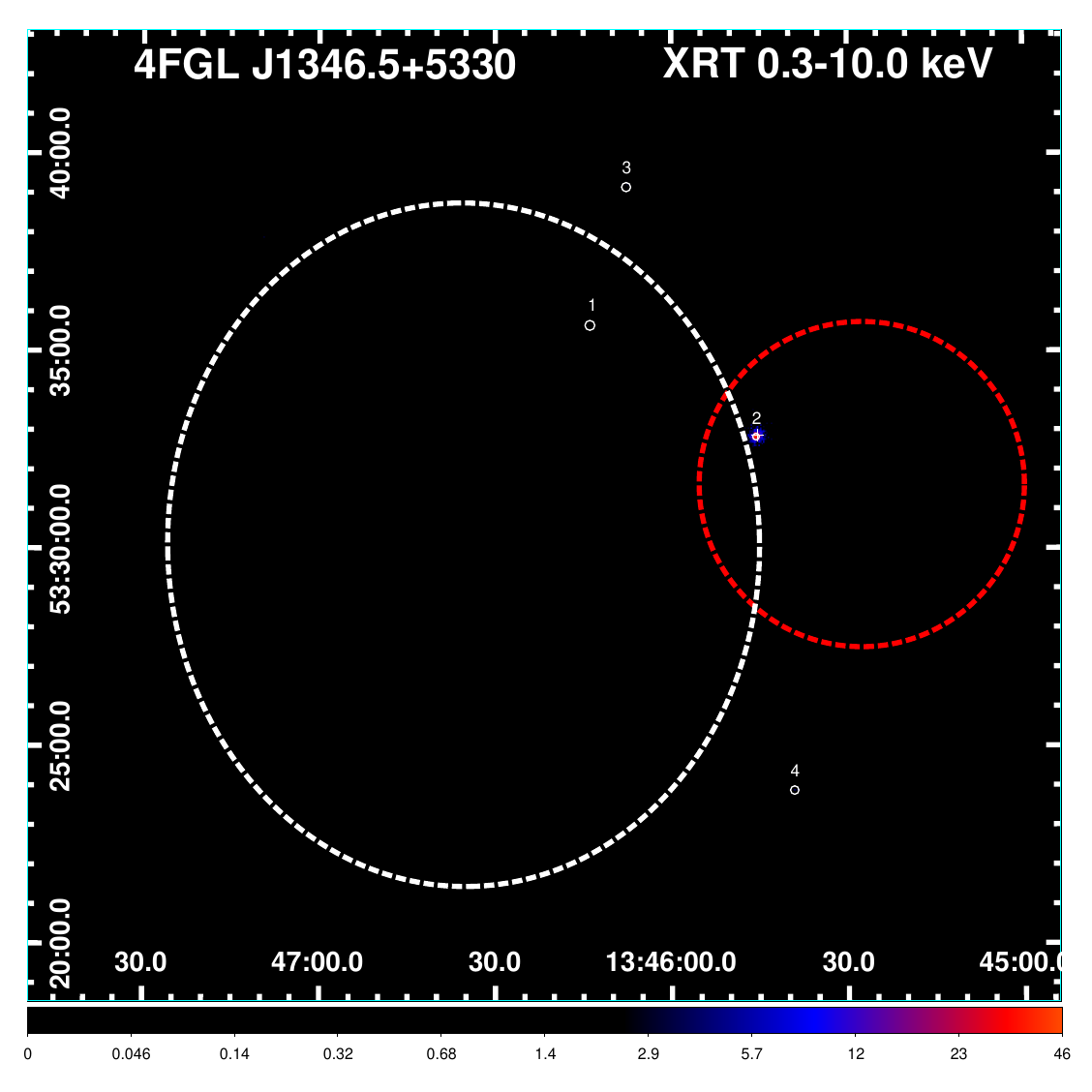}
		\includegraphics[scale=0.35]{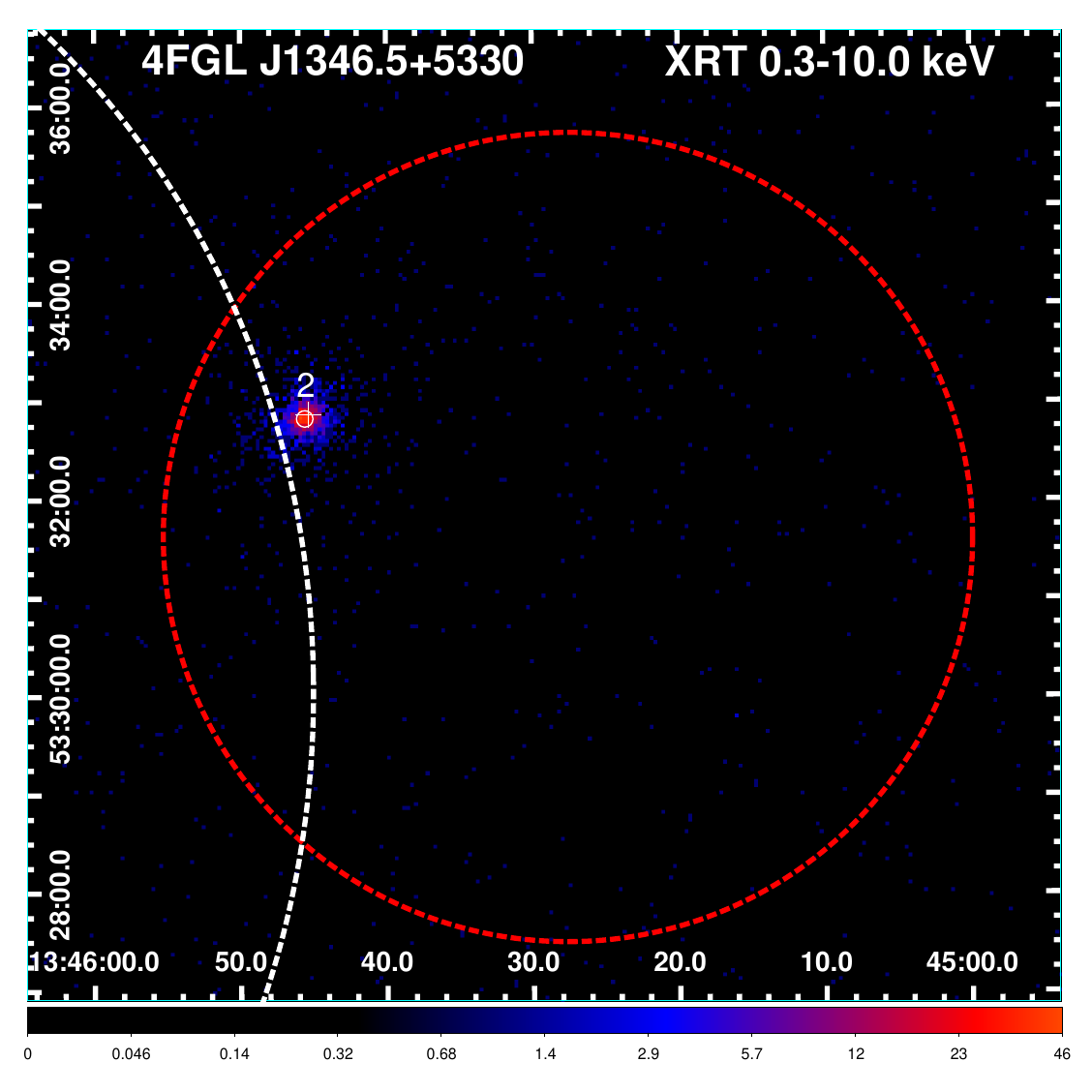}\\        
		\includegraphics[scale=0.35]{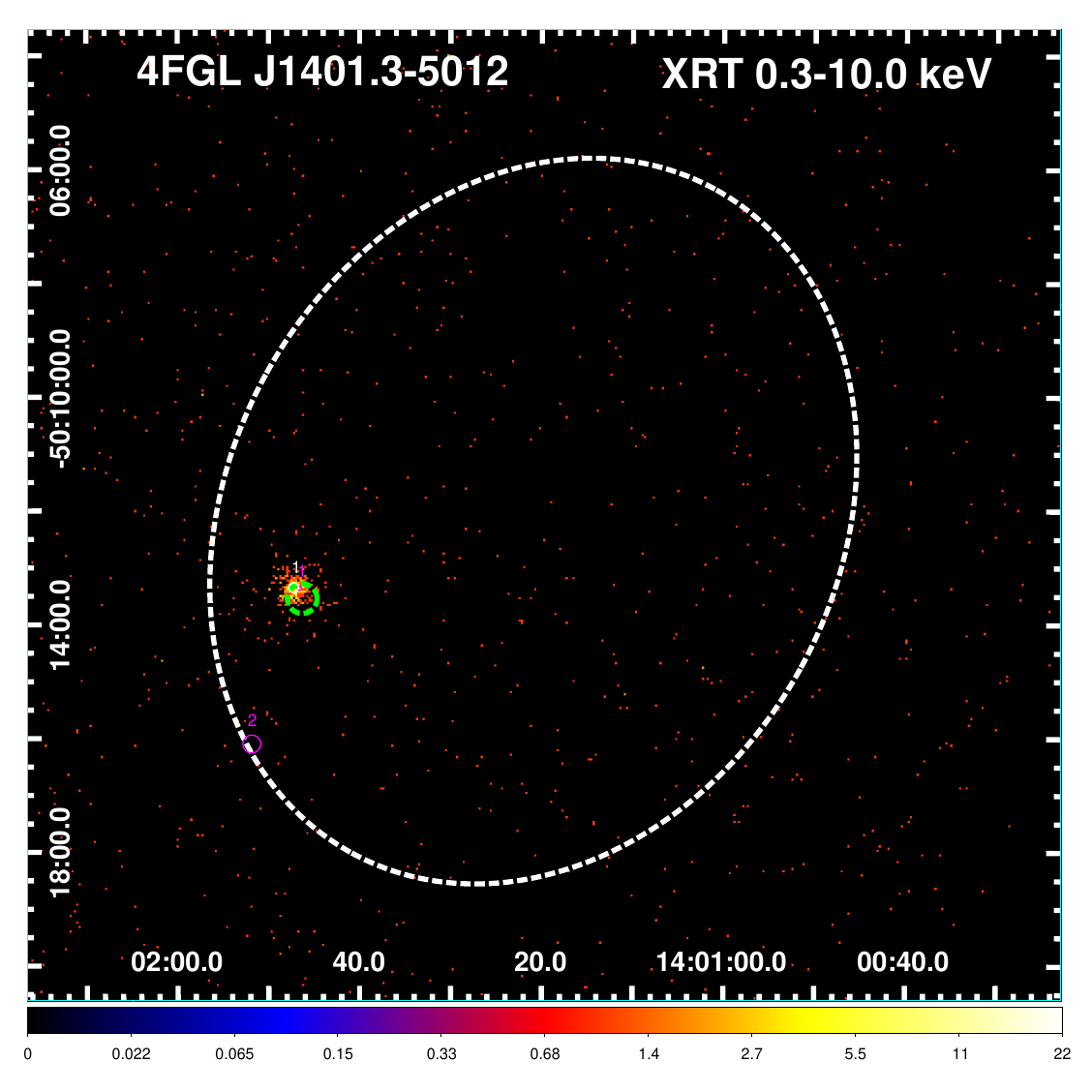}
		\includegraphics[scale=0.35]{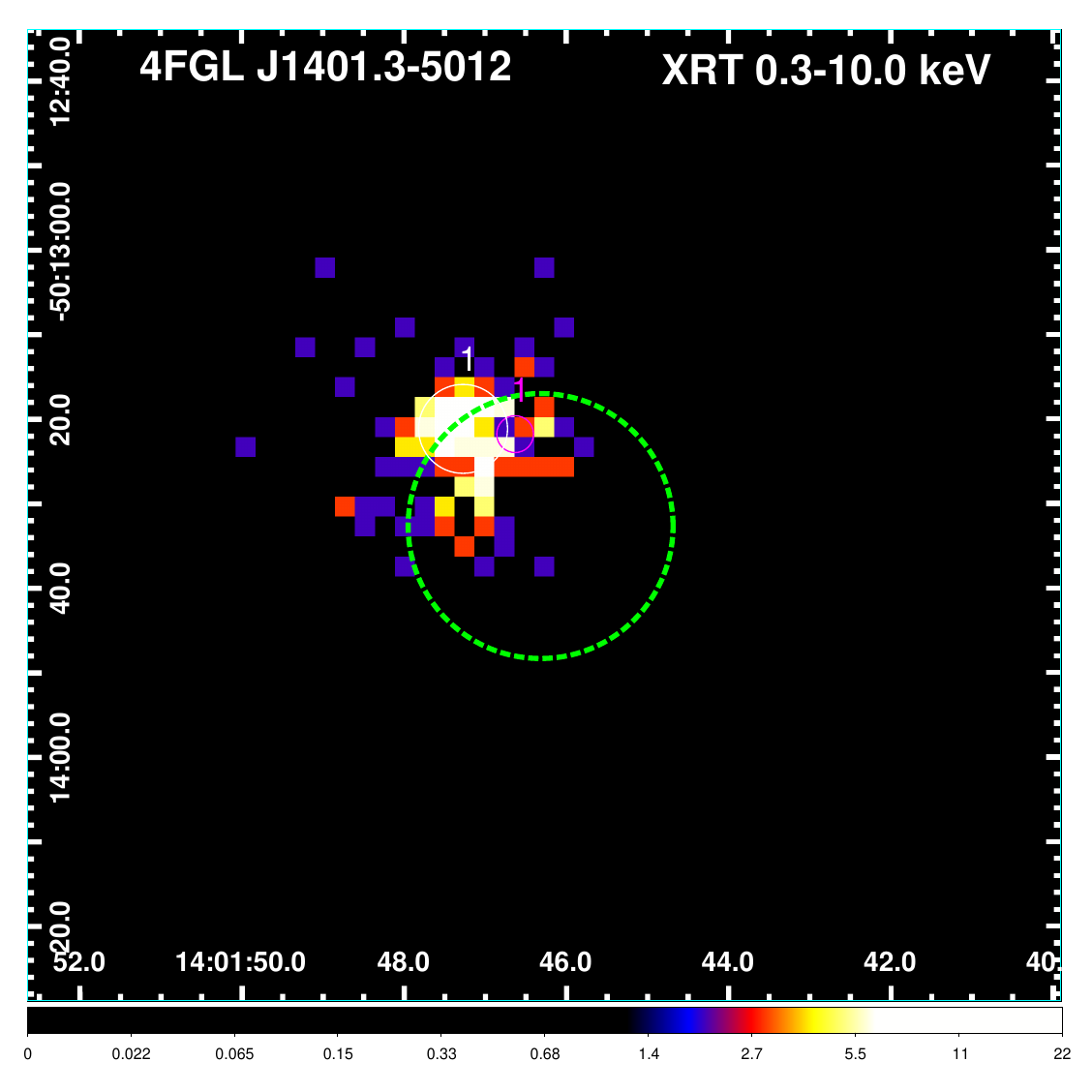}\\
		\includegraphics[scale=0.35]{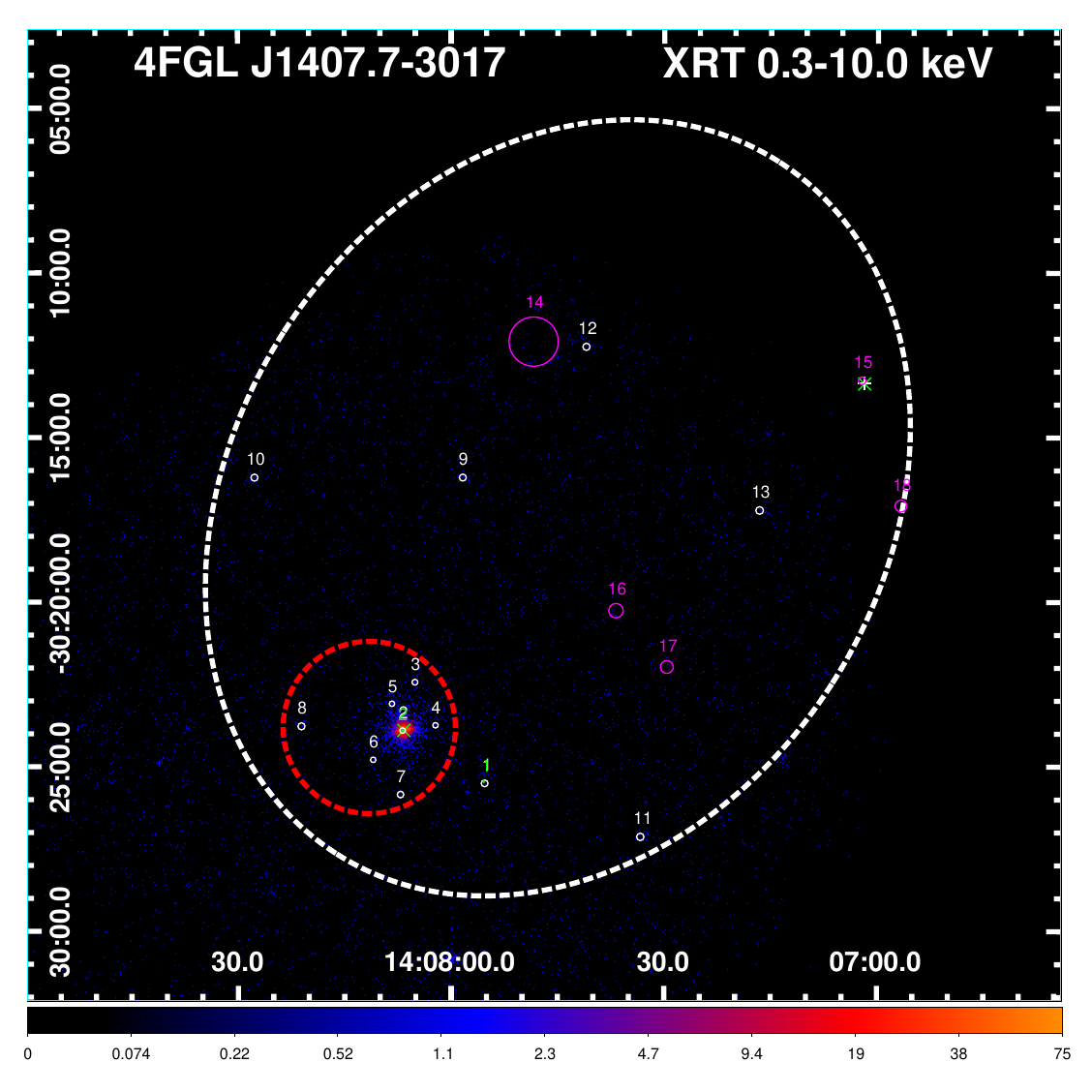}
		\includegraphics[scale=0.35]{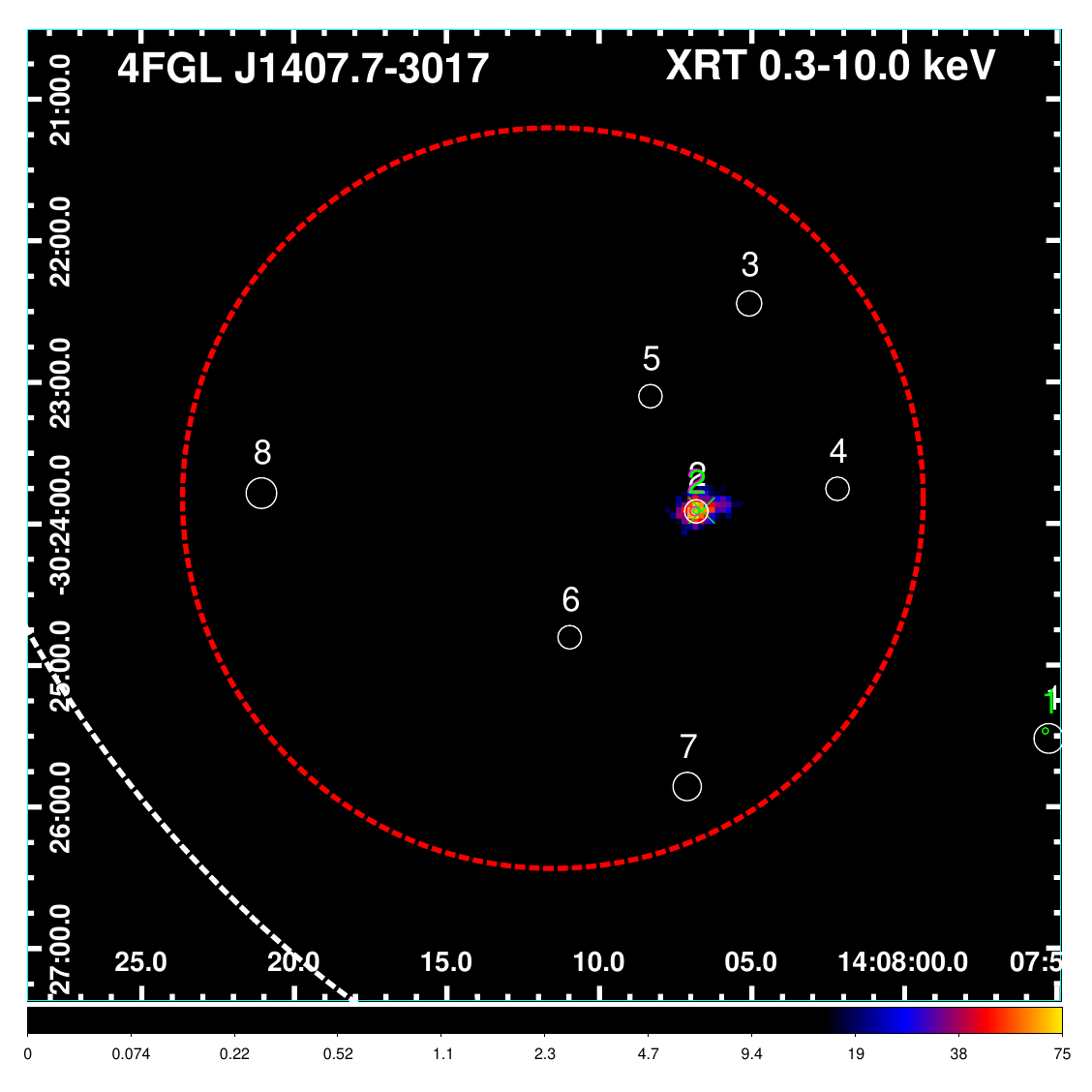}
		\caption*{Fig. 2: Continued.}
	\end{figure*}
	
	\begin{figure*}
		\centering
		\includegraphics[scale=0.35]{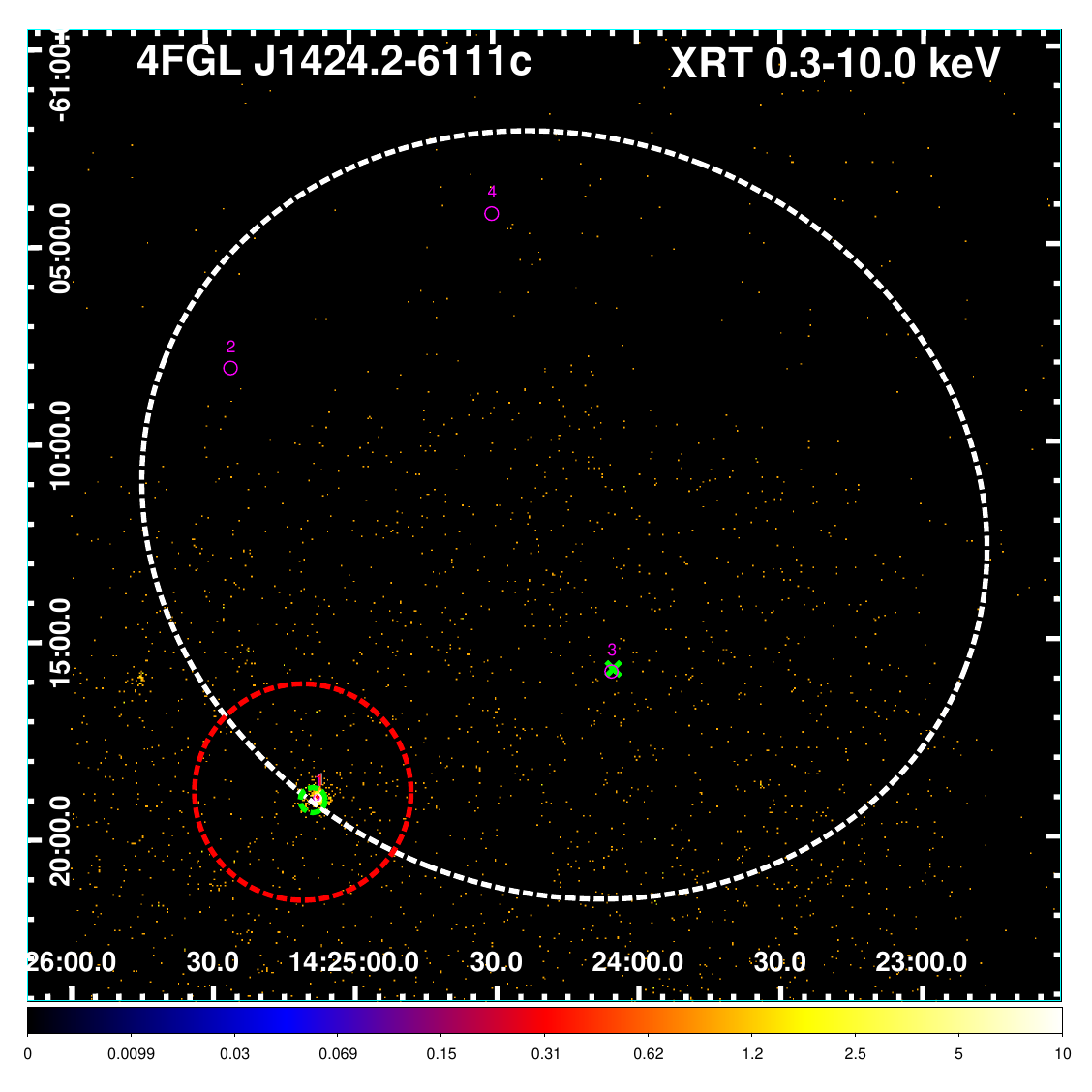}
		\includegraphics[scale=0.35]{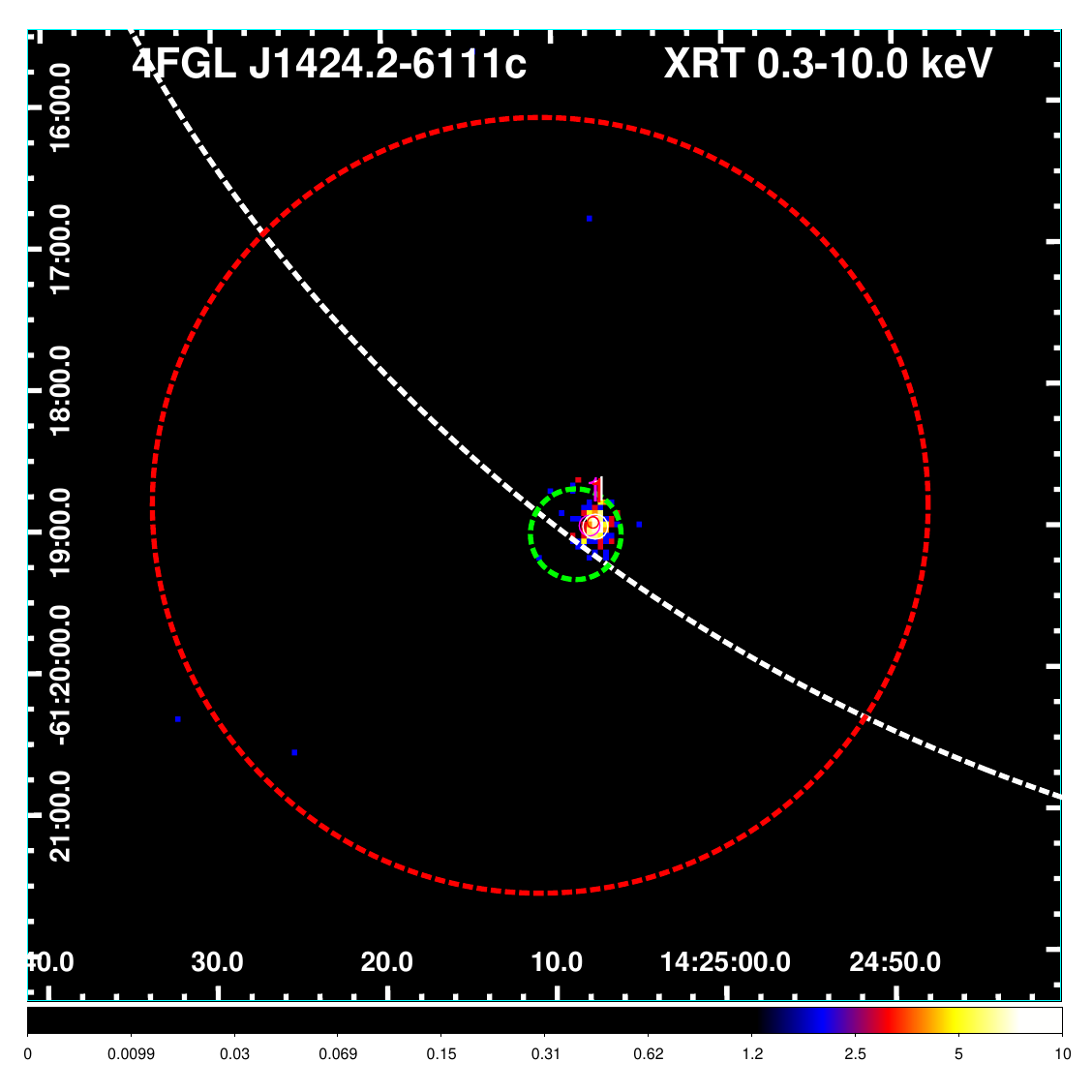}\\        
		\includegraphics[scale=0.35]{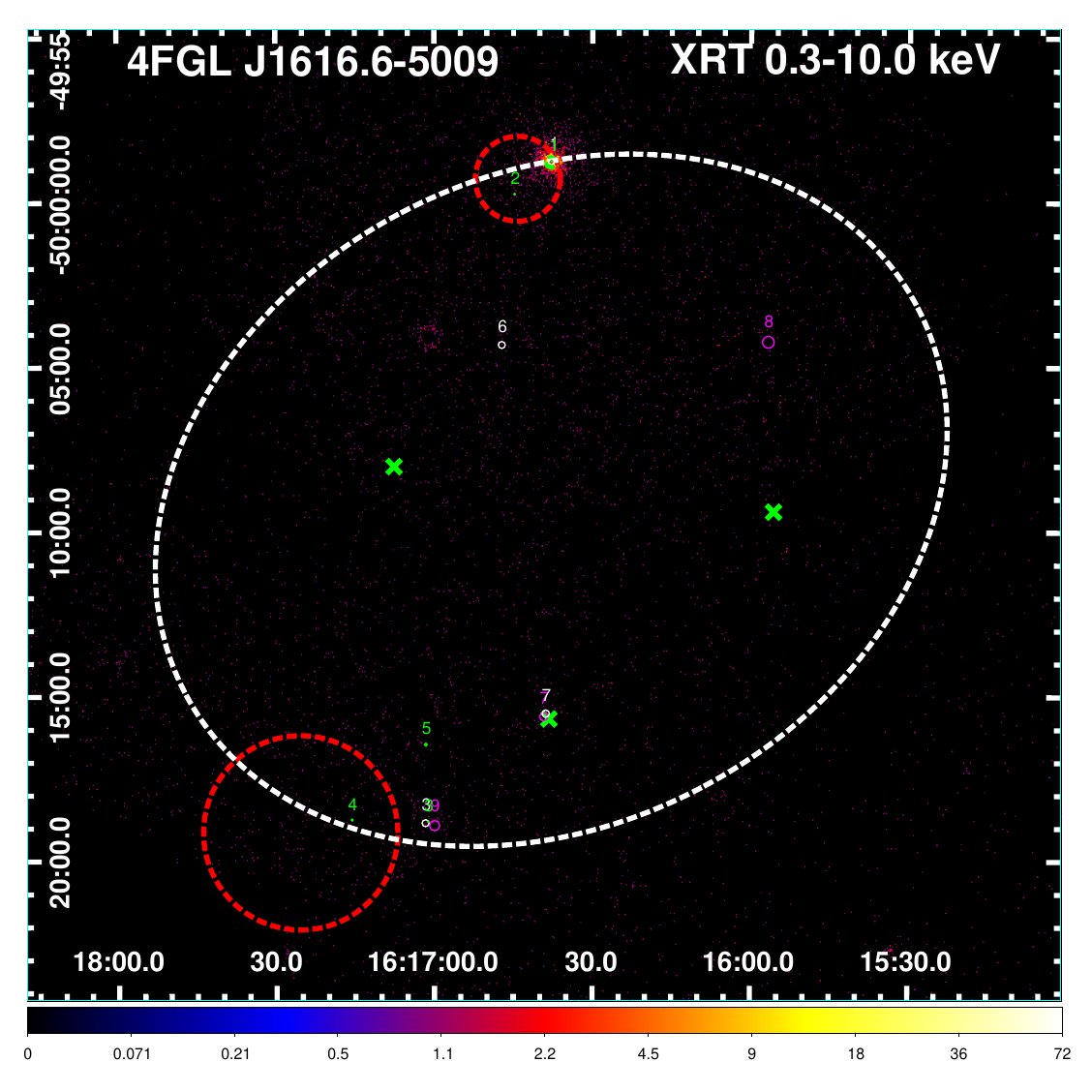}
		\includegraphics[scale=0.35]{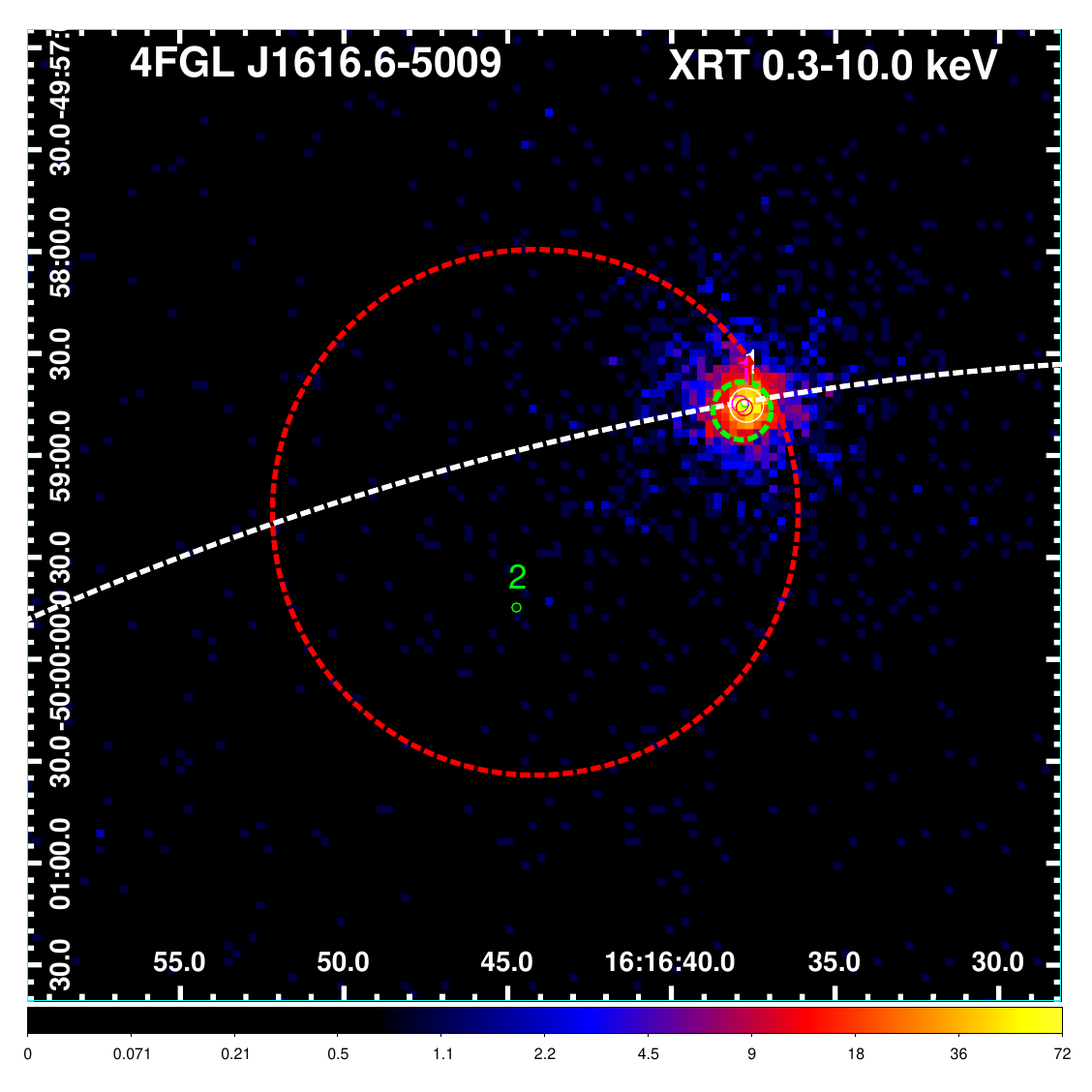}\\
		\includegraphics[scale=0.35]{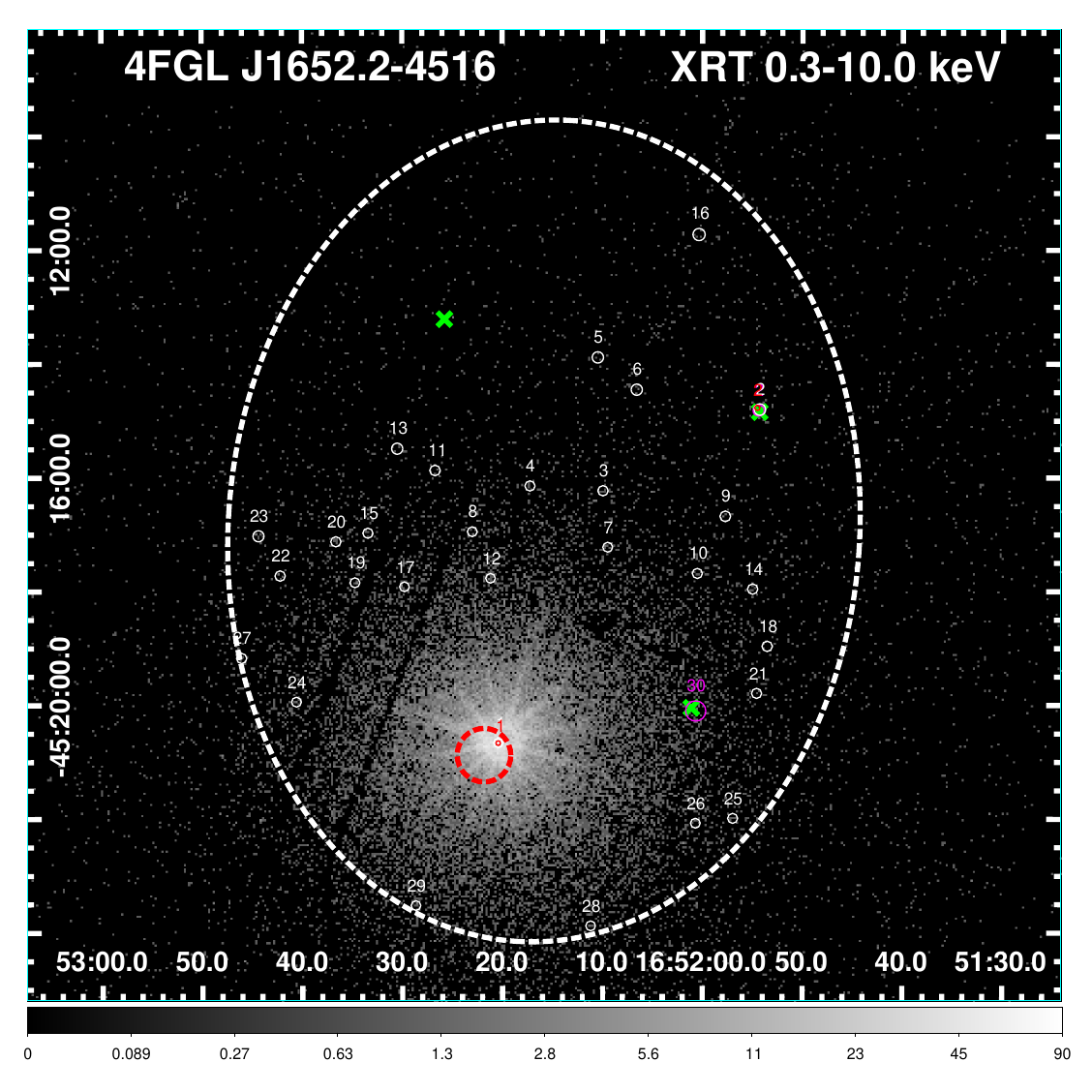}
		\includegraphics[scale=0.35]{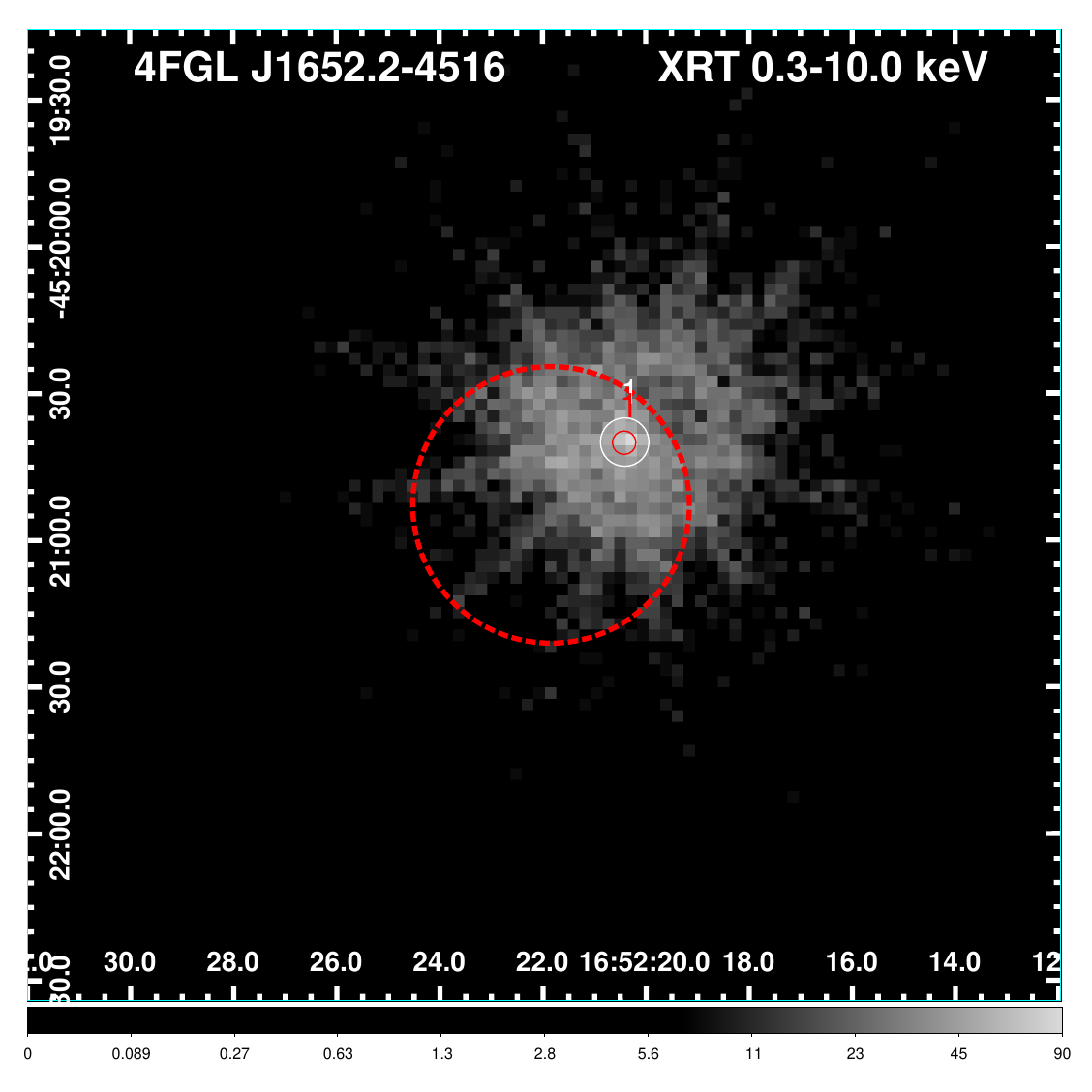}
		\caption*{Fig. 2: Continued.}
	\end{figure*}
	
	\begin{figure*}
		\centering
		\includegraphics[scale=0.35]{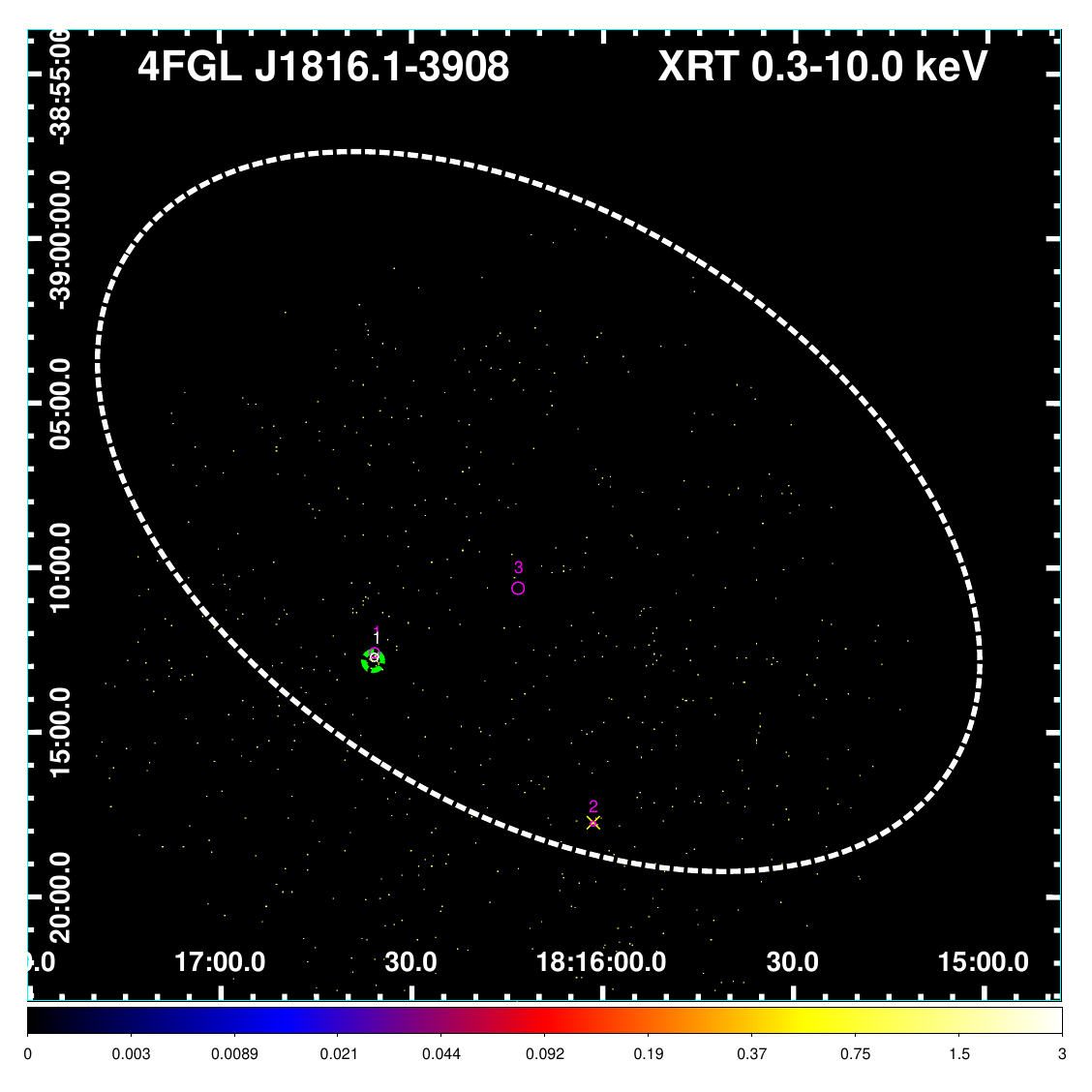}
		\includegraphics[scale=0.35]{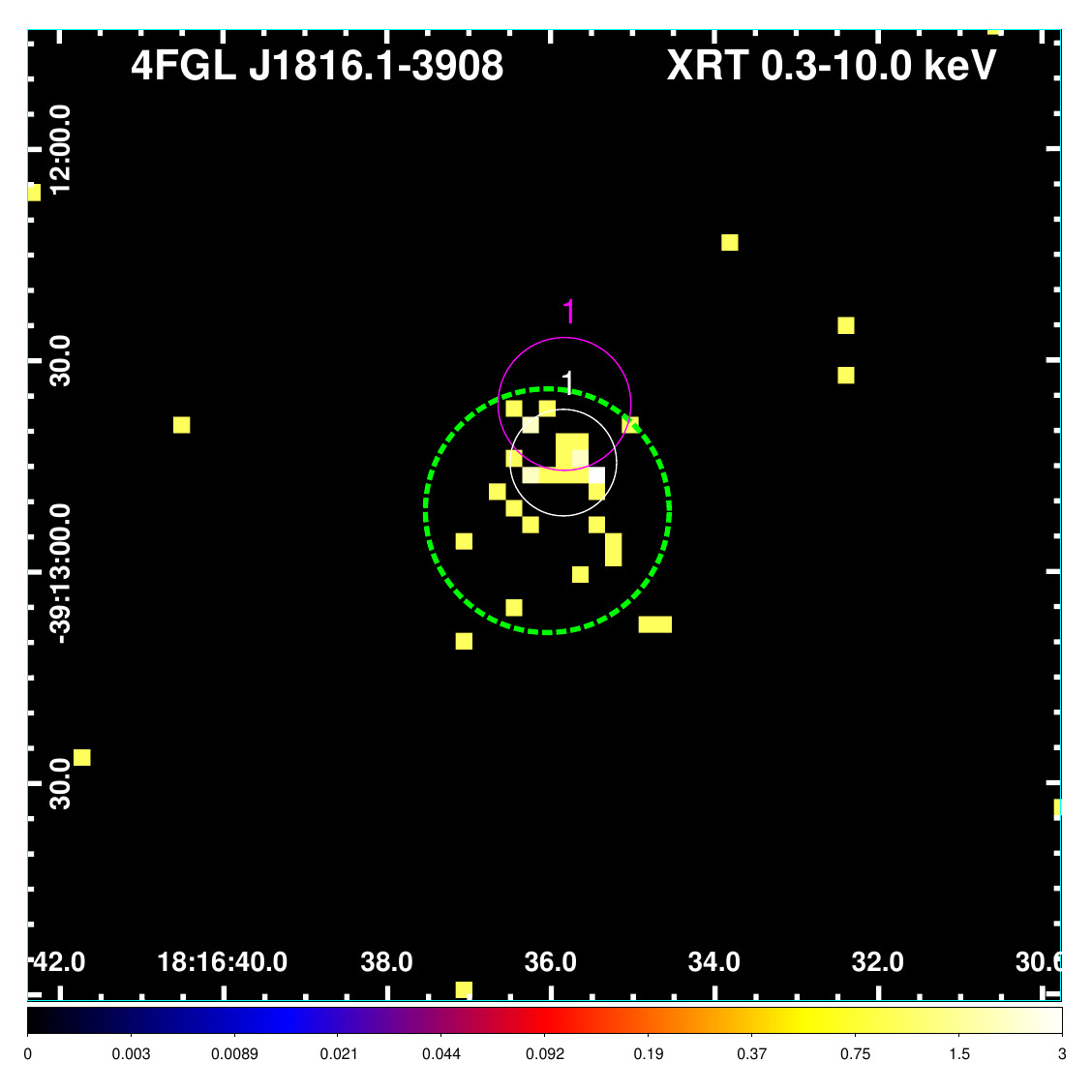}\\        
		\includegraphics[scale=0.35]{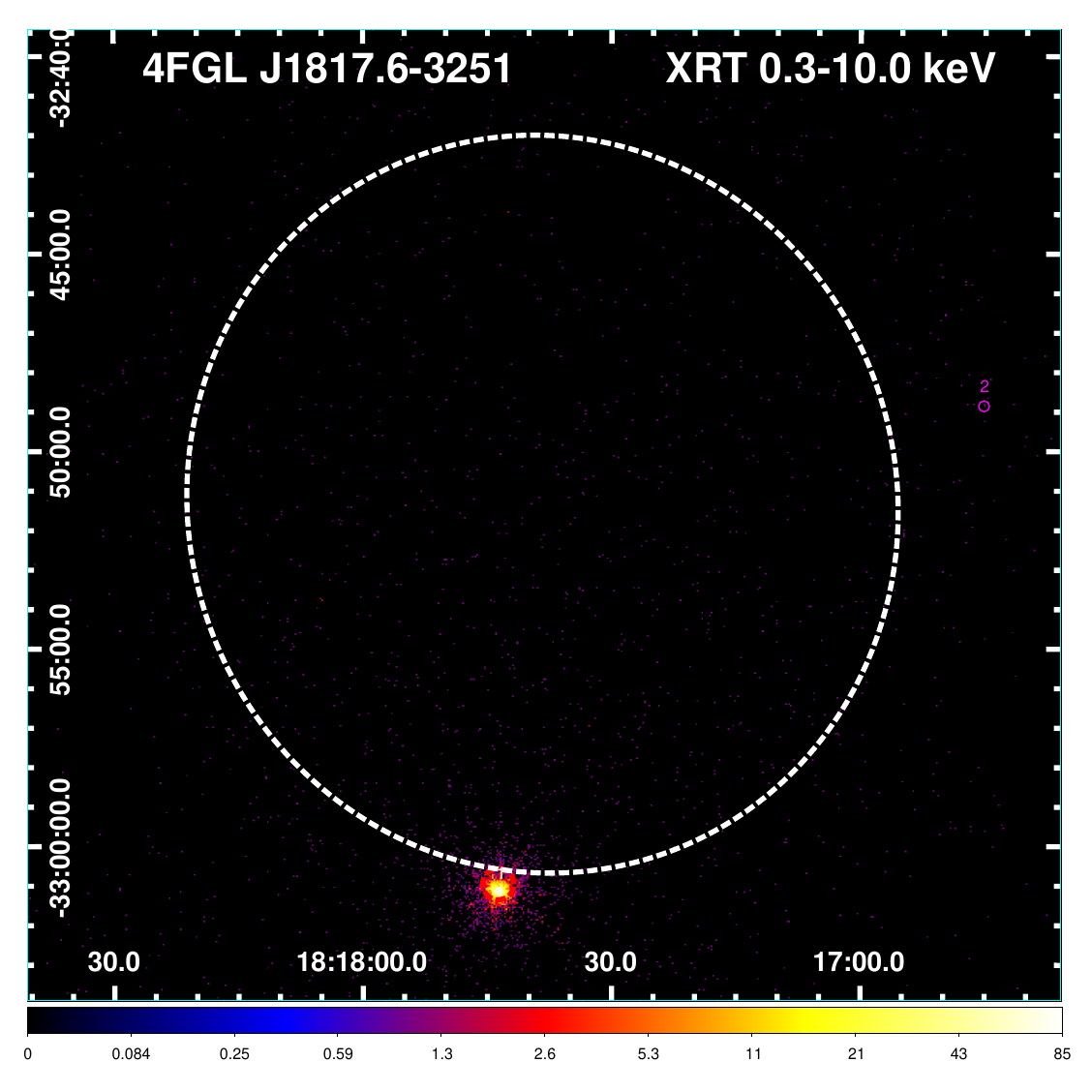}
		\includegraphics[scale=0.35]{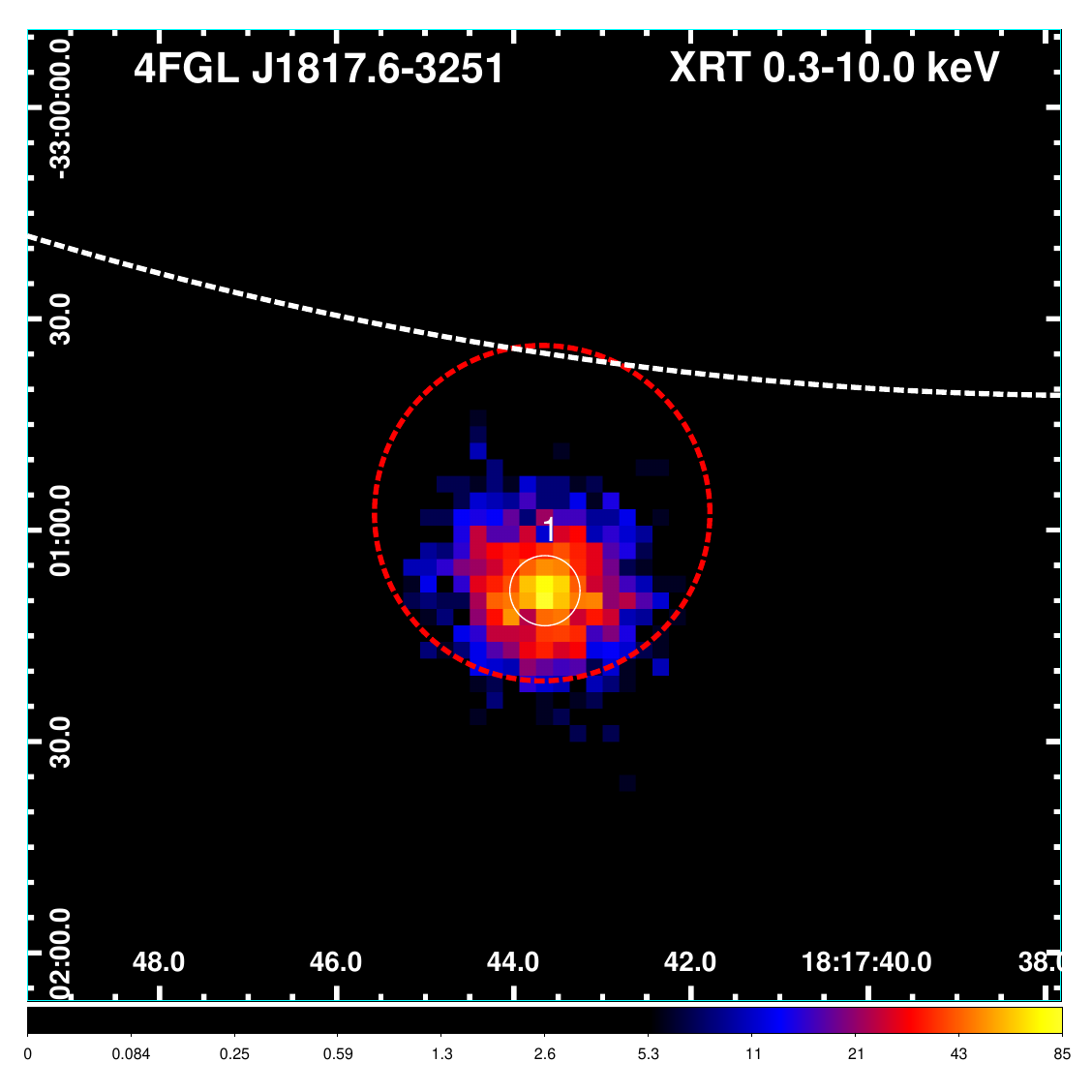}\\
		\includegraphics[scale=0.35]{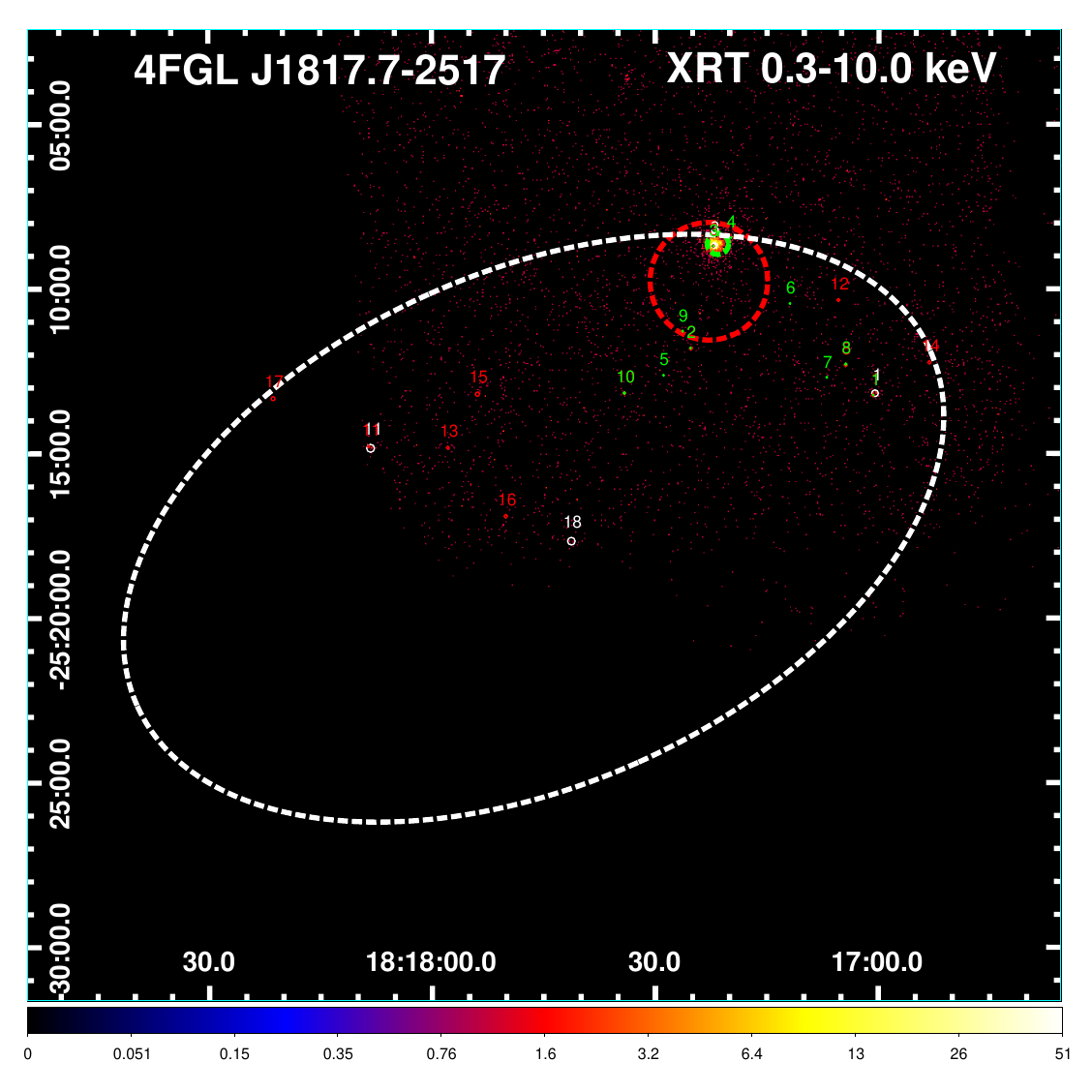}
		\includegraphics[scale=0.35]{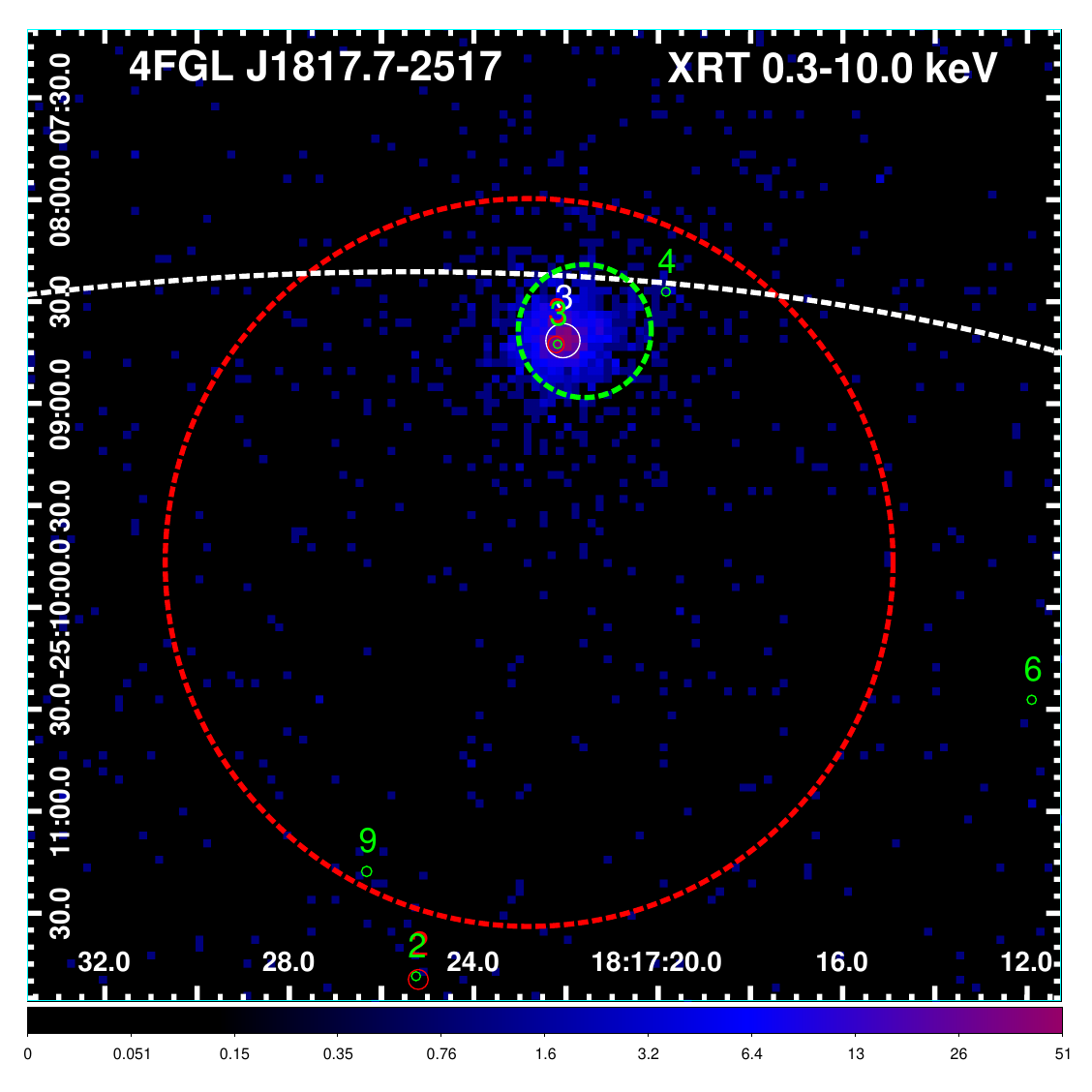}
		\caption*{Fig. 2: Continued.}
	\end{figure*}

	\begin{figure*}
		\centering
		\includegraphics[scale=0.35]{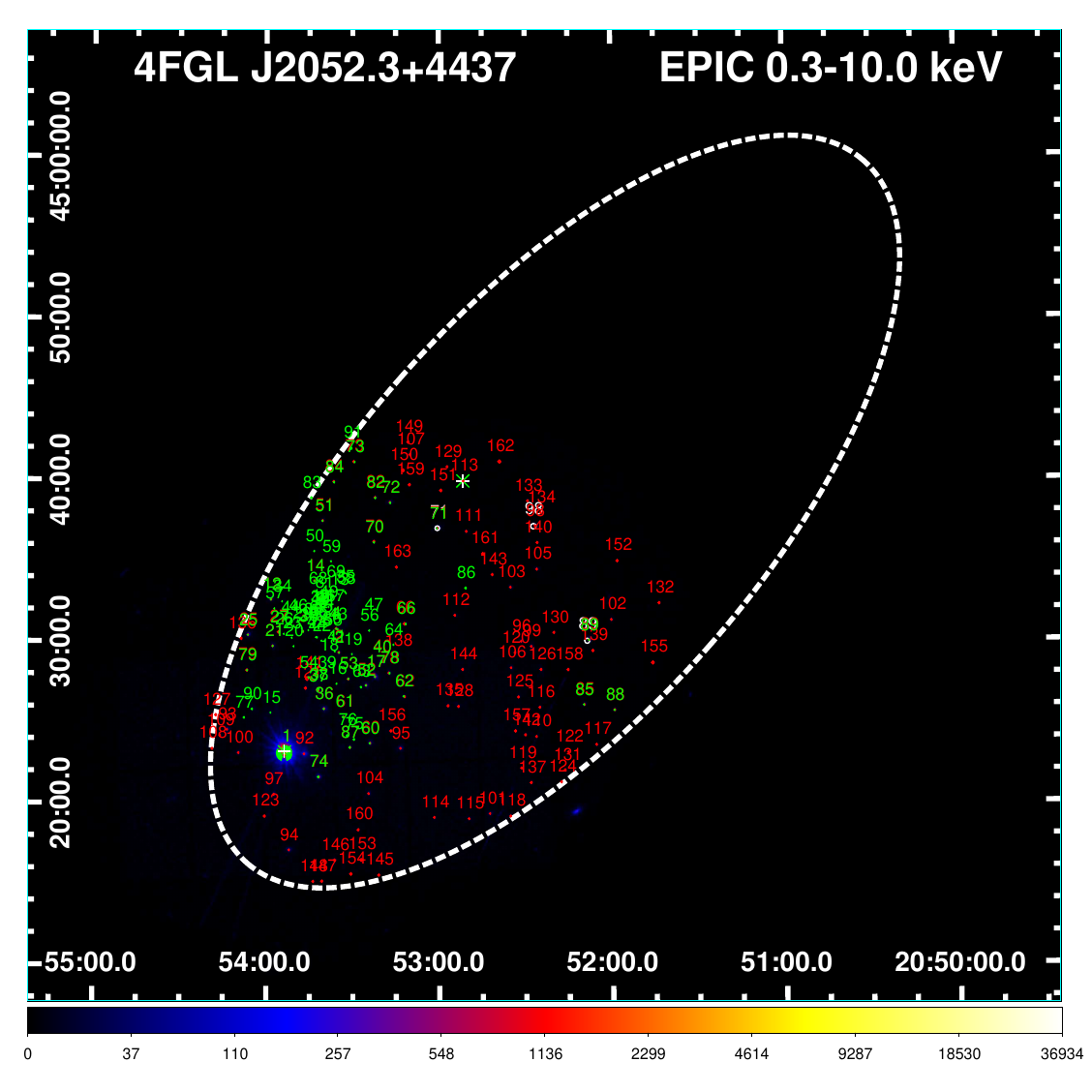}
		\includegraphics[scale=0.35]{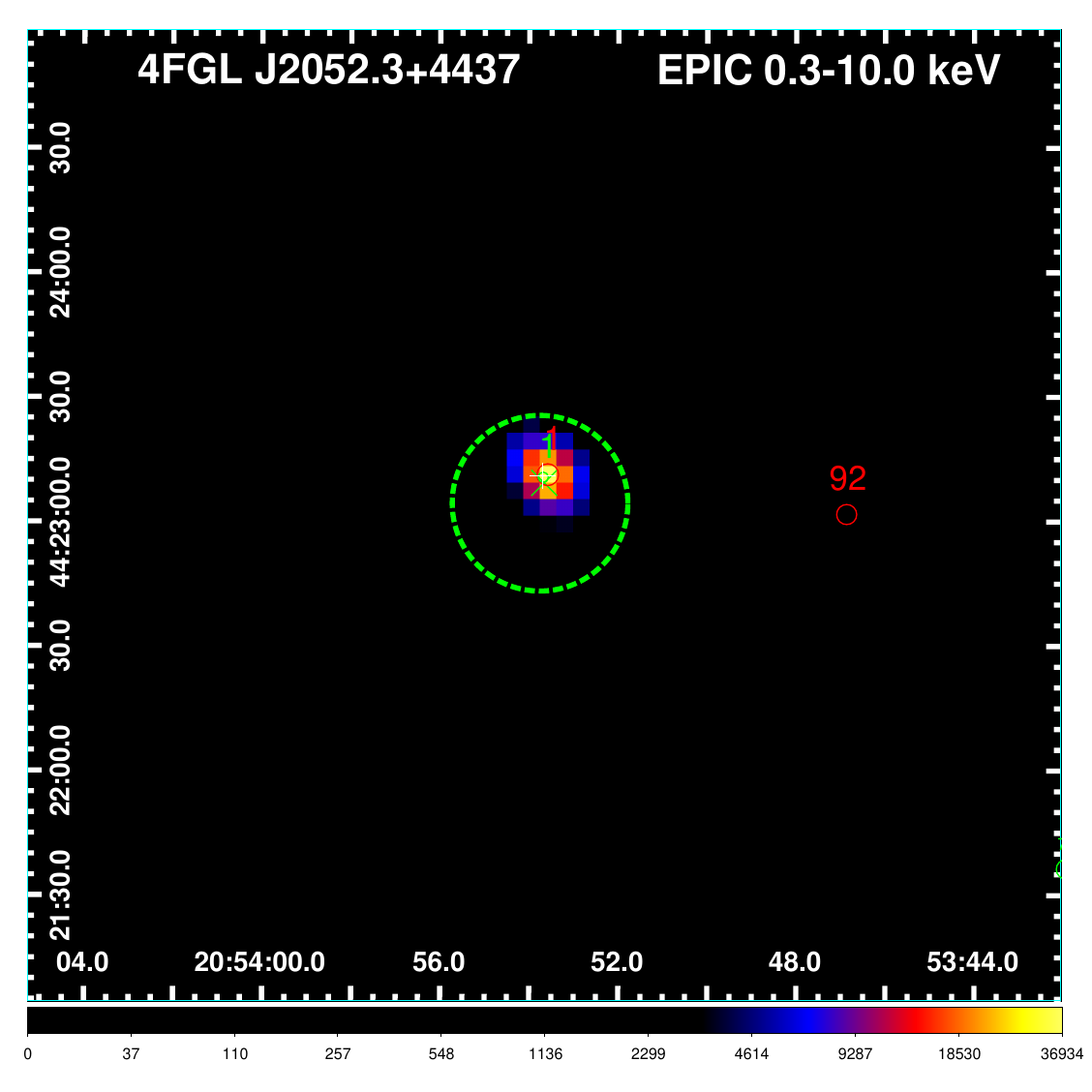}\\
		\includegraphics[scale=0.35]{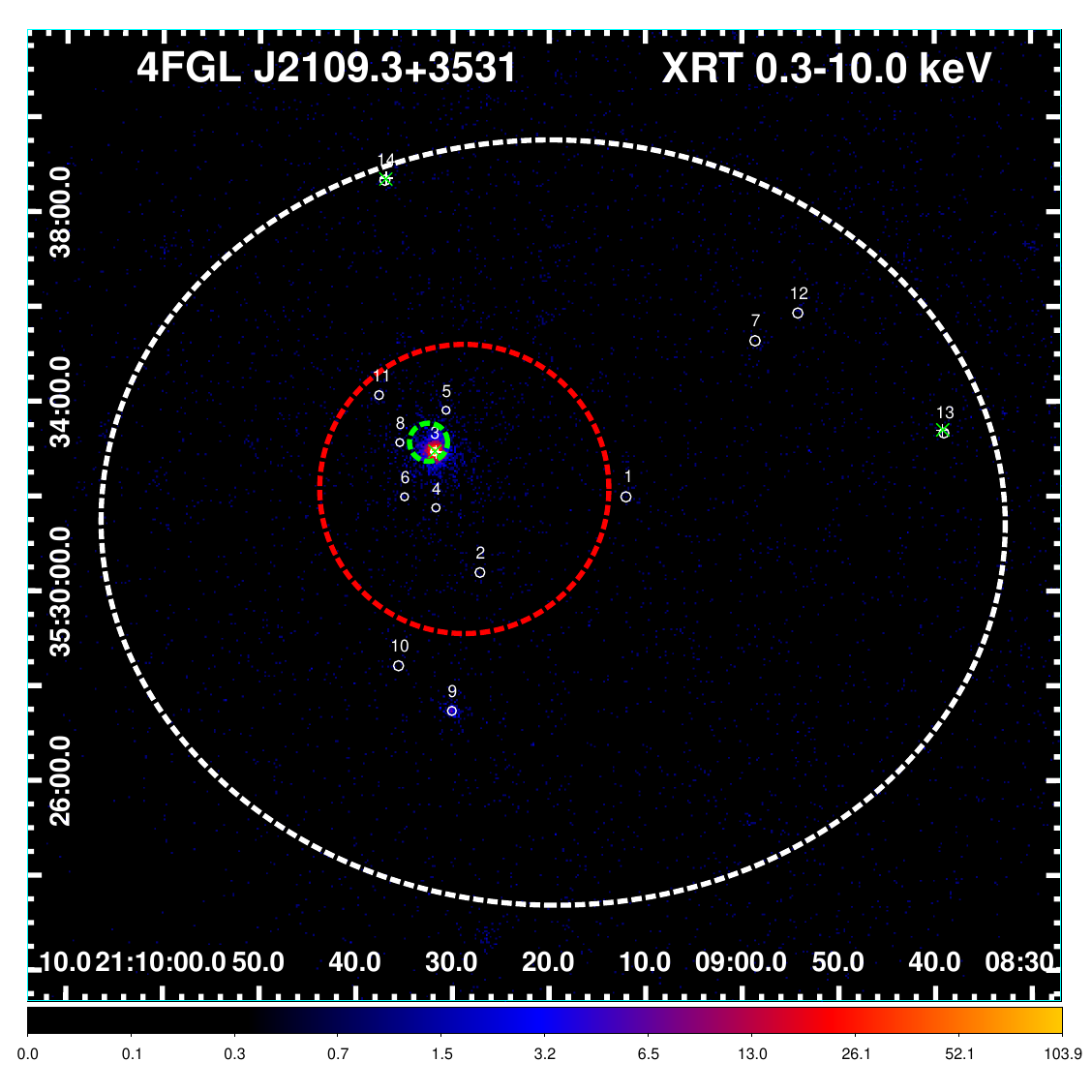}
		\includegraphics[scale=0.35]{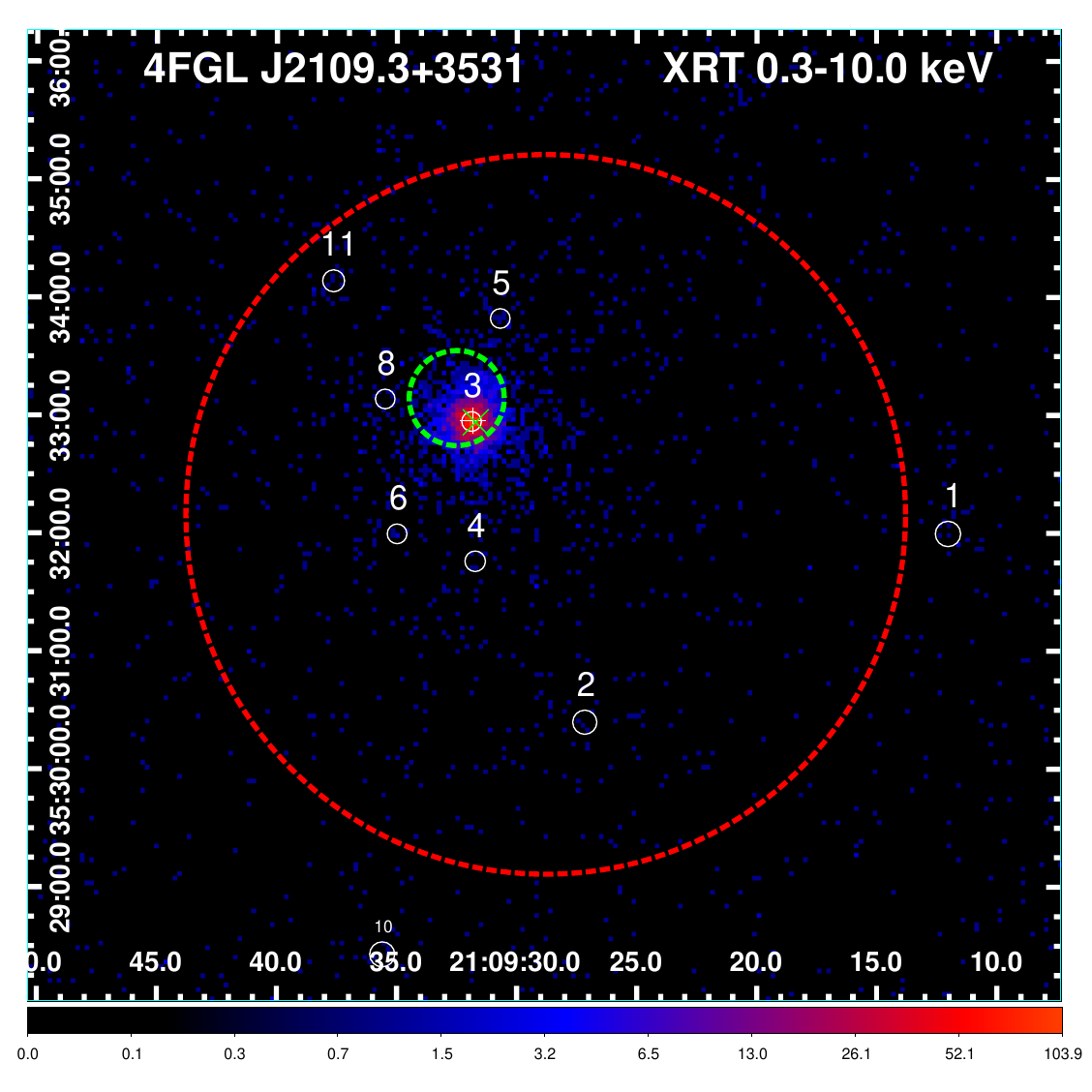}
		\caption*{Fig. 2: Continued.}
	\end{figure*}

\end{document}